\documentclass{jfm}

\usepackage{graphicx}
\usepackage{natbib}
\usepackage{amsmath}
\usepackage{gensymb}
\usepackage[dvipsnames]{xcolor}
\usepackage{multirow}
\usepackage{wasysym}
\usepackage{tikz}
\usepackage{float}
\begin{document}

\newtheorem{lemma}{Lemma}
\newtheorem{corollary}{Corollary}

\shorttitle{Numerical study of cavitation regimes in flow over a circular cylinder} %for header on odd pages
\shortauthor{F. L. Brandao et al.} %for header on even pages

\title{Numerical study of cavitation regimes in flow over a circular cylinder}

\author
 {
 Filipe L. Brandao\aff{1}
  , 
  Mrugank Bhatt\aff{1}
  \and 
  Krishnan Mahesh\aff{1}
  \corresp{\email{kmahesh@umn.edu}}}

\affiliation
{
\aff{1}
Department of Aerospace Engineering and Mechanics, University of Minnesota, Minneapolis, MN 55455, USA
}

\maketitle

\begin{abstract}

Cavitating flow over a circular cylinder is investigated over a range of cavitation numbers ($\sigma=5$ to $0.5$) for both laminar (at $Re=200$) and turbulent (at $Re=3900$) regimes.
We observe non--cavitating, cyclic and transitional cavitation regimes with reduction in freestream $\sigma$.
The cavitation inside the K\'arm\'an vortices in the cyclic regime, is significantly altered by the onset of ``condensation front" propagation in the transitional regime. At the transition, an order of magnitude jump in shedding Strouhal number is observed as the dominant frequency shifts from periodic vortex shedding in the cyclic regime, to irregular--regular vortex shedding in the transitional regime. In addition, a peak in pressure fluctuations, and a maximum in $St$ versus $\sigma$ based on cavity length are observed at the transition. Shedding characteristics in each regime are discussed using dynamic mode decomposition (DMD). A numerical method based on the homogeneous mixture model, fully compressible formulation and finite rate mass transfer developed by Gnanaskadan \& Mahesh (\textit{Intl. J. Multiphase Flows}, vol. 70, 2015, pp. 22--34) is extended to include the effects of non--condensable gas (\textit{NCG}). 
It is demonstrated that the condensation fronts observed in the transitional regime are supersonic (referred as ``condensation shocks''). In the presence of \textit{NCG}, multiple condensation shocks in a given cycle are required for complete cavity condensation and detachment, as compared to a single condensation shock when only vapor is present. This is explained by the reduction in pressure ratio across the shock in the presence of \textit{NCG} effectively reducing its strength. In addition, at $\sigma=0.85$ (near transition from the cyclic to the transitional regime), presence of \textit{NCG} suppresses the low frequency irregular--regular vortex shedding. 
Vorticity transport at $Re=3900$, in the transitional regime, indicates that the region of attached cavity is nearly two--dimensional with very low vorticity, affecting K\'arm\'an shedding in the near wake. Majority of vortex stretching/tilting and vorticity production is observed following the cavity trailing edge. In addition, the boundary--layer separation point is found to be strongly dependent on the amounts of vapor and gas in the freestream for both laminar and turbulent regimes. 

\end{abstract}

\section{Introduction}\label{intro}

Cavitation refers to phase change of liquid into vapor as the liquid pressure drops below vapor pressure. It is often encountered in hydrodynamic applications such as marine propulsors, hydrofoils and rotating turbomachinery. Cavitation can be a major source of noise, vibrations and material damage in such systems. 
\cite{Fry} characterizes cavitation over a circular cylinder as belonging to cyclic and transitional regimes. As the cavitation number ($\sigma = \frac{p_{\infty} - p_v}{\frac{1}{2}\rho_{\infty}U_{\infty}^2}$, where $p_{\infty}$, $\rho_{\infty}$ and $U_{\infty}$ are pressure, density and velocity in the freestream respectively) in the freestream is sufficiently dropped, cavities develop inside the core of the vortices shed from either side of the cylinder, which is referred to as cyclic cavitation. With further reduction in $\sigma$, in the transitional regime, the cavities grow larger in size and begin to interact with each other, causing the cavity shedding to become irregular, until they coalesce to form a single cavity fixed to the cylinder. It is observed that noise and erosion rates peak at the transition from the cyclic to the transitional regime as the cavities begin to interact \citep{Fry}. In the present investigation, we study cavitation over a circular cylinder over a range of $\sigma$ spanning non--cavitating, cyclic and transitional cavitation regimes.  
We observe that cavitation inside the K\'arm\'an vortices in the cyclic regime, is significantly altered at the onset of ``condensation shock" propagation in the transitional regime. At the transition, an order of magnitude jump in shedding Strouhal number, a peak in pressure fluctuations and a maximum in $St$ versus $\sigma$ based on a cavity length are observed. Hence, changes in shedding characteristics in these regimes and condensation shock propagation are studied. It is also known that, \textit{NCG} can change acoustic properties such as sound speed, acoustic impedance and consequently the shock propagation. This motivates the study of \textit{NCG} effects on cavity shedding and condensation shock propagation. 

It is known that the sound speed of the two--phase water--vapor mixture is orders of magnitude smaller than the speed of sound of its constituent phases \citep{franc_book}. If the sound speed becomes comparable to the magnitude of the velocities in the flow, it can lead to the formation of shock waves. Observation of shock waves in bubbly mixtures have been made as early as $1964$ in the head breakdown process in cavitating inducers \citep{Jakobsen}, although, ``condensation shock" propagation as a mechanism for partial cavity shedding has been shown only recently \citep{Harish}. Note that ``condensation shocks" refer to shock waves associated with a retracting partial cavity, typically have a weak discontinuity in pressure (order of few $kpa$) and involve phase change \citep{Budichwedge}. Subsequent to \cite{Harish}, various computational and experimental studies have considered condensation shock propagation as a mechanism in context of sheet to cloud cavitation \citep{chahinewedge,Terwisgawedge,Budichwedge,Mrugankwedge,jahangir}. In these studies, at sufficiently small $\sigma$, the sheet to cloud transition is observed by the propagation of condensation shocks, instead of the classically observed re--entrant jet mechanism \citep{Laberteaux}. Similarly, in the present work involving bluff body cavitation, with a significant reduction in $\sigma$ (moving from the cyclic to the transitional regime), we observe that the condensation shock propagation rather than a periodic cavitation inside K\'arm\'an vortices dominate the cavity shedding. 

Presence of \textit{NCG} can influence cavitating flows in various ways \citep{BRIANCON-MARJOLLET,Kawakami,Orley,Makiharju,Vennig,Brandao,Trummler}.  
Influence of dissolved and injected \textit{NCG} in partial cavitation over a wedge has been considered by \cite{Makiharju}. They found that injection of \textit{NCG} into the cavity suppressed vapor formation altering the dynamics of condensation shock formation. \cite{Vennig} in the experimental investigation on a flow over a hydrofoil observed that for the flow rich in vapor/\textit{NCG} nuclei, multiple shock waves are necessary for complete condensation and detachment of the cavity. \cite{Trummler} concluded that gas present in vapor bubbles would lead to stronger rebound and dampen the emitted shockwaves. Similar damping effect due to presence of gas in the medium was also observed by \cite{Brandao}. In such flows it is also important to note the influence of nuclei content of vapor/\textit{NCG}. At a pressure lower than the vapor pressure, cavitation is triggered by imperfections in water, that are mostly small \textit{NCG} or vapor bubbles (known as cavitation nuclei) that initiates the liquid breakdown \citep{franc_book}. In addition, \textit{NCG} can behave differently from vapor in response to pressure variations; gas can only experience volume change due to expansion/compression, while vapor can in addition, undergo phase change due to evaporation/condensation. Since the \textit{NCG} does not undergo phase change, the flow is more sensitive to the nuclei content of initially present \textit{NCG} in the system than vapor.  
Numerical studies involving fully compressible formulation and a homogeneous mixture approach often use relatively high values of freestream nuclei \citep{Saito, Seo, Aswinwedge, AswinJFM, Mrugankwedge}, to avoid extremely small time s.pdf due to low Mach numbers in water. The studies have shown that good agreement with experiment is observed for large regions of vapor and developed cavitation regimes \citep{Saito, Aswinwedge, Mrugankwedge}. Although, cavitation inception and incipient cavitation are known to be highly sensitive to the nuclei size and their distribution \citep{Hsiao}. Hence, in the present work, we also consider the effect of freestream nuclei of vapor/\textit{NCG} on the cylinder wake. 

Single phase flow over circular cylinders have been studied extensively in the past. Limited studies exist on the cavitating flow over a cylinder \citep{Rao,Ramamurthy,Fry,Seo,AswinJFM,Kumar,KumarIJMF}.
\cite{Fry} investigated cavity dynamics in the cylinder wake by measuring noise spectra. The author observes a peak in pressure fluctuations as the cyclic cavitation inside the periodic vortex shedding, transitions (with the reduction in $\sigma$) to irregular--regular vortex shedding, and eventually to a fixed cavity. \cite{Seo} studied cavitating flow at $Re=200$ and observed that the shock waves generated by the coherent collapse of the vapor cloud significantly change the aerodynamic noise characteristics. \cite{KumarIJMF} studied the cavitating structures of the near--wake of a circular cylinder for a subcritical Reynolds number and concluded that the cavities originate primarily in the free shear layer, not in the wake or in the attached boundary layer. For the cyclic cavitation, \cite{AswinJFM} explained the reduction in K\'arm\'an shedding frequency with the reduction in freestream $\sigma$, using the increase in the vorticity dilatation term due to cavitation. At lower $\sigma$, they observe condensation front propagation for the transitional regime, but do not discuss the nature of the front or the alteration in shedding characteristics. Effects of \textit{NCG} are not discussed in any of these works.

The objectives of this paper are to (i) investigate cavitating flow over a circular cylinder over a range of $\sigma$ spanning non--cavitating, cyclic and transitional cavitation regimes, (ii) discuss the changes in shedding characteristics over the regimes (eg. significant drop in shedding frequency, condensation front propagation, peak in $St$ versus $\sigma$ and pressure fluctuations) using numerical results and dynamic mode decomposition, (iii) using the Rankine--Hugoniot jump conditions, discuss the condensation shock propagation in the transitional regime, (iv) extend the numerical method of \cite{AswinJFM} based on a fully compressible formulation and a homogeneous mixture to include the \textit{NCG}. Study the effect of \textit{NCG} on the shedding characteristics, condensation shock propagation and effect of the freestream nuclei content, (v) study the turbulent cavitating flow at $Re=3900$ and compare it to the past work and laminar flow simulations at $Re=200$.  

The paper is organized as follows. Section \ref{method} discusses the physical model, the governing equations and the extension of the numerical method to account for \textit{NCG}. The problem setup and the simulation details are given in section \ref{prob}. Results and discussions are provided in section \ref{results}. The paper is summarized in section \ref{Summ}. The appendix is devoted to the derivation of an equation for the speed of a upstream moving front using Rankine--Hugoniot jump conditions and a discussion about temperature ratio across the condensation shock.

\section{Physical model and numerical method}
\label{method}
Numerical methods which include the effects of \textit{NCG} were often based on incompressible Navier--Stokes equations \citep{kunz, singhal, BinJi, Lu}. More recently, fully compressible formulations have been employed \citep{Orley, Mithun}. In the present work, the numerical method of \cite{AswinIJMF} based on fully compressible formulation for the vapor--water mixture is extended to account for \textit{NCG}. The ideal gas equation of state is used for \textit{NCG} and is coupled with the stiffened equation of state for water and ideal gas equation for vapor, to derive the mixture equation of state. The mixture sound speed is obtained from the mixture equation of state and Gibbs equation. Transport equations for the non--condensable gas and the vapor mass fraction are solved along with the compressible Navier--Stokes equations for the mixture quantities. Both vapor and \textit{NCG} are uniformly introduced in the freestream in terms of volume fraction. However, separate transport equations for vapor and gas allow both to evolve in a different manner depending upon the local flow conditions, which allows study of their distribution in the wake of the cylinder.

\subsection{Homogeneous mixture approach}
We use the homogeneous mixture approach where the mixture of water, vapor and \textit{NCG} is considered as a single compressible medium. We assume mechanical equilibrium (i.e. each phase has the same pressure as the pressure of the cell and slip velocity between the phases is not considered) and thermal equilibrium (i.e. temperature of each phase is same as the cell temperature). Surface tension effects are assumed small and hence neglected. The governing equations are the compressible Navier--Stokes equations for the mixture quantities along with transport equations for vapor and \textit{NCG}. Different from the works of \cite{Orley} and \cite{Mithun} where the homogeneous equilibrium barotropic model is employed, here we assume finite mass transfer rate between vapor and water, which is explicitly modeled through source terms. These equations are Favre--averaged and spatially filtered to perform LES. The subgrid terms are modeled with the dynamic Smagorinsky model. Details can be found in \cite{AswinIJMF}. The unfiltered governing equations are:

\begin{equation}
\label{compressible NS}
\begin{aligned}
\frac{\partial \rho}{\partial t} &= - \frac{\partial }{\partial x_{j}}(\rho u_{j}), \\
\frac{\partial \rho u_{i}}{\partial t} &= - \frac{\partial}{\partial x_{j}}(\rho u_{i} u_{j} + p \delta_{ij}-\sigma_{ij}), \\ 
\frac{\partial \rho e_{s}}{\partial t} &= -\frac{\partial }{\partial x_{j}}(\rho e_{s} u_{j} - Q_{j}) -p\frac{\partial u_{j}}{\partial x_{j}} + \sigma_{ij}\frac{\partial u_{i}}{\partial x_{j}} , \\
\frac{\partial \rho Y_v}{\partial t} &= -\frac{\partial}{\partial x_j}(\rho Y_v u_j) + S_e - S_c \quad \textrm{and}\\
\frac{\partial \rho Y_g}{\partial t} &= -\frac{\partial}{\partial x_j}(\rho Y_g u_j). \\
\end{aligned}
\end{equation}
Here $\rho$, $u_i$, $e_s$ and $p$ are density, velocity, internal energy and pressure of the mixture respectively. $Y_v$ is the vapor mass fraction and $Y_g$ is the \textit{NCG} mass fraction. The mixture density is defined as
\begin{equation}
\label{mixture density}
\rho = \rho_l (1 - \alpha_v - \alpha_g) + \rho_v \alpha_v + \rho_g \alpha_g,
\end{equation}
where $\rho_l$, $\rho_v$ and $\rho_g$ are densities of liquid, vapor and gas respectively. $\alpha_v$ and $\alpha_g$ are the volume fractions of vapor and \textit{NCG} respectively. Volume fractions of each constituent phase are related to their respective mass fractions as 
\begin{equation}
\label{mass volume fractions}
\rho_l(1 - \alpha_v - \alpha_g) = \rho(1 - Y_v - Y_g), \quad \rho_v \alpha_v= \rho Y_v \quad \textrm{and} \quad  \rho_g \alpha_g = \rho Y_g.
\end{equation}
Internal energy of the mixture is obtained by mass weighted average of its constituent phases:
\begin{equation}
\label{mixture internal energy}
\begin{aligned}
\rho e_s &= \rho (1 - Y_v -Y_g) e_l + \rho Y_v e_v + \rho Y_g e_g , \quad \textrm{where} \\
e_l &= C_{vl} T + \frac{P_c}{\rho_l}, \\ 
e_v &= C_{vv} T \quad \textrm{and} \\ 
e_g &= C_{vg} T. 
\end{aligned}
\end{equation}
Here, $e_l$, $e_v$ and $e_g$ are the internal energies of liquid, vapor and \textit{NCG} respectively and $C_{vl}$, $C_{vv}$ and $C_{vg}$ are their specific heats at constant volume respectively. The system is closed using a mixture equation of state obtained using stiffened equation of state for the liquid and ideal gas equation of state for both vapor and \textit{NCG}:
\begin{equation}
\label{eos}
p = Y_v \rho R_v T + Y_g \rho R_g T + (1 - Y_v - Y_g) \rho K_l T \frac{p}{p + P_c},
\end{equation}
where $R_v= 461.6 J/(Kg K)$, $R_g= 286.9 J/(Kg K)$, $K_l =2684.075 J/(Kg K)$ and $P_c= 786.333 \times 10^6 Pa$ are the constants associated with equation of state of the mixture. Parameters for the stiffened equation of state used for water are derived by \cite{AswinIJMF} to match speed of sound in liquid at a given density. Parameters for the gas and vapor equations of state are taken from \cite{white} and \cite{Saito} respectively. Hence, the current approach accurately predicts the liquid speed of sound and density variation as shown in \cite{AswinIJMF}, although the specific heat at constant volume is under predicted ($1500 J/Kgk$ as compared to the NIST value of $4157.4. J/Kgk$). This however, is not considered as a serious drawback, considering the isothermal nature of the current problem as discussed in the Appendix B. In addition, numerical studies of \cite{AswinIJMF} have demonstrated validation of the numerical method using stiffened equation of state for variety of flow problems for the study of hydrodynamic cavitation.

\begin{figure}
   \centering
   \includegraphics[width=13pc,trim={0 0.1cm 0.1cm 0}, clip]{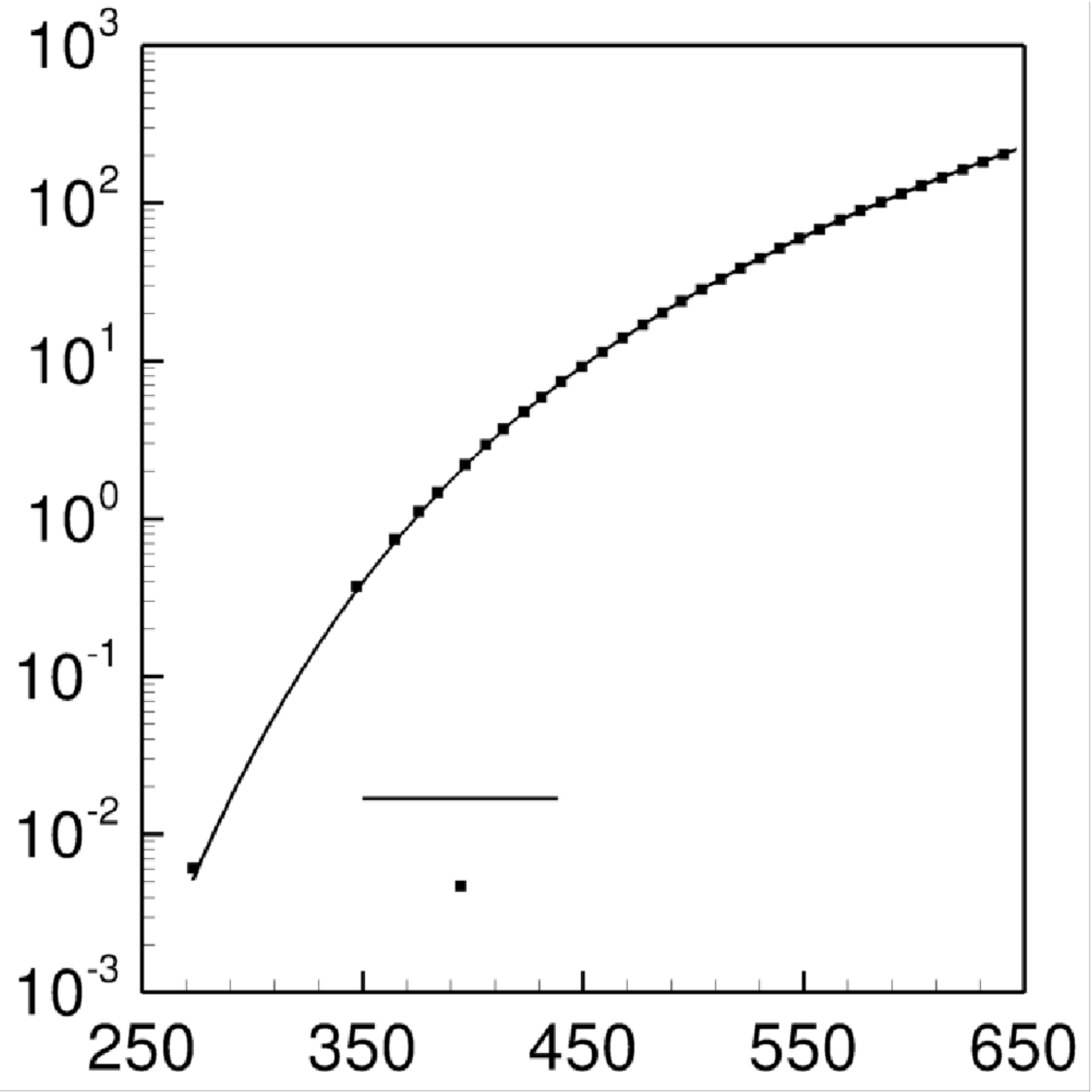}
   \put(-85,-10){$T(K)$}
   \put(-190,75){$p_v(Pa)$}
   \put(-73,38){Equation (\ref{vapor pressure})}
   \put(-73,25){NIST data}
   \caption{Vapor pressure variation with temperature.}
 \label{saturation_curve} 
\end{figure}
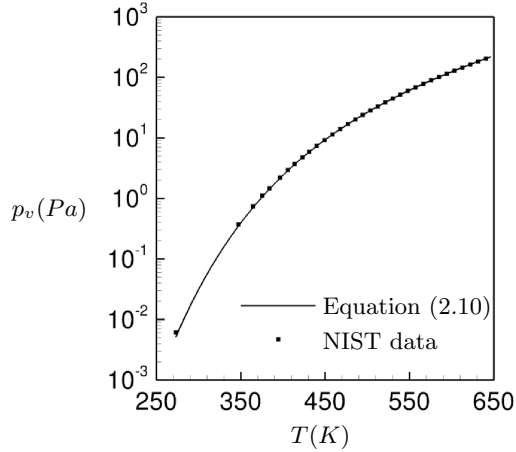

The viscous stress tensor ($\sigma_{ij}$) and heat flux vector ($Q_j$) are given by
\begin{equation}
\label{viscous stress and heat flux}
\sigma_{ij} = \mu(\frac{\partial u_i}{\partial x_j} + \frac{\partial u_j}{\partial u_i} - \frac{2}{3}\frac{\partial u_k}{\partial x_k}\delta_{ij}) \quad \textrm{and} \quad Q_j = k \frac{\partial T}{\partial x_j},
\end{equation}
where the mixture thermal conductivity is defined as a volume average between the conductivities of the individual constituent phases. For the mixture viscosity, we follow \citep{Beattie} and assume that the effective dynamic viscosity of the liquid--vapor--gas mixture satisfies a quadratic law with a maximum in the two--phase region. The mixture thermal conductivity and viscosity are given in equation (\ref{viscosity and thermal conductivity}) as
\begin{equation}
\label{viscosity and thermal conductivity}
\begin{aligned}
\mu &= \mu_l(1-\alpha_v - \alpha_g)(1 + 2.5(\alpha_v + \alpha_g)) + \mu_v \alpha_v + \mu_g \alpha_g \quad \textrm{and} \\
k &= k_l (1 - \alpha_v - \alpha_g) + k_v \alpha_v + k_g \alpha_g .
\end{aligned}
\end{equation}
In equation (\ref{viscosity and thermal conductivity}), $k_l$, $k_v$, and $k_g$ are thermal conductivities of water, vapor and \textit{NCG} respectively while $\mu_l$, $\mu_v$ and $\mu_g$ are the dynamic viscosities of water, vapor and \textit{NCG} respectively. Note that $\mu_l$$>>$$\mu_v$,$\mu_g$. A simple volume average would give maximum in the liquid region for $\alpha_v + \alpha_g = 0$ (i.e. in liquid), while a quadratic dependence in \cite{Beattie} yields an initial increase in the mixture viscosity, moving from liquid to the mixture. The mixture viscosity is maximum in the two--phase region near liquid. Molecular dynamics simulations confirm this behavior \citep{Ilja}.
They are related to the temperature of the mixture as
\begin{equation}
\label{viscosity}
\begin{aligned}
\mu_l = C_{0l} \times 10^{\frac{C_{1l}}{T- C_{2l}}}, \quad
\mu_v = C_{0v} \bigg(\frac{T}{T_{0v}}\bigg)^{n_{v}}, \quad
\mu_g = C_{0g} \bigg(\frac{T}{T_{0g}}\bigg)^{n_{g}}, 
\end{aligned}
\end{equation}
where the constants in equation (\ref{viscosity}) and their references are given in table \ref{table_viscosity}. Thermal conductivity in a constituent phase ($k_l$, $k_v$ and $k_g$ respectively in liquid, vapor and \textit{NCG}) is obtained from Prandtl number ($Pr$) in each phase. Since the maximum observed values of vapor mass fractions in the cases considered are orders of magnitude smaller than unity, latent heat of vaporization can be neglected \citep{AswinIJMF} and it was not considered in the present work. $S_e$ and $S_c$ are the source terms due to evaporation of water and condensation of vapor and are given by 
\begin{table}
\begin{center}
  \begin{tabular}{ccc}
       $\mu_l = C_{0l} \times 10^{\frac{C_{1l}}{T- C_{2l}}}$ & $\mu_v = C_{0v} \bigg(\frac{T}{T_{0v}}\bigg)^{n_{v}}$ & $\mu_g = C_{0g} \bigg(\frac{T}{T_{0g}}\bigg)^{n_{g}}$ \\[3pt]
       $C_{0l} = 2.414\times 10^{-5} Pa \cdot s$ & $C_{0v} = 1.78 \times 10^{-5} Pa \cdot s$ & $C_{0g} = 1.71 \times 10^{-5} Pa \cdot s$ \\
       $C_{1l} = 247.8K$ & $T_{0v} = 288K$ & $T_{0g} = 273K$ \\
       $C_{2l} = 140K$ & $n_v=0.76$ & $n_g = 0.7$ \\
       \cite{AswinJFM} & \cite{AswinJFM} & \cite{Lagumbay} \\
  \end{tabular}
  \caption{Constants for species viscosity.}
  \label{table_viscosity}
 \end{center}
\end{table}
\begin{equation}
\label{source terms}
\begin{aligned}
S_e &= C_e (\alpha_v + \alpha_g)^2 (1 - \alpha_v - \alpha_g)^2 \frac{\rho_l}{\rho_v} \frac{max((p_v - p),0)}{\sqrt{2 \pi R_v T_s}} \quad \textrm{and} \\
S_c &= C_c (\alpha_v + \alpha_g)^2 (1 - \alpha_v - \alpha_g)^2 \frac{max((p - p_v),0)}{\sqrt{2 \pi R_v T_s}}.
\end{aligned}
\end{equation}
Here $T_s$ is a reference temperature. $C_e$ and $C_c$ are empirical constants based on the interfacial area per unit volume and their values are taken to be equal to 0.1 $m^{-1}$ as described by \cite{Saito}. They have shown that the solution is not sensitive to the value of empirical constants using cavitating flow over hemispherical/cylindrical bodies. $p_v$ is vapor pressure which is related to temperature as 
\begin{equation}
\label{vapor pressure}
p_v = p_k exp((1 - \frac{T_k}{T})(a + (b -cT)(T-d)^2)),
\end{equation}
where $p_k = 22.130 MPa$, $T_k = 647.31K$, $a = 7.21$, $b= 1.152 \times 10^{-5}$, $c = -4.787 \times 10^{-9}$ and $d = 483.16$ \citep{Saito}. Vapor pressure variation with temperature obtained from equation (\ref{vapor pressure}) is compared to the National Institute of Standards and Technology (NIST) data in figure \ref{saturation_curve} showing excellent agreement.

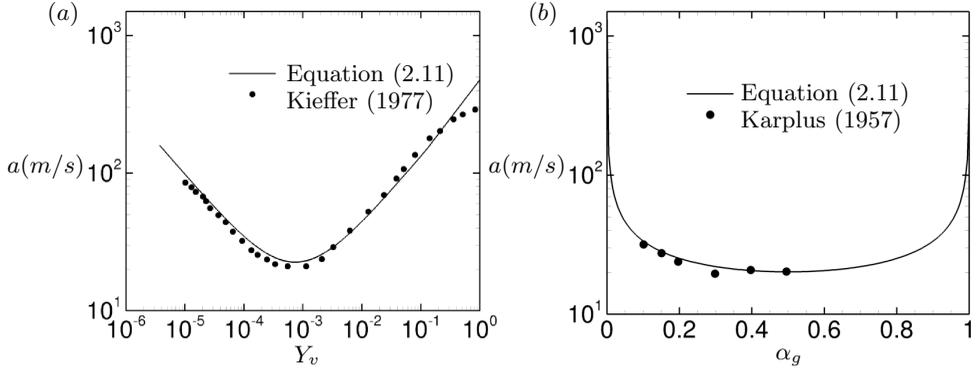
\begin{figure}
\centering
\begin{minipage}{13pc}\hspace{1pc}
\includegraphics[width=13pc,trim={0 0.1cm 0.1cm 0}, clip]{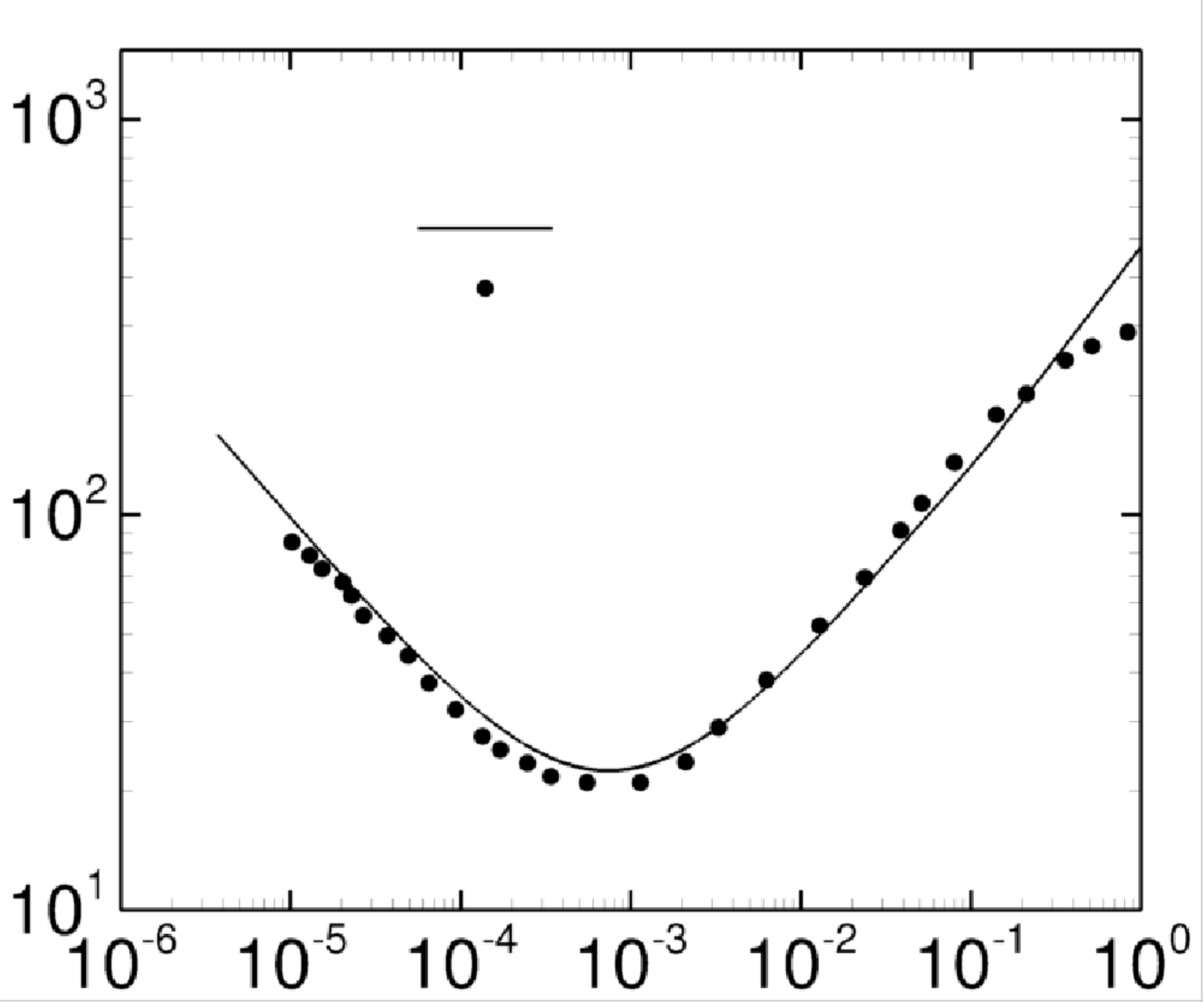}
\put(-185,63){$a (m/s)$}
\put(-80,97){Equation (\ref{soseq})}
\put(-80,87){\cite{Kieffer}}
\put(-77,-8){$Y_v$}
\put(-170,120){$(a)$}
\end{minipage}\hspace{3pc}
\begin{minipage}{13pc}
\includegraphics[width=13pc,trim={0 0.1cm 0.1cm 0}, clip]{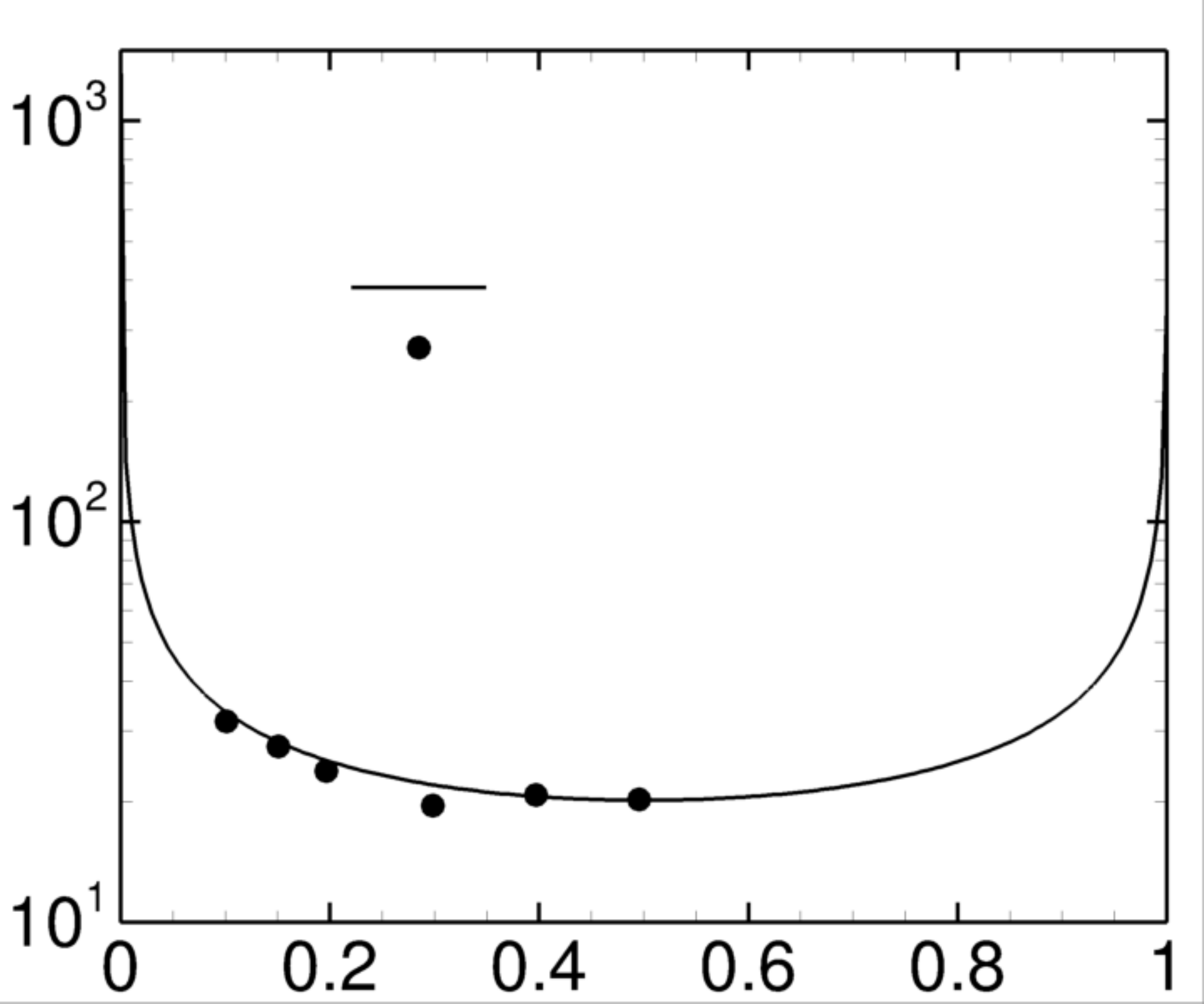}
\put(-77,-7){$\alpha_g$}
\put(-185,63){$a (m/s)$}
\put(-90,90){Equation (\ref{soseq})}
\put(-90,80){\cite{Karplus}}
\put(-170,120){$(b)$}
\end{minipage}%\hspace{4pc}
\caption{Speed of sound in water--vapor mixture $(a)$ and in water--\textit{NCG} mixture $(b)$.}
\label{sosf}
\end{figure}

The expression for the speed of sound in the mixture is obtained from equation (\ref{eos}) and the Gibbs equation and is given by
\begin{eqnarray}
\label{soseq}
\begin{aligned}
a^2 &= \frac{C_1 T}{C_0 - C_1/C_{pm}}, \quad \text{where} \\
C_1 &= (Y_v R_v + Y_g R_g)(p + P_c) + (1 - Y_v - Y_g)K_l p,\\
C_0 &= 2p + P_c -\rho T (Y_v R_v + Y_g R_g)- (1 -Y_v - Y_g) \rho K_ l T \quad \textrm{and}\\
C_{pm} &= Y_g C_{pg} + Y_v C_{pv} + (1 - Y_v - Y_g) C_{pl}.
\end{aligned}
\end{eqnarray}
Here, $C_{pv}$, $C_{pg}$ and $C_{pl}$ are the specific heats at constant pressure for vapor, \textit{NCG} and liquid respectively. Speed of sound obtained from equation (\ref{soseq}) is compared to the experimentally available data for water--vapor mixture as shown in figure \ref{sosf}(a) and water--air mixture in figure \ref{sosf}(b). The speed of sound derived in the present work does not consider mass transfer effects, and hence is a frozen speed of sound.

\subsection{Numerical method}

The numerical method of \cite{AswinIJMF} is extended to include the effects of \textit{NCG}. The algorithm has been extensively validated for the water--vapor mixture over a variety of problems by \cite{AswinIJMF}. The algorithm was used successfully to simulate both re--entrant jet \citep{Aswinwedge} and condensation shock regimes \citep{Mrugankwedge} in sheet to cloud cavitation over a wedge. Thus we only focus on the details pertaining to the inclusion of \textit{NCG}. 

The algorithm uses a predictor--corrector approach. In the predictor step, the governing equations are spatially discretized using a symmetric non--dissipative finite volume scheme. 
The viscous fluxes are split into compressible and incompressible contributions and treated separately. Once the fluxes are obtained, a predicted value $\hat{\bold{q}}_{j}^{n+1}$ is computed using an explicit Adams--Bashforth time integration.
The corrector step uses characteristic based filtering to compute the final solution $\bold{q}_{j}^{n+1}$ from the predicted value $\hat{\bold{q}}_{j}^{n+1}$ as 
\begin{equation}
\label{corrector}
\bold{q}_{j,cv}^{n+1} = \hat{\bold{q}_{j,cv}^{n+1}} - \frac{\Delta t}{V_{cv}} \sum_{faces} (\bold{F}_f^* n_f)A_f,
\end{equation}
where $F_f^*$ is the filter numerical flux of the following form:
\begin{equation}
\label{filter flux}
\bold{F}_{f_c}^* = \frac{1}{2} R_{f_c} \bold{\Phi_{fc}^*}.
\end{equation}
Here $R_{f_c}$ is the matrix of right eigenvectors at the face computed using the Roe average of the variables from left and right cell--centered values. $\bold{\Phi_{fc}^*}$ is a vector, $l$th component of which, $\phi^{*l}$, is given by
\begin{equation}
\label{phi}
\phi_{fc}^{*l} = k \theta_{fc}^{l} \phi_{fc}^{l},
\end{equation}
where $k$ is an adjustable parameter and $\theta_{fc}$ is Harten's switch function, given by
\begin{equation}
\label{harten}
\theta_{fc} = \sqrt{0.5(\hat{\theta}_{icv1}^2 +\hat{\theta}_{icv2}^2)}, \quad \hat{\theta}_{icv1} = \frac{|\beta_{fc}| - |\beta_{f1}|}{|\beta_{fc}| + |\beta_{f1}|}, \quad \hat{\theta}_{icv2} = \frac{|\beta_{f2}| - |\beta_{fc}|}{|\beta_{f2}| + |\beta_{fc}|}.
\end{equation}
Here, $\beta_f = R_f^{-1}(\bold{q}_{icv2} - \bold{q}_{icv1})$ is the difference between characteristic variables across the face. For $\phi_l$, the Harten--Yee total variation diminishing (TVD) form is used as suggested by \cite{Yee}:
\begin{equation}
\label{TVD}
\begin{aligned}
\phi_{fc}^{l} &= \frac{1}{2} \Psi (a_{fc}^l)(g_{icv1}^{l} + g_{icv2}^{l}) - \Psi (a_{fc}^{l} + \gamma_{fc}^{l}) \beta_{fc}^{l}, \\
\gamma_{fc}^{l} &= \frac{1}{2} \frac{\Psi (a_{fc}^{l}) (g_{icv2}^{l} - g_{icv1}^{l}) \beta_{fc}^{l}}{(\beta_{fc}^{l})^{2} + .pdfilon}, \\
\end{aligned}
\end{equation}
where $.pdfilon = 10^{-7}$, $\Psi(z) = \sqrt{\delta + z^{2}} $ ($\delta$ being $1/16$) is introduced for entropy fixing and $a_{fc}^{l}$ is an eigenvalue of the Jacobian matrix. The limiter function $g_{icv}$ is computed using the minmod limiter as described by \cite{Noma} on unstructured grids.

\cite{Noma} proposed a modification to Harten's switch $\theta_{fc}$ to accurately represent under--resolved turbulence for single--phase flows by multiplying $\theta_{fc}$ by $\theta_{fc}^{*}$ as
\begin{equation}
\label{modified harten noma}
\begin{aligned}
\theta_{fc} &= \theta_{fc} \theta_{fc}^{*}, \\
\theta_{fc}^{*} &= \frac{1}{2}(\theta_{icv1}^{*} + \theta_{icv2}^{*}) , \\
\theta_{icv1}^{*} &= \frac{(\nabla \cdot \bold{u})_{icv1}^{2}}{(\nabla \cdot \bold{u})_{icv1}^{2} + \Omega_{icv1}^{2} + .pdfilon }. \\
\end{aligned}
\end{equation}

\cite{AswinIJMF} modified it for the multiphase mixture of water and vapor to avoid the non--monotonic behavior in the regions of flow cavitation as the single phase switch, equation (\ref{modified harten noma}), reaches extremely small values due to high vorticity. This is given by
\begin{equation}
\label{modified harten aswin}
\begin{aligned}
\theta_{fc}^{*} &= \frac{1}{2}(\theta_{icv1}^{*} + \theta_{icv2}^{*}) + |(\alpha_{v_{icv2}} - \alpha_{v_{icv1}})|. \\
\end{aligned}
\end{equation}

\subsection{Modifications to the multiphase switch}
\begin{figure}
\centering
\begin{minipage}{12pc}
\includegraphics[width=12pc, trim={0 0.1cm 0.1cm 0}, clip]{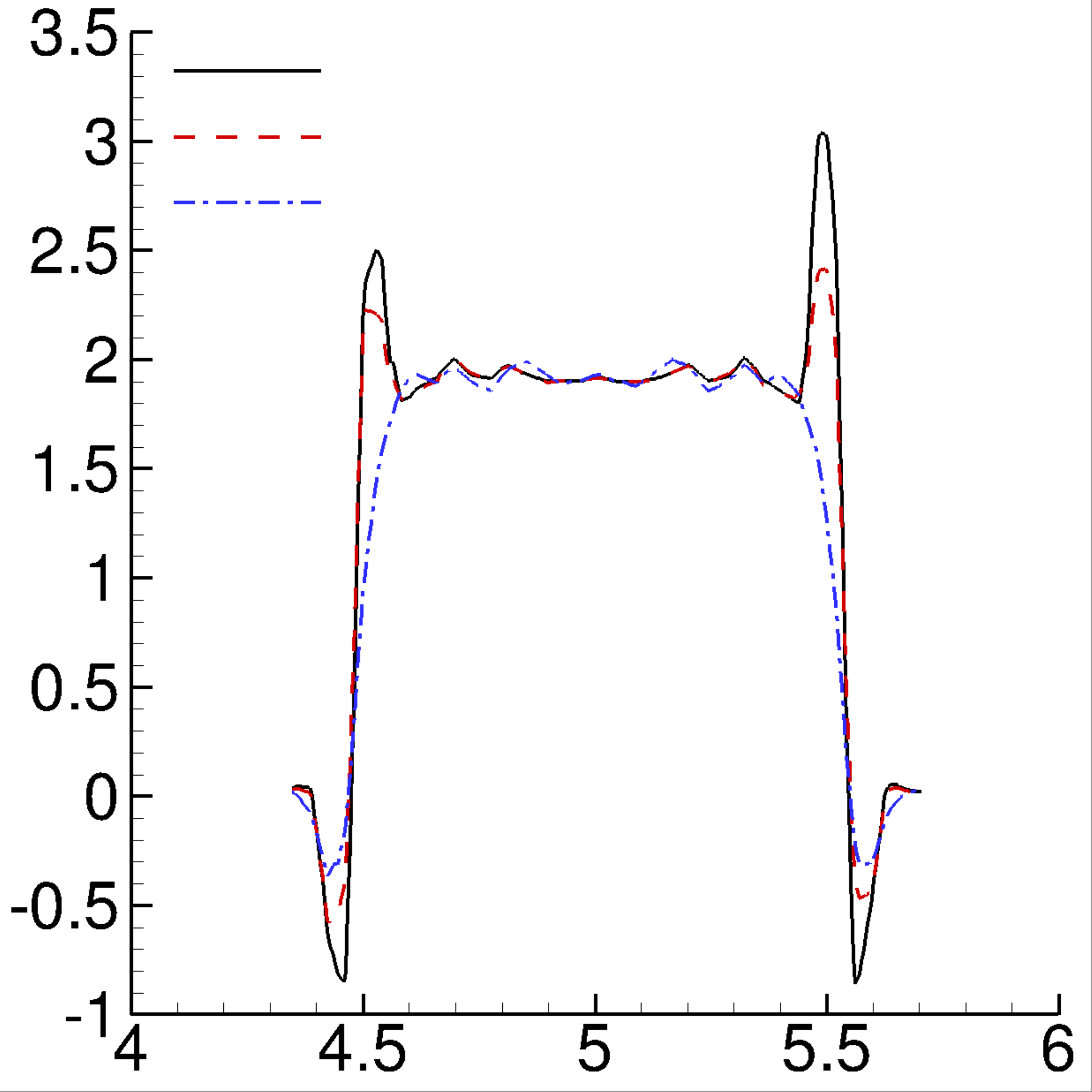}
\put(-70,-7){$x$}
\put(-170,75){$div(\vec{u})$}
\put(-95,132){Eq. (\ref{modified harten noma})}
\put(-95,122){Eq. (\ref{modified harten aswin})}
\put(-95,112){Eq. (\ref{modified harten gas})}
\end{minipage}
\caption{Line extracted along the vortex center-line showing absence of ``non-monotonic behavior'' with the modified switch.}
\label{switch}
\end{figure}
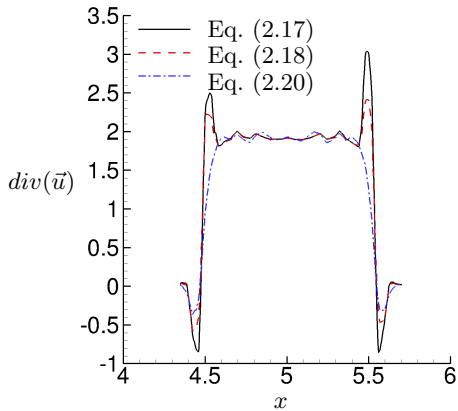

While the modification proposed by \cite{AswinIJMF} works well for water--vapor mixture, it still does not prevent the non--monotonic behavior when \textit{NCG} is present. We illustrate this by considering a cavitating inviscid vortex. We consider a square domain of size $10R \times 10R$. The flow is initialized with the following velocity field:
\begin{equation}
\label{inviscid vortex}
\begin{aligned}
u &= -\frac{C(y-y_c)}{R^{2}} exp(\frac{-r^2}{2}) \quad \textrm{and} \\
v &= \frac{C(x-x_c)}{R^{2}} exp(\frac{-r^2}{2}), 
\end{aligned}
\end{equation}
where $r = \sqrt{(x-x_c)^2 + (y-y_c)^2}/R$, $R=1.0$, $C=5.0$  and $x_c = y_c = 5R$. As we march in time, pressure inside the vortex core drops leading to flow cavitation and \textit{NCG} expansion. We see non--monotonic behavior in the solution as illustrated by the flow velocity divergence in figure \ref{switch}. As a remedy, an additional term due to the \textit{NCG} volume fraction is added to the multiphase switch as
\begin{equation}
\label{modified harten gas}
\begin{aligned}
\theta_{fc}^{*} &= \frac{1}{2}(\theta_{icv1}^{*} + \theta_{icv2}^{*}) + |(\alpha_{v_{icv2}} - \alpha_{v_{icv1}})| +  |(\alpha_{g_{icv2}} - \alpha_{g_{icv1}})|. \\
\end{aligned}
\end{equation}
This additional term prevents non--monotonic behavior due to the expansion of \textit{NCG} in the low pressure regions as shown in the figure \ref{switch}. This term goes to zero in the absence of non--condensable gas. Hence, $\theta_{fc}$ as defined by equation (\ref{modified harten gas}) is used for the computation of $\theta_{fc}^{l}$ in equation (\ref{phi}).

\section{Problem Setup}\label{prob}

Table \ref{table_problem} lists the flow conditions considered for the simulations. The cavitation number in the freestream is $\sigma = \frac{p_{\infty} - p_v}{\frac{1}{2}\rho_{\infty}U_{\infty}}$, where $p_{\infty}$, $\rho_{\infty}$ and $U_{\infty}$ are pressure, density and velocity in the freestream respectively. Cavitation number in the flow is varied from non--cavitating conditions to the cloud shedding regime. The Reynolds number, defined as $Re = \frac{\rho_{\infty} U_{\infty} D}{\mu} $ where $D$ is the cylinder diameter, used here are $Re=200$ and $Re=3900$ as considered by \cite{AswinJFM} for investigation of cavitation in near wake of the cylinder for water--vapor mixture. 
The simulations are initialized with a spatially uniform void fraction of vapor ($\alpha_{v0}$) that nucleates the cavitation. \textit{NCG} ($\alpha_{g0}$) is introduced in the freestream similar to the vapor nuclei in a spatially uniform manner. Different amounts of freestream vapor and gas volume fraction are used in this study. Details are provided in the table \ref{table_problem}, with the corresponding $\sigma$ and $Re$.  

\begin{table}
 \begin{center}
  \begin{tabular}{ccc}
     Freestream nuclei & Cavitation number ($\sigma$) & Reynolds number ($Re$) \\[3pt]
     Case A3900: $\alpha_{v0}=1.0 e^{-9}, \alpha_{g0}=1.0 e^{-6}$ & $1.0, 0.7$ & $3900$\\
     Case A200: $\alpha_{v0}=1.0 e^{-9}, \alpha_{g0}=1.0 e^{-6}$ & $5.0, 1.0, 0.85, 0.75, 0.7$ & $200$\\
     Case B: $\alpha_{v0}=1.0 e^{-2}, \alpha_{g0}=1.0 e^{-2}$ & $1.5, 1.0, 0.85, 0.75, 0.7, 0.5$ & $200$\\
     Case C: $\alpha_{v0}=1.0 e^{-2}, \alpha_{g0}=0.0$ & $1.5, 1.0, 0.85, 0.75, 0.7, 0.5$ & $200$\\
  \end{tabular}
  \caption{Cases showing flow conditions chosen for the problem.}
  \label{table_problem}
 \end{center}
\end{table}

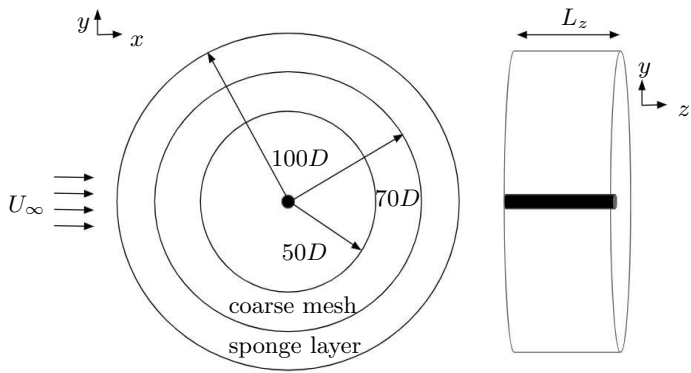
\begin{figure}
   \centering{\includegraphics[width=20pc, height=12pc]{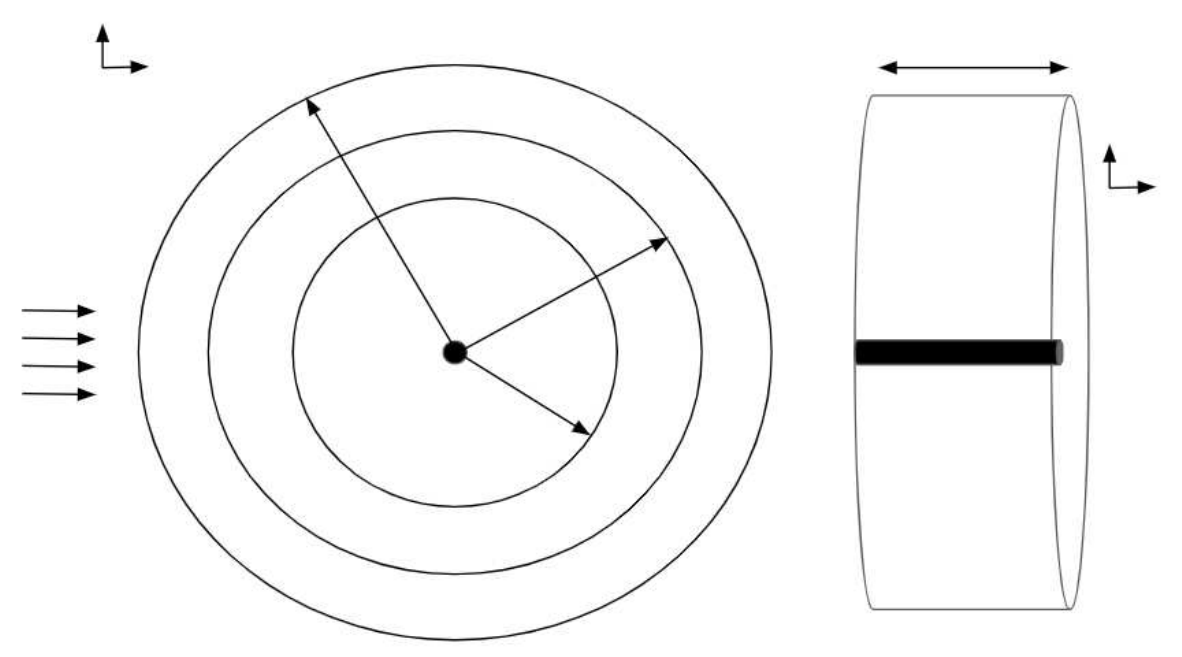}}
   \put(-115,65){$70D$}
   \put(-150,45){$50D$}
   \put(-154,80){$100D$}
   \put(-207,125){$x$}
   \put(-227,132){$y$}
   \put(-170,25){coarse mesh}
   \put(-170,10){sponge layer}
   \put(-253,63){$U_\infty$}
   \put(-45,133){$L_z$}
   \put(-1,99){$z$}
   \put(-16,115){$y$}
   \caption{Domain illustrating sponge layer and region of coarse mesh (not to scale).}
 \label{cyl_domain} 
\end{figure}

Figure \ref{cyl_domain} shows the schematic of the problem. The grid is 2D and 3D for the $Re=200$ and $Re=3900$ simulations, respectively. The domain size and mesh used in the present work is same as the finer grid and larger domain size used by \cite{AswinJFM}. They performed a grid refinement study and showed that time evolution of lift/drag coefficient as well as the profiles of mean and fluctuations in the void fraction show good agreement between their chosen grids. The computational domain is cylindrical with the origin at the center of the cylinder. The domain is extended radially until $100D$ and covers a distance of $2\pi$ and $\pi$ in the spanwise direction for the 2D and 3D simulations respectively. The freestream direction is in the positive $x$ direction as indicated by the arrows in the figure \ref{cyl_domain}. Freestream conditions are imposed on all the farfield boundaries. Collapse of cavitation clouds produces strong pressure waves which propagate over the entire domain. In order to avoid reflection of these pressure waves from the boundaries, we apply acoustically absorbing sponge layers at the boundaries as shown in figure \ref{cyl_domain}. This introduces an additional term in the governing equations (\ref{compressible NS}) given by, $\Gamma(q - q_{ref})$.
Here `$q$' denotes the vector of conservative variables and the subscript `$ref$' denotes the reference solution to which the flow is damped to, which is freestream values in the cases considered. `$\Gamma$' denotes the amplitude of the forcing. In addition, the grid is coarsened in the far field to further reduce any reflections. 

The mesh spacing considered near the cylinder surface is $0.005D \times 0.01D$ in the radial and azimuthal directions, which stretches to $0.03D \times 0.03D$ at approximately $2D$ downstream and then further stretches to $0.07D \times 0.07D$ at a distance of $5D$ downstream. For the 3D grid required at $Re=3900$, 80 points are used in the spanwise direction while the same resolution as the 2D grid is maintained in the $xy$ plane.

\section{Results}\label{results}

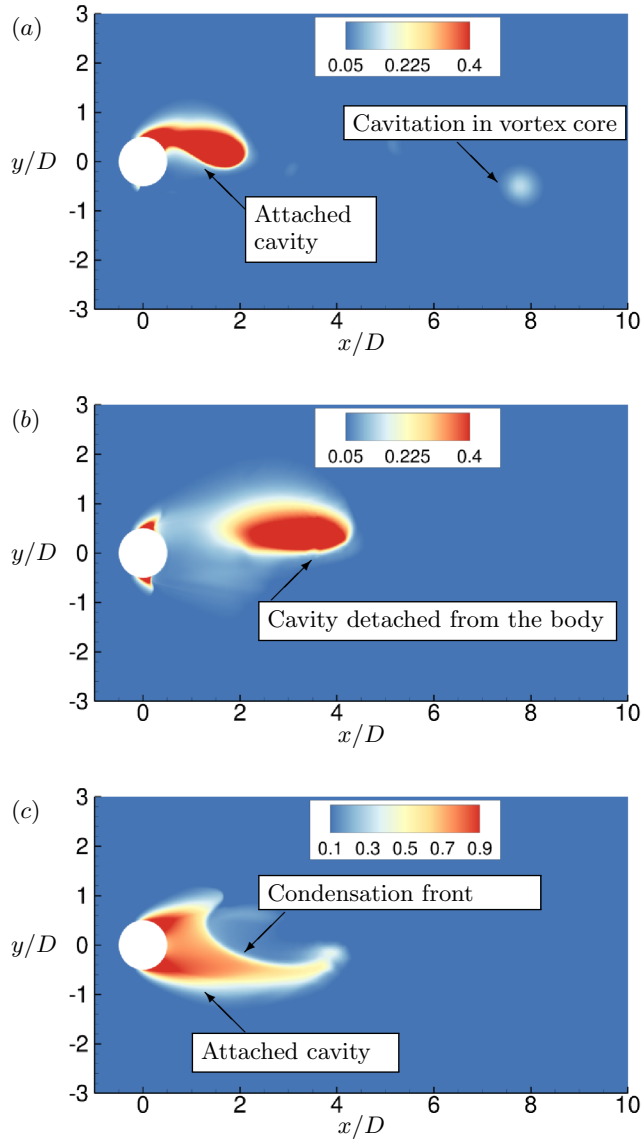
\begin{figure}
\centering
\begin{minipage}{18pc}
\includegraphics[width=18pc,trim={0.2cm 0.2cm 0.1cm 0}, clip]{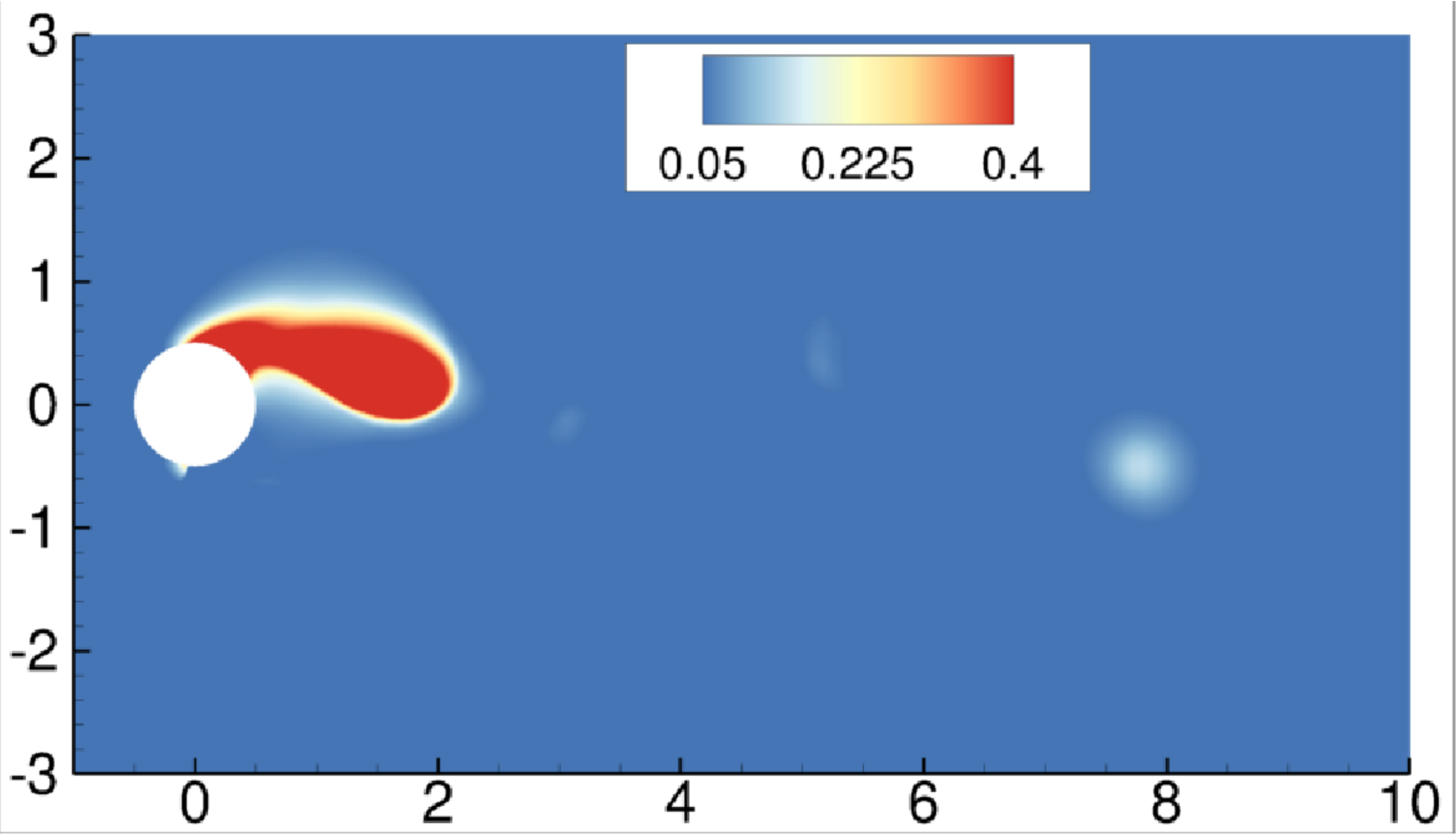}
\put(-238,110){$(a)$}
\put(-238,60){$y/D$}
\put(-115,-8){$x/D$}
\put(-150,35){\fcolorbox{black}{white}{\parbox{15mm}{\footnotesize{Attached cavity}}}}
\put(-150,45){{\textcolor{black}{\vector(-1,1){15}}}}
\put(-113,75){\fcolorbox{black}{white}{\parbox{35mm}{\footnotesize{Cavitation in vortex core}}}}
\put(-70,70){{\textcolor{black}{\vector(1,-1){15}}}}
\end{minipage}\vspace{1pc}
\begin{minipage}{18pc}
\includegraphics[width=18pc,trim={0.2cm 0.2cm 0.1cm 0}, clip]{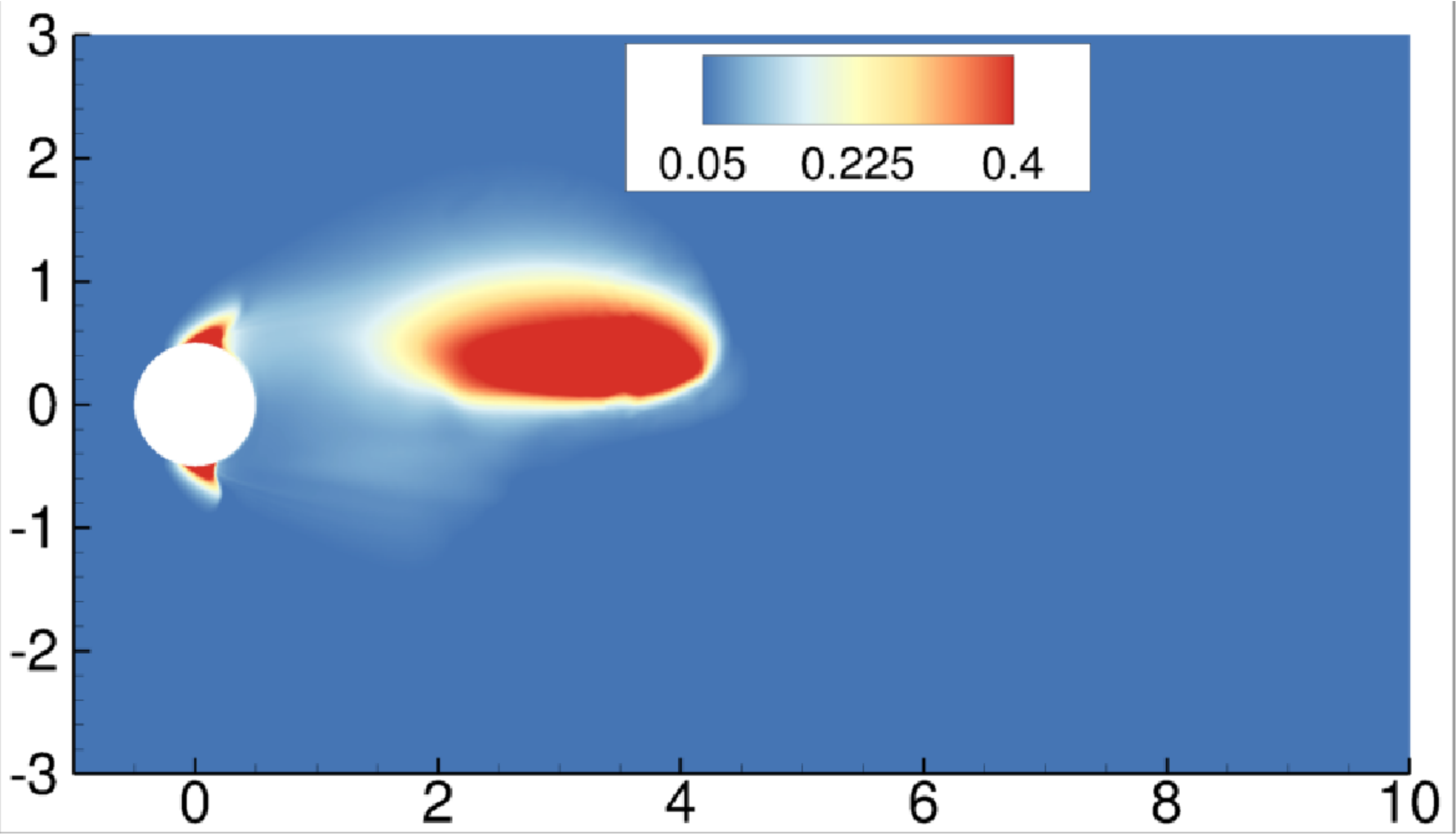}
\put(-238,110){$(b)$}
\put(-238,60){$y/D$}
\put(-115,-8){$x/D$}
\put(-145,35){\fcolorbox{black}{white}{\parbox{45mm}{\footnotesize{Cavity detached from the body}}}}
\put(-140,45){{\textcolor{black}{\vector(1,1){15}}}}
\end{minipage}\vspace{1pc}
\begin{minipage}{18pc}
\includegraphics[width=18pc,trim={0.2cm 0.2cm 0.1cm 0}, clip]{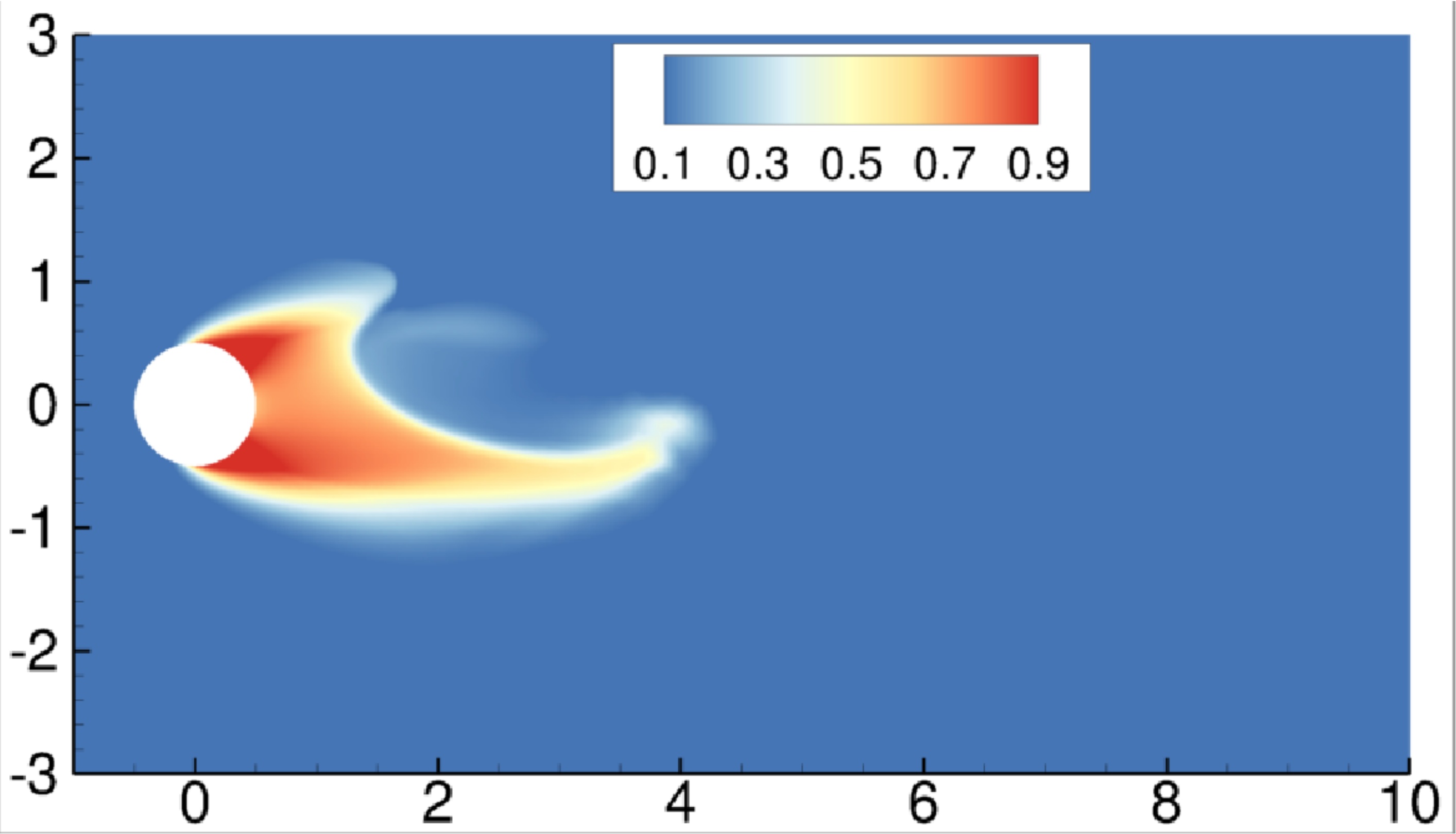}
\put(-238,110){$(c)$}
\put(-238,60){$y/D$}
\put(-115,-8){$x/D$}
\put(-170,20){\fcolorbox{black}{white}{\parbox{25mm}{\footnotesize{Attached cavity}}}}
\put(-150,30){{\textcolor{black}{\vector(-1,1){15}}}}
\put(-145,80){\fcolorbox{black}{white}{\parbox{35mm}{\footnotesize{Condensation front}}}}
\put(-135,75){{\textcolor{black}{\vector(-1,-1){15}}}}
\end{minipage}\vspace{1pc}
\caption{Instantaneous total void fraction (vapor $+$ \textit{NCG} volume fraction) contour for the cyclic regime $(a)$ and for the transitional regime $(b,c)$ for Case B.}
\label{sketch}
\end{figure}

Over the range of $\sigma$ studied, we observe two types of cavitation regimes as described by \cite{Fry}: cyclic and transitional. The cyclic cavitation regime is observed for high values of $\sigma$, which is characterised by periodic shedding of the cavitating vortices originating at the surface and is illustrated in figure \ref{sketch}$(a)$. These cavitating vortices collapse as they move downstream into the region of high pressure, producing pressure waves. As $\sigma$ is reduced, the flow enters the transitional cavitation regime. Here, the cavity shedding process alternates between two phenomena. The first is similar to cyclic cavitation in the vortex cores; the difference however is that these vortex cores cavitate further downstream and the cavity thus formed is not attached to the cylinder. During this part of the cycle, the cylinder surface and immediate wake remain cavitation free as shown in figure \ref{sketch}$(b)$. This is followed by the second phenomenon where the instantaneous pressure in the immediate wake drops below vapor pressure, where a cavity forms symmetrically spanning the entire aft--body of the cylinder as shown in figure \ref{sketch}$(c)$. Then, a pressure wave generated after the collapse of a vortex core impinges on the attached cavity, condensing it as displayed in figure \ref{sketch}$(c)$. This is called a condensation front, or condensation shock if it moves at supersonic speed. Once this front hits the cylinder, it will lead to cavity detachment. Details of the shedding characteristics in these regimes are discussed in section \ref{sec:freq}. In section \ref{topology} we discuss the mean flow characteristics including vapor/\textit{NCG} distribution in the cylinder wake and the boundary--layer. Effects of freestream void fraction on the boundary--layer separation and vapor/gas distribution are also discussed. In section \ref{bubbly} we show that the condensation fronts responsible for the cavity detachment travel at supersonic speed and that the condensation shocks are weakened by the \textit{NCG} as they propagate towards the cylinder. Finally, large--eddy simulation of cavitating cylinder is presented and discussed in section \ref{LES}.

\subsection{Shedding characteristics}\label{sec:freq}

For single--phase flow over a bluff body, vortices shed periodically from the surface forming the classical primary K\'arm\'an vortex street in the near wake. This is followed by a transition in the intermediate wake to a two--layered vortex street (e.g. \cite{Jiang}). The first vortex street transition was explained by \cite{Durgin} using a model in which a concentration of vorticity is strained into an elliptical shape by the nearby vortices in the street. This distorted vortex is then rotated, aligning its major axis with the streamwise direction. This process eventually results in distorted vortices merging and becoming shear layers on either side of the street. An important parameter that indicates the straining of the vortices and their merging is the spacing ratio, defined as the ratio between the cross--wake distance of different sign vortices to the longitudinal distance between same sign vortices. In the experiments of \cite{Durgin}, the authors found a spacing ratio greater than $0.366$ to be indicative of the transition, which was later confirmed by \cite{Karasudani}.

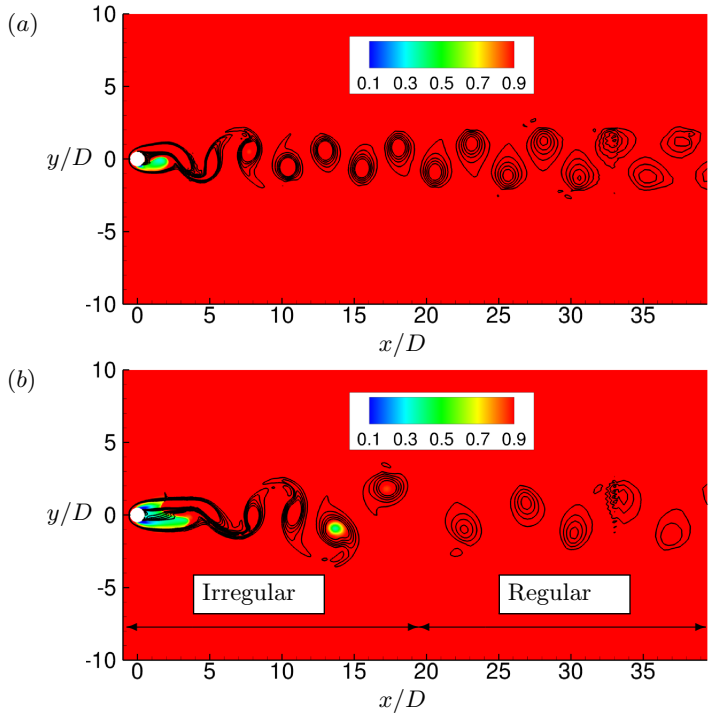
\begin{figure}
\centering
\begin{minipage}{20pc}
\includegraphics[width=20pc, trim={0.1cm 0.1cm 0.1cm 0}, clip]{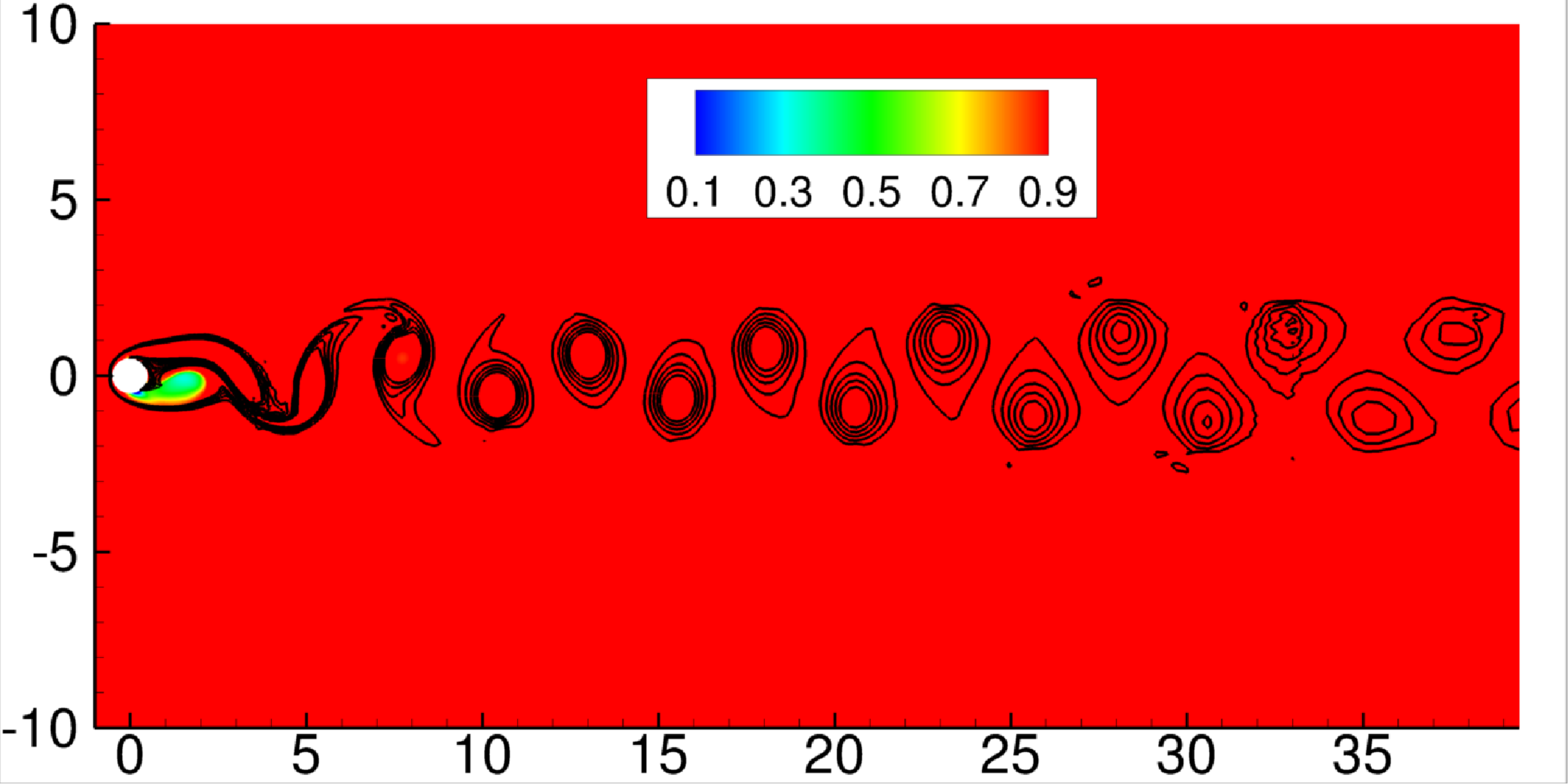}
\put(-270,110){$(a)$}
\put(-130,-10){$x/D$}
\put(-255,60){$y/D$}
\end{minipage}
\begin{minipage}{20pc}
\includegraphics[width=20pc, trim={0.1cm 0.1cm 0.1cm 0}, clip]{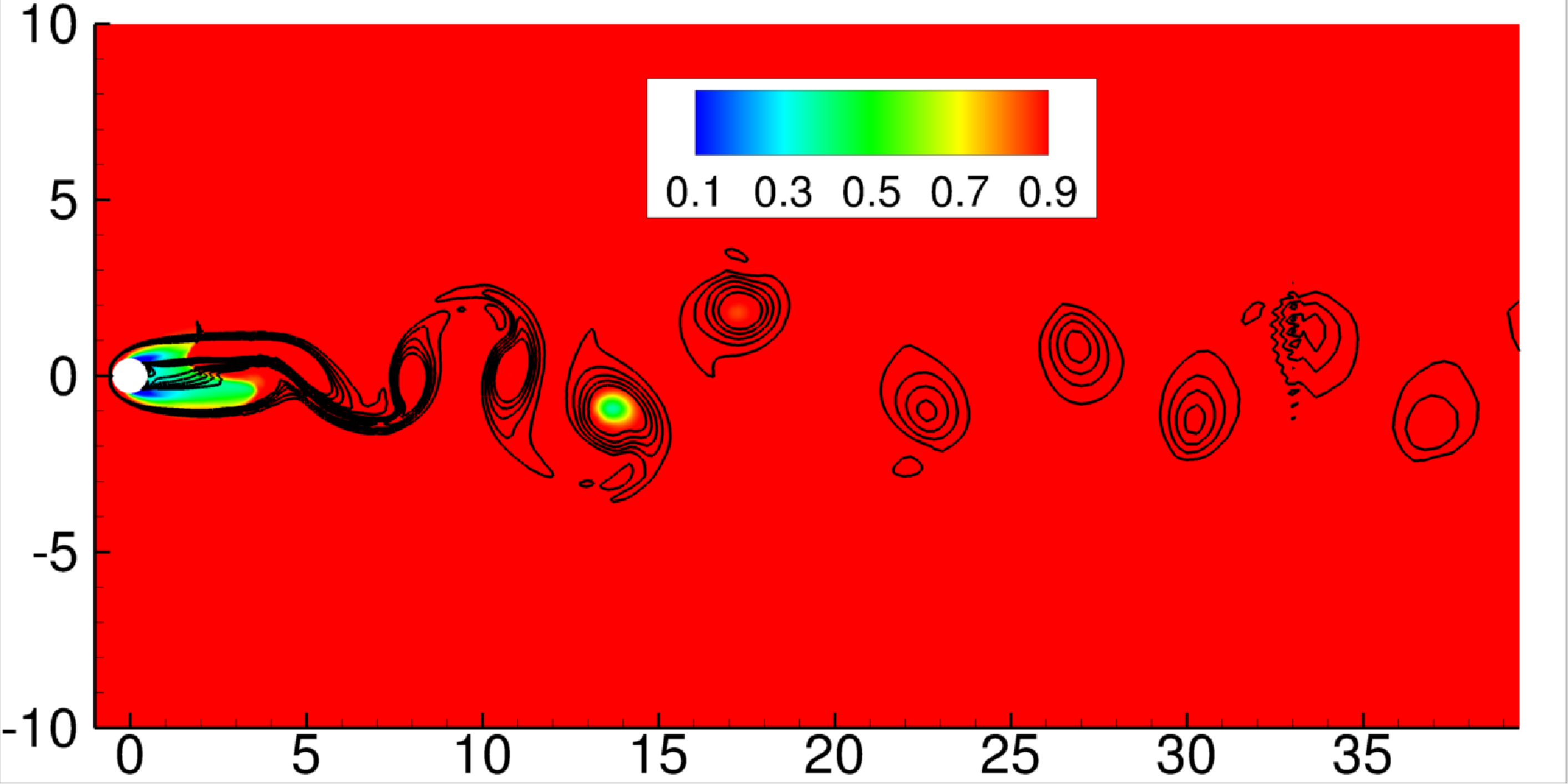}
\put(-270,110){$(b)$}
\put(-130,-10){$x/D$}
\put(-255,60){$y/D$}
\put(-200,30){\fcolorbox{black}{white}{\parbox{15mm}{\footnotesize{Irregular}}}}
\put(-90,20){\vector(-1,0){25}}
\put(-90,20){\vector(1,0){83}}
\put(-85,30){\fcolorbox{black}{white}{\parbox{15mm}{\footnotesize{Regular}}}}
\put(-200,20){\vector(-1,0){25}}
\put(-200,20){\vector(1,0){85}}
\end{minipage}
\caption{Instantaneous solution indicating vortex shedding at $\sigma=1.0$ $(a)$ and $\sigma=0.7$ $(b)$ for Case B. Lines indicate constant vorticity and coloured countours represent density.}
\label{vortex-shedding}
\end{figure}

For cavitating flows, analysis of flow variables in the near wake can reveal both the vortex and/or cavity shedding frequency. In the cyclic regime (figure \ref{vortex-shedding}$(a)$), the dominant shedding frequency is that of a single cavitated vortex from the surface into the wake. Note the regular vortex shedding from top and bottom of the cylinder in this regime (figure \ref{vortex-shedding}$(a)$).
In the transitional regime (figure \ref{vortex-shedding}$(b)$), we observe that this regular vortex shedding is disrupted at the onset of condensation front propagation, which occurs as the entire aft--body of the cylinder cavitates due to lower $\sigma$. Consequently, the cylinder wake exhibits irregular and regular vortex shedding periodically (figure \ref{vortex-shedding}$(b)$). The dominant shedding frequency in the transitional regime indicates the cavity shedding after the passage of the condensation front and the recurrence of irregular and regular vortex shedding processes. The frequency of individual vortex shedding from the surface becomes secondary.

\subsubsection{$St$ versus $\sigma$}  
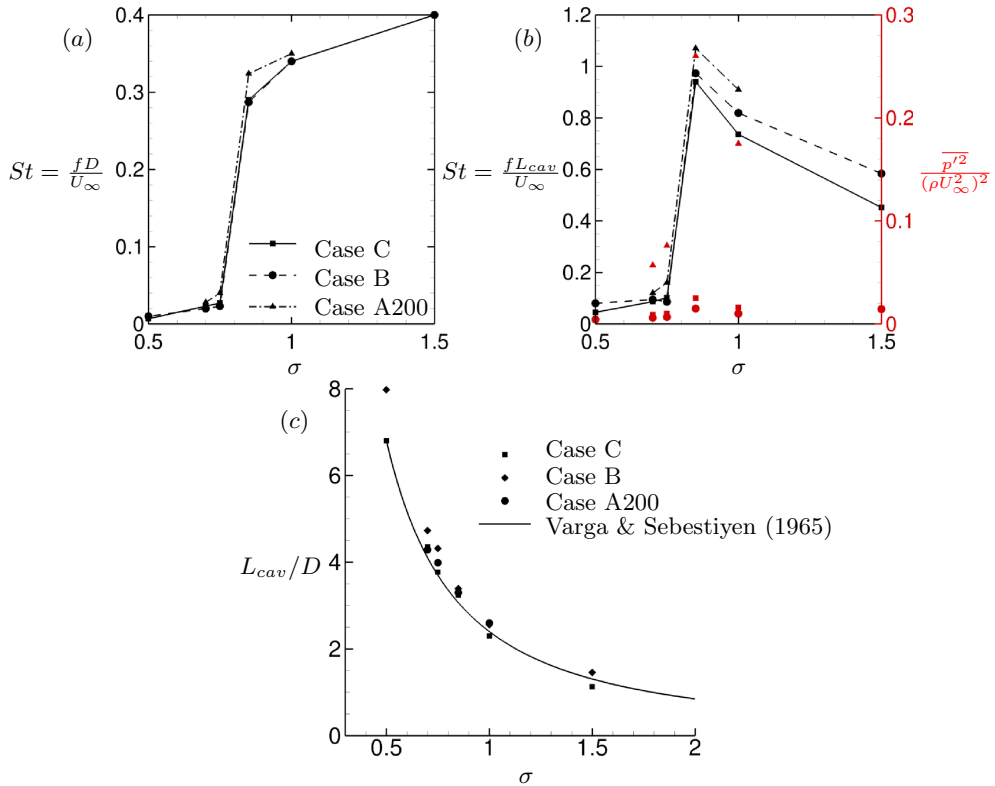
\begin{figure}
\centering
\begin{minipage}{25pc}
\includegraphics[width=25pc, trim={0.1cm 0.1cm 0.2cm 0.1cm}, clip]{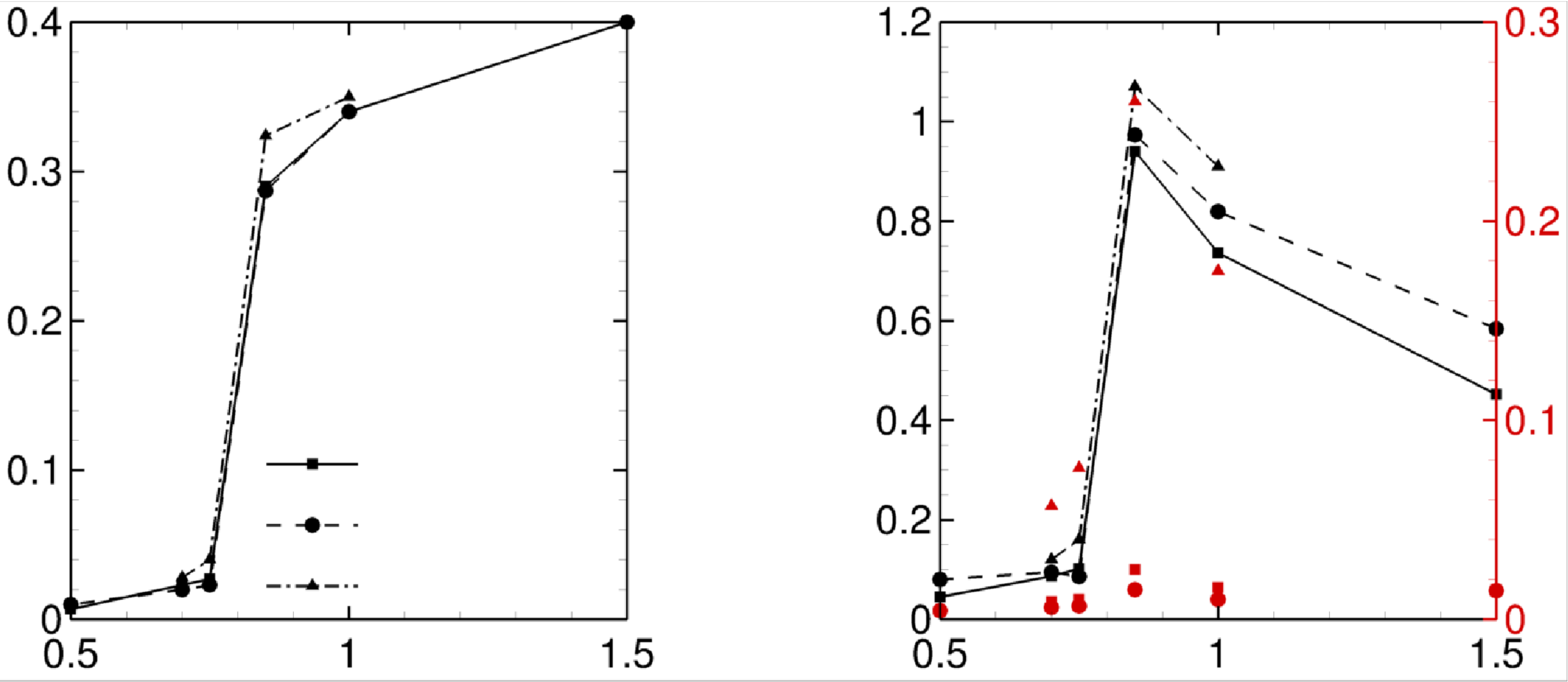}
\put(-320,115){$(a)$}
\put(-235,-8){$\sigma$}
\put(-340,65){$St=\frac{fD}{U_{\infty}}$}
\put(-225,36){Case C}
\put(-225,25){Case B}
\put(-225,14){Case A200}
\put(-150,115){$(b)$}
\put(-68,-8){$\sigma$}
\put(-177,65){$St=\frac{fL_{cav}}{U_{\infty}}$}
\put(2,65){\textcolor{red}{$\frac{\overline{p'^2}}{(\rho U_{\infty}^2)^2}$}}
\end{minipage}
\begin{minipage}{12pc}
\includegraphics[width=12pc, trim={0 0.1cm 0.1cm 0}, clip]{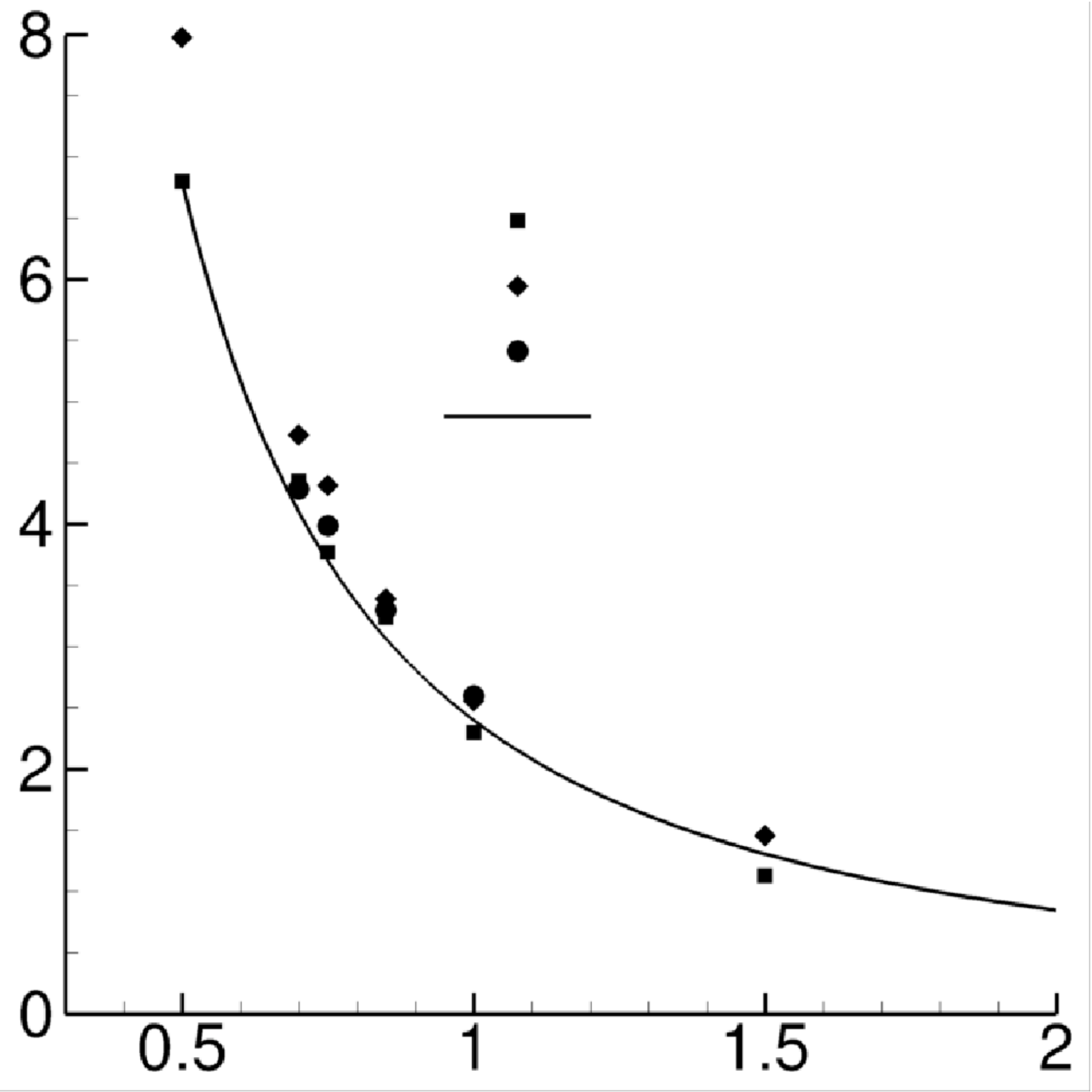}
\put(-160,125){$(c)$}
\put(-70,-8){$\sigma$}
\put(-175,70){$L_{cav}/D$}
\put(-60,114){Case C}
\put(-60,104){Case B}
\put(-60,94){Case A200}
\put(-60,85){\cite{Varga}}
\end{minipage}
\caption{$St$ for different $\sigma$ based on the cylinder diameter $(a)$ and cavity length $(b)$. Non--dimensional pressure fluctuations are represented by red symbols in $(b)$ for Case A200 ({\color{red}$\blacktriangle$}), Case B ({\color{red}$\bullet$}) and Case C ({\color{red}$\blacksquare$}). Cavity length normalized by cylinder diameter for different $\sigma$ $(c)$.} 
\label{St-Lcav} 
\end{figure}

We define Strouhal number ($St = fL/U$) to characterize the shedding frequency and plot it over a range of cavitation numbers spanning the cyclic and the transitional regime as shown in figure \ref{St-Lcav}. Here, $f$ is the cavity shedding frequency obtained from drag history, $U$ is the freestream velocity and two length scales, $D$ and $L_{cav}$, are chosen for $L$ and plotted respectively in figure \ref{St-Lcav}$(a)$ and \ref{St-Lcav}$(b)$. $L_{cav}$ is the cavity length defined as the position along the wake centerline where the total void fraction decreases to a value lower than 0.05 \citep{Harish,GaneshSNH}. The values used to compute $St$ at different $\sigma$ in figure \ref{St-Lcav}$(b)$ are shown in figure \ref{St-Lcav}$(c)$ and compared to the experimental fit from \cite{Varga} showing good agreement. We note that $St$ computed for non--cavitating conditions is 0.385. With reduction in $\sigma$, $St$ decreases in the cyclic regime. \cite{AswinJFM} explained this behavior through vorticity dilatation due to cavitation. However, the authors did not consider the sharp jump in $St$ moving through the transitional regime with further reduction in $\sigma$ (figure \ref{St-Lcav}). \cite{Fry} characterized the transition from the cyclic to the transitional regime by a peak in pressure fluctuations along the wake. Exactly at this transition, we observe that $St$ drops by an order of magnitude with $\sigma$ ($St = 0.285$ at $\sigma =0.85$ to $St = 0.018$ at $\sigma=0.75$) as the dominant frequency in the cyclic regime due to the periodic vortex shedding shifts to the frequency of irregular--regular vortex shedding in the transitional regime.

In addition, a maxima in the $St$ versus $\sigma$ plot (figure \ref{St-Lcav}$(b)$) is observed at $\sigma=0.85$ when $L_{cav}$ is chosen as a reference length. \cite{Young} and \cite{GaneshSNH} for flow over a triangular prism reported similar maximum in $St$ versus $\sigma$. They considered a configuration with flow confinement (top/bottom walls) and considered the base of the prism as the length scale. 

\subsubsection{Effect of \textit{NCG}} 

\begin{figure}
\centering
\begin{minipage}{12pc}
\includegraphics[width=12pc,trim={0 0.1cm 0.1cm 0}, clip]{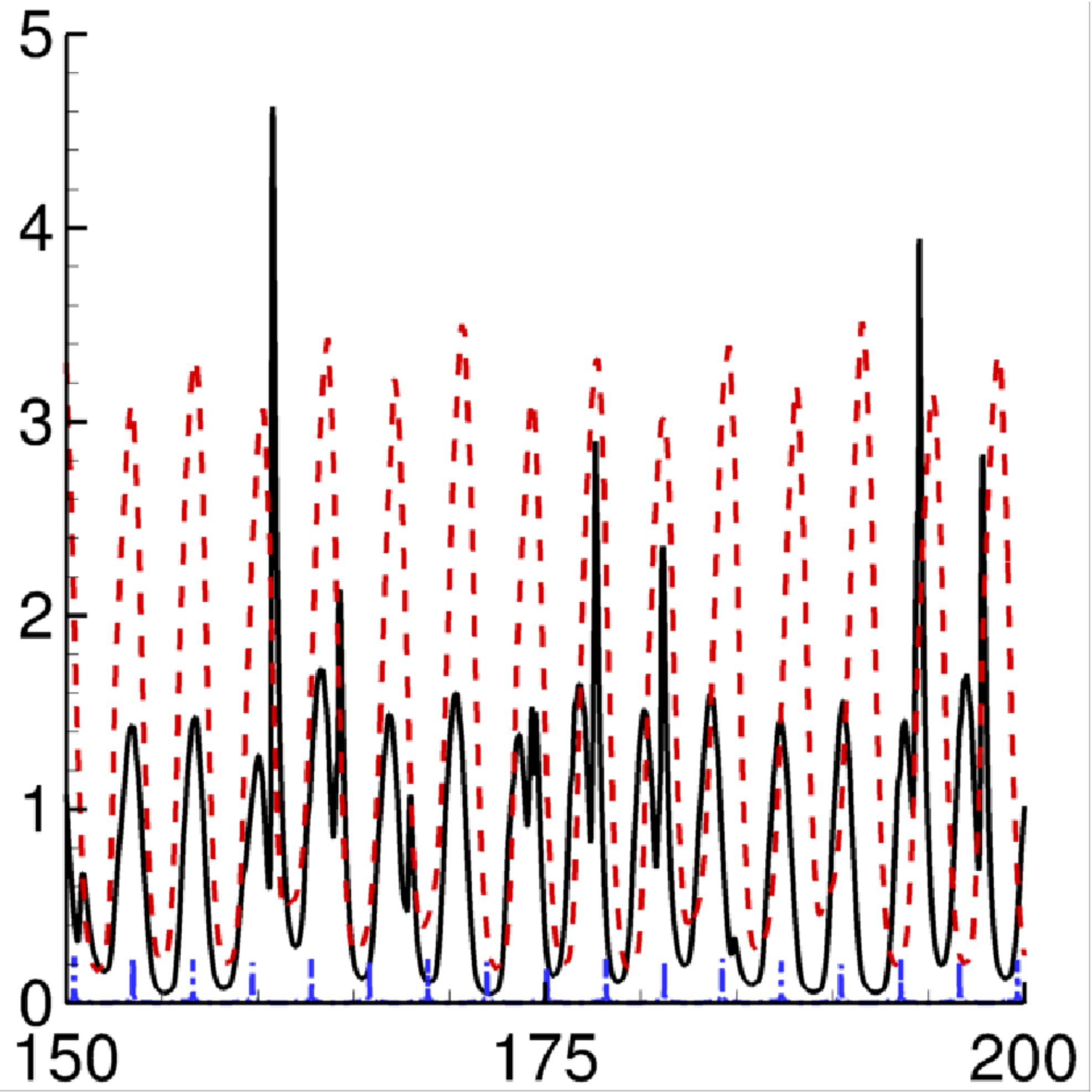}
\put(-160,140){$(a)$}
\put(-175,75){$\frac{p \times 10^{3}}{\rho_{\infty} a_{\infty}^2}$}  
\put(-80,-10){$tU/D$}
\end{minipage}\hspace{2pc}
\begin{minipage}{12pc}
\includegraphics[width=12pc,trim={0 0.1cm 0.1cm 0}, clip]{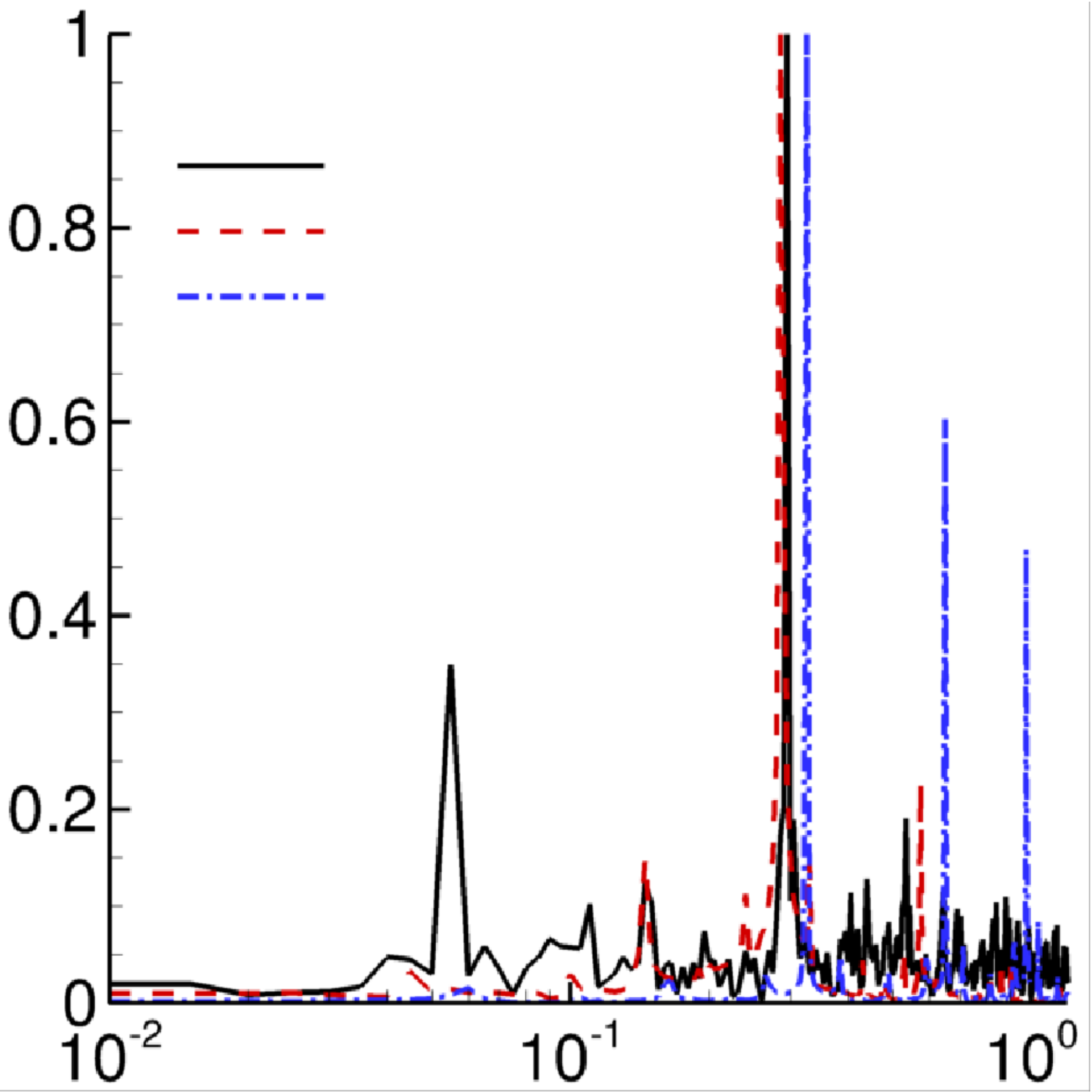}
\put(-160,140){$(b)$}
\put(-98,122){Case C}
\put(-98,112){Case B}
\put(-98,102){Case A200}
\put(-170,75){$FFT$}
\put(-100,60){\vector(1,-1){15}}
\put(-40,85){\vector(1,1){15}}
\put(-130,70){\fcolorbox{black}{white}{\parbox{22mm}{\footnotesize{low frequency \\ (transitional)}}}}
\put(-50,110){\fcolorbox{black}{white}{\parbox{22mm}{\footnotesize{high frequency \\ (cyclic)}}}}
\put(-80,-10){$St=\frac{fD}{U_{\infty}}$}
\end{minipage}
\caption{$(a)$ Pressure signal at $x=2.5D$ in the cylinder wake scaled with freestream density and speed of sound at $\sigma=0.85$ and $(b)$ FFT of the corresponding signal scaled with its maximum value for better visualization.} 
\label{pressure hist transit}
\end{figure}
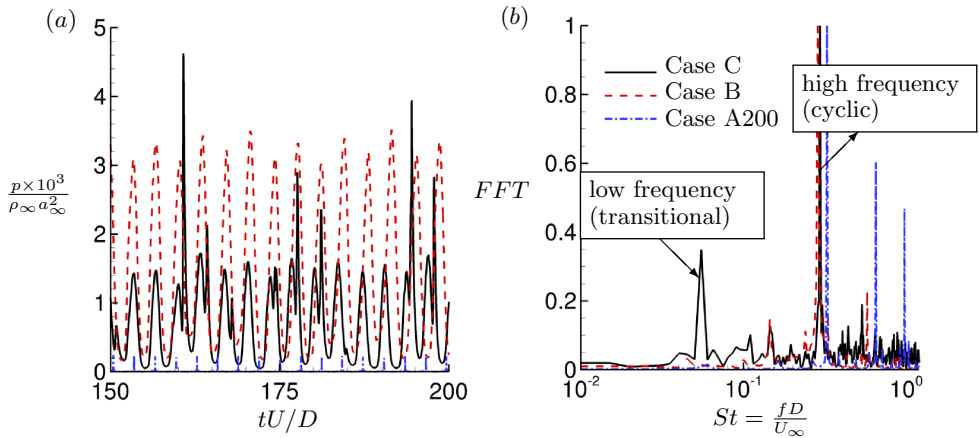

Note that $St$ corresponding to the dominant frequency of shedding plotted in figure \ref{St-Lcav} shows that the trend observed is not sensitive to the freestream nuclei content of vapor and \textit{NCG}. However, a small amount of $NCG$ does influence the secondary shedding process, which is explained through the frequency components of the pressure history in the wake of the cylinder as shown in figure \ref{pressure hist transit}. We consider $\sigma =0.85$ corresponding to the cyclic cavitation regime near transition. The pressure history of flow without \textit{NCG} exhibits both cyclic and transitional behavior; the dominant frequency corresponds to cyclic shedding. In the presence of \textit{NCG}, regardless of its freestream nuclei content, the low frequency due to regular--irregular vortex shedding (figure \ref{pressure hist transit}$(b)$) of the transitional regime is completely suppressed. Thus, the presence of \textit{NCG} can delay the transition from cyclic to transitional shedding.

\subsubsection{Dynamic mode decomposition}\label{DMD}

A detailed analysis of the behavior behind the dominant frequencies in both the cyclic and the transitional, as well as in the non--cavitating regime, is considered by performing dynamic mode decomposition (DMD) and examining the corresponding modes. DMD is a data--driven modal decomposition technique that identifies a set of modes from multiple snapshots of the observable vectors. An eingenvalue is assigned to each of these modes, which denotes the growth/decay rate and oscillation frequency of the mode. The obtained modes and their eigenvalues capture the system dynamics. We use a novel DMD algorithm developed by \cite{Anantharamu} that has low computational cost and low memory requirements.
The basic idea behind DMD is that the set of observable vectors (snapshot vectors of flow variables) $\{\psi_i\}_{i=1}^{N-1}$ can be written as a linear combination of DMD modes $\{\phi_i\}_{i=1}^{N-1}$ as

\begin{equation}
\label{dmd_equation}
\psi_i = \sum_{j=1}^{N-1} c_j \lambda_j \phi_j; i=1,...,N-1,
\end{equation}
where $\lambda_j$ are the eigenvalues of the projected linear mapping and $c_j$ are the $j^{th}$ entry of the first vector $\psi_1$. The complete derivation of the algorithm can be seen in \cite{Anantharamu}. For the cyclic and non--cavitating regime around $N=200$ snapshots of the flow field were taken with $\Delta t/(D/u_{\infty})=0.1$ between them, while $N=400$ snapshots with $\Delta t/(D/u_{\infty})=0.5$ were taken for the transitional regimes. We consider (i) the delay of K\'arm\'an vortex street transition to the two layer vortices moving from non--cavitating to the cyclic cavitation regime and (ii) comparison of mode shapes in the cyclic and the transitional regime. 

\begin{figure}
\centering
\begin{minipage}{25pc}
\includegraphics[width=25pc, trim={0 0.1cm 2.0cm 0.1cm}, clip]{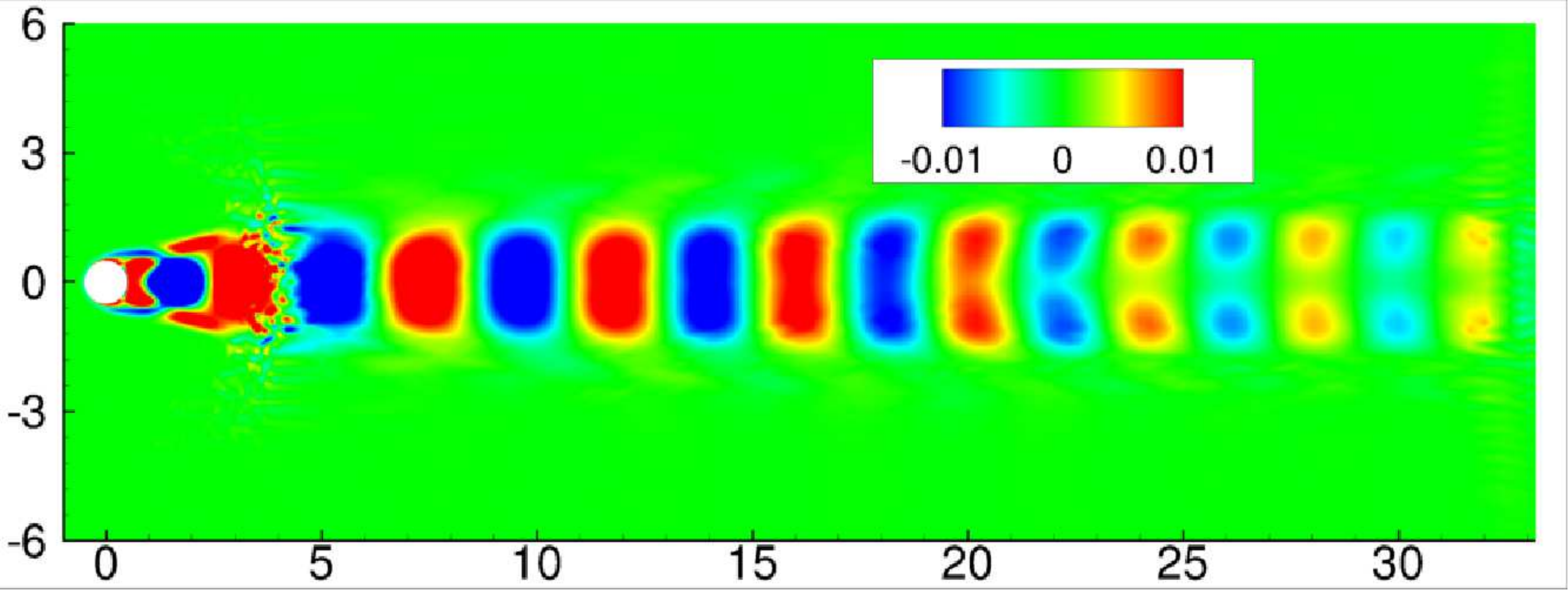}
\put(-315,105){$(a)$}
\put(-150,-10){$x/D$}
\put(-320,57){$y/D$}
\put(-240,20){\vector(-1,0){35}}
\put(-240,20){\vector(1,0){135}}
\put(-260,30){\fcolorbox{black}{white}{\parbox{40mm}{\footnotesize{Primary K\'arm\'an shedding}}}}
\put(-90,20){\vector(-1,0){5}}
\put(-90,20){\vector(1,0){85}}
\put(-98,30){\fcolorbox{black}{white}{\parbox{30mm}{\footnotesize{Two--layered vortices}}}}
\end{minipage}
\begin{minipage}{25pc}
\includegraphics[width=25pc, trim={0 0.1cm 2.0cm 0.1cm}, clip]{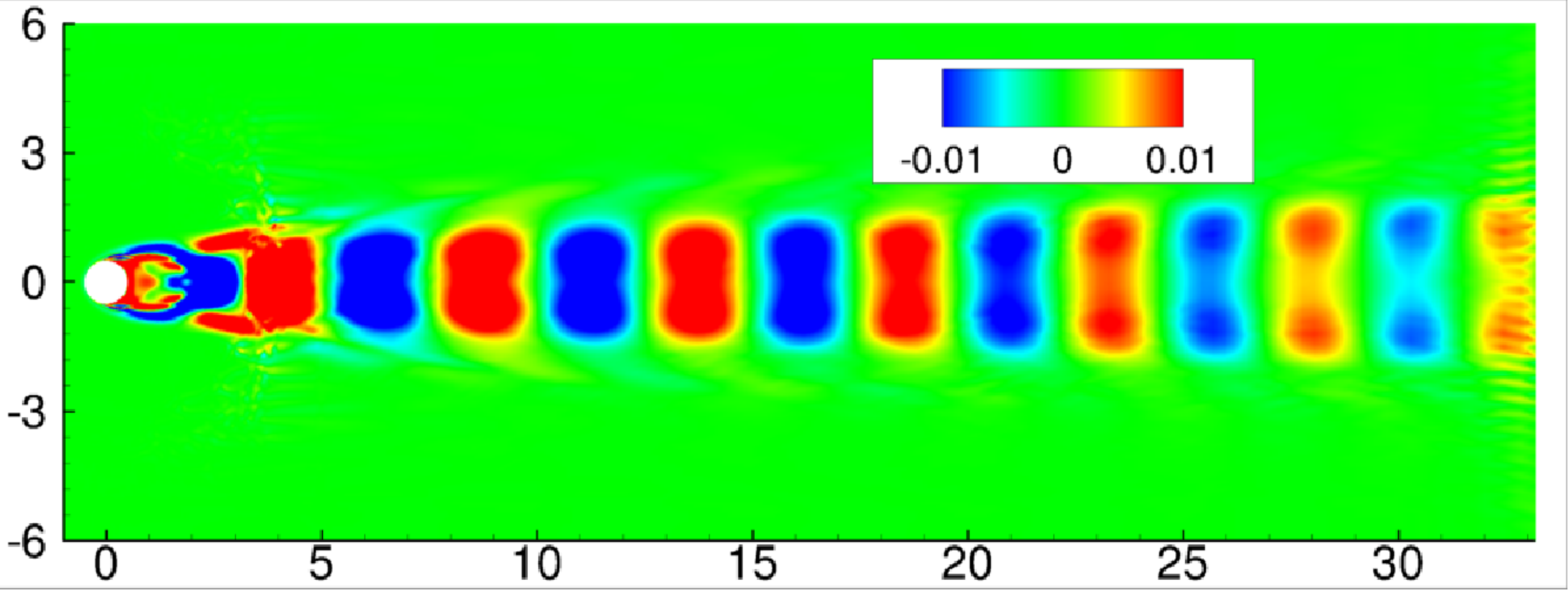}
\put(-315,105){$(b)$}
\put(-150,-10){$x/D$}
\put(-320,57){$y/D$}
\end{minipage}
\begin{minipage}{25pc}
\includegraphics[width=25pc, trim={0 0.1cm 2.0cm 0.1cm}, clip]{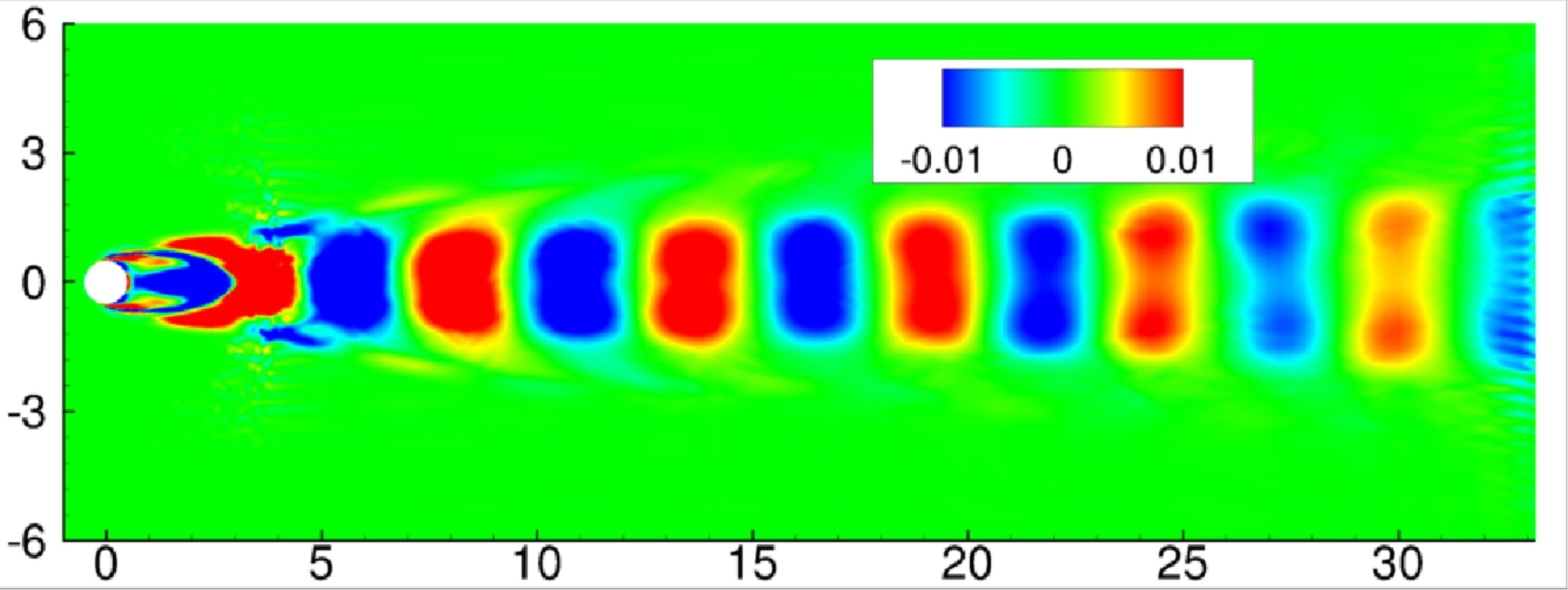}
\put(-315,105){$(c)$}
\put(-150,-10){$x/D$}
\put(-320,57){$y/D$}
\end{minipage}
\caption{Most energetic modes colored by spanwise vorticity at $\sigma=5.0$ (a), $\sigma=1.0$ (b) and $\sigma=0.85$ (c).}
\label{shedding-modes}
\end{figure}

The most dominant mode for Case A200 at non--cavitating ($\sigma=5.0$) and cavitating conditions in the cyclic regime ($\sigma=1.0$ and $\sigma=0.85$), corresponds to the dominant frequencies in lift spectra and are shown in figure \ref{shedding-modes} colored by spanwise vorticity. These dominant frequencies in lift spectra indicate the shedding frequency of individual vortices. The dominant mode in figure \ref{shedding-modes}$(a)$ clearly reveals the primary K\'arm\'an vortex street and its transition to a two--layered vortex street. 
The streamwise position of this transition is $Re$ dependent and is observed at about $x=23D$ for the non--cavitating case in figure \ref{shedding-modes}$(a)$. Comparison to the cavitating cases in figures \ref{shedding-modes}$(b)$ and $(c)$ reveals that this transition is delayed to $x=30D$ for $\sigma=1.0$ and to even farther distances at $\sigma=0.85$. This indicates that cavitation delays the first transition of the K\'arm\'an vortex street and that its distance from the cylinder grows with decreasing cavitation number.

\begin{figure}
\centering
\begin{minipage}{25pc}
\includegraphics[width=25pc, trim={0 0.1cm 0.1cm 0}, clip]{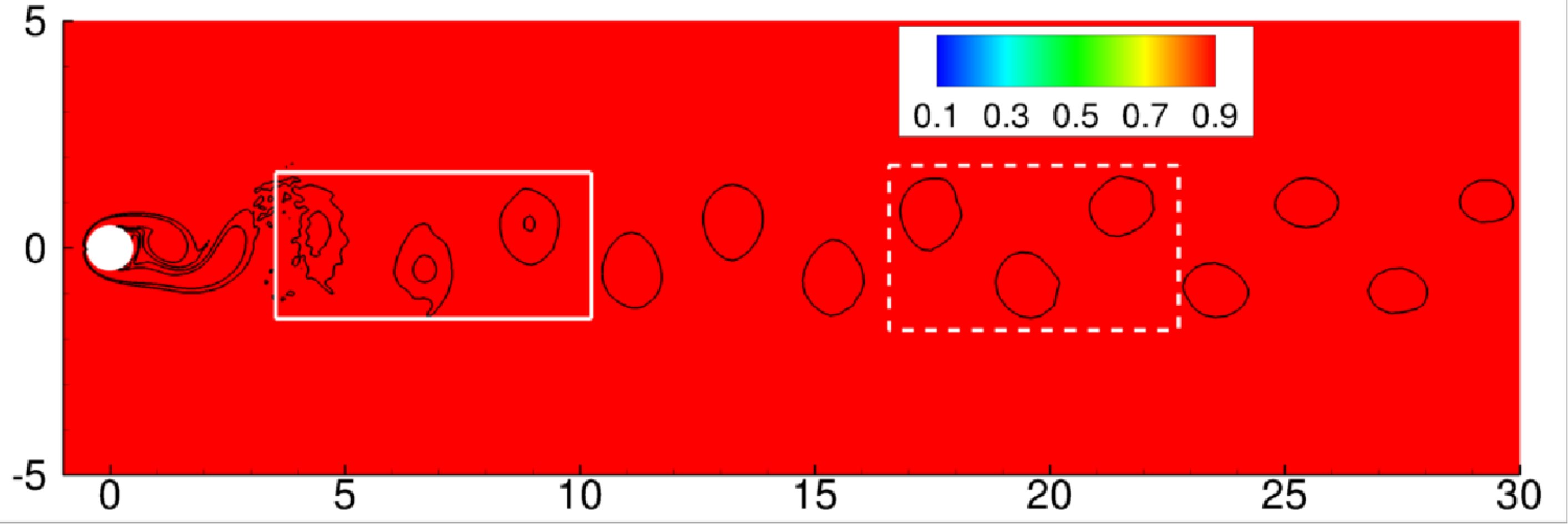}
\put(-310,100){$(a)$}
\put(-155,-10){$x/D$}
\put(-315,55){$y/D$}
\put(-200,30){{\textcolor{white}{\textbf{box 1}}}}
\put(-100,28){{\textcolor{white}{\textbf{box 2}}}}
\end{minipage}
\begin{minipage}{25pc}
\includegraphics[width=25pc, trim={0 0.1cm 0.1cm 0}, clip]{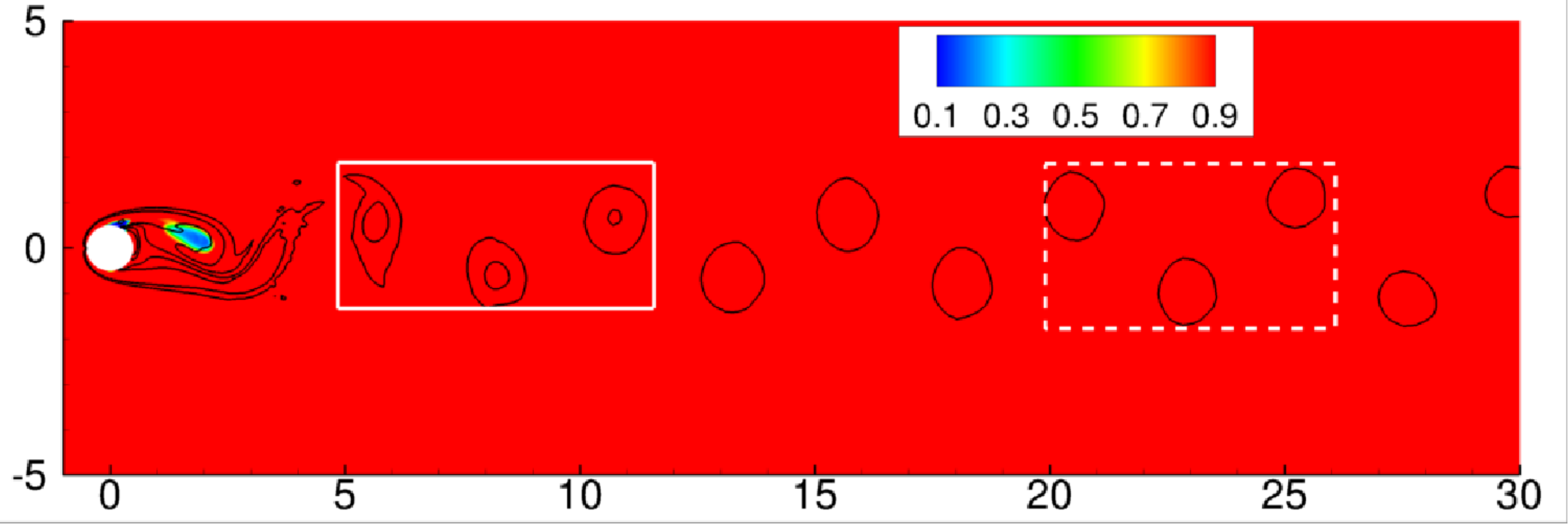}
\put(-310,100){$(b)$}
\put(-155,-10){$x/D$}
\put(-315,55){$y/D$}
\put(-200,30){{\textcolor{white}{\textbf{box 1}}}}
\put(-100,27){{\textcolor{white}{\textbf{box 2}}}}
\end{minipage}
\caption{Vortex street colored by density for $\sigma=5.0$ (a), $\sigma=1.0$ (b). White boxes indicate the region where spacing ratio is computed.}
\label{vortex_street_low_alpha}
\end{figure}

Figures \ref{vortex_street_low_alpha}$(a)$ and $(b)$ display the vortex street at $\sigma=5.0$ and $\sigma=1.0$ respectively. The vortices inside the boxes are used to compute the spacing ratio, as defined in section \ref{sec:freq}, at two streamwise positions: the first position is the closest possible to the cylinder and the second is just before the vortex street transition. Table \ref{table_vortex} shows that for $\sigma=5.0$, the spacing ratio more than doubles over a small distance, quickly surpassing the $0.366$ threshold estimated by \cite{Durgin}. Meanwhile, the spacing ratio for $\sigma=1.0$ grows slowly with streamwise distance and it is just slightly higher than the limit before the transition initiates. In order for the spacing ratio to be larger in the non--cavitating case, either the cross--wake distance $(h)$ has to be higher or the longitudinal distance $(a)$ has to be smaller. Table \ref{table_vortex} reveals that it is the longitudinal distance between same sign vortices that is smaller for $\sigma=5.0$ at the two different streamwise positions. This parameter is inversely proportional to the shedding frequency of individual vortices, which is reduced from $0.193$ to $0.175$, based on lift history, when cavitation number is lowered from $\sigma=5.0$ to $\sigma=1.0$. Thus, we can conclude that the reduction of shedding frequency due to cavitation plays a major role in delaying the first vortex street transition.

\begin{table}
 \begin{center}
  \begin{tabular}{ccccc}
         & & $h$ & $a$ & $h/a$ \\[3pt]
       \multirow{2}{4em}{$\sigma=5.0$} & box 1 & $0.845$ & $4.421$ & $0.191$ \\
                                       & box 2 & $1.633$ & $4.027$ & $0.406$ \\
       \hline
       \multirow{2}{4em}{$\sigma=1.0$} & box 1 & $1.147$ & $5.026$ & $0.228$ \\
                                       & box 2 & $1.751$ & $4.724$ & $0.371$ \\
  \end{tabular}
  \caption{Cross wake distance between different sign vortices $(h)$, longitudinal distance between same sign vortices $(a)$ and their ratio at two different wake positions for $\sigma=5.0$ and $\sigma=1.0$.}
  \label{table_vortex}
 \end{center}
\end{table}

\begin{figure}
\centering
\begin{minipage}{25pc}
\includegraphics[width=25pc, trim={0 0.1cm 2.0cm 0.1cm}, clip]{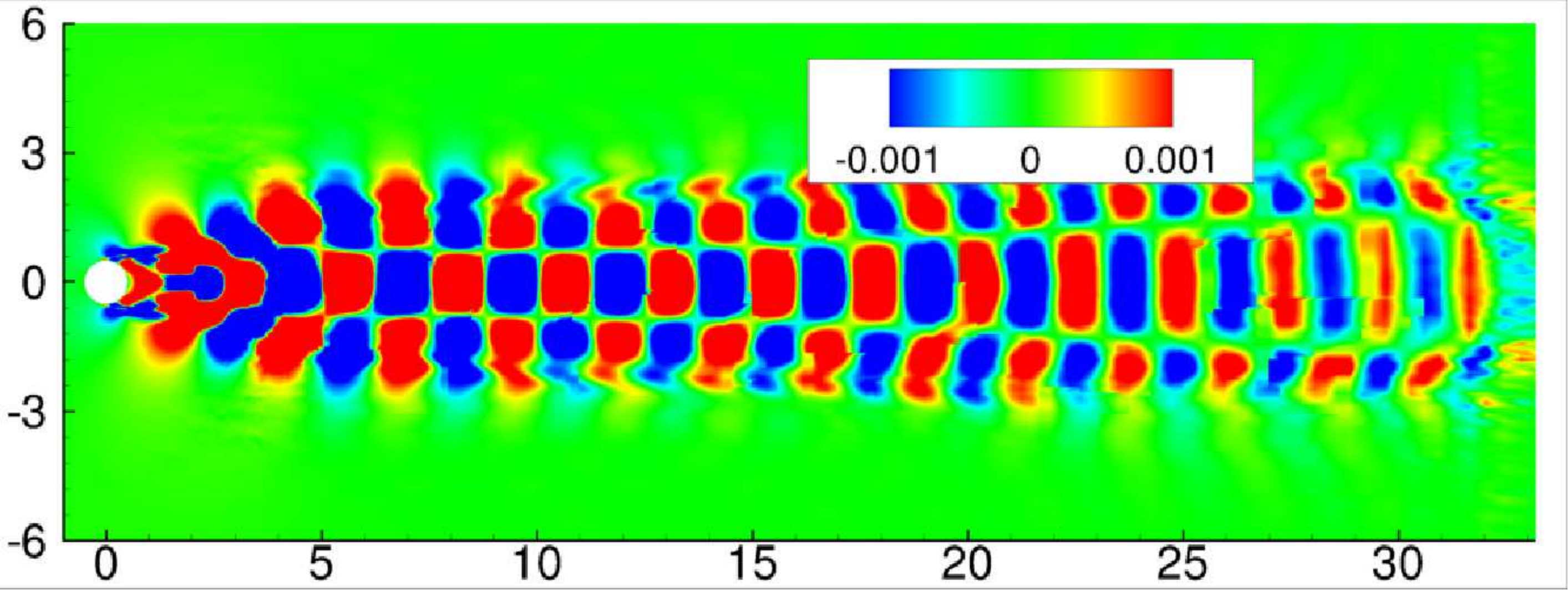}
\put(-315,105){$(a)$}
\put(-150,-10){$x/D$}
\put(-320,57){$y/D$}
\end{minipage}
\begin{minipage}{25pc}
\includegraphics[width=25pc, trim={0 0.1cm 2.0cm 0.1cm}, clip]{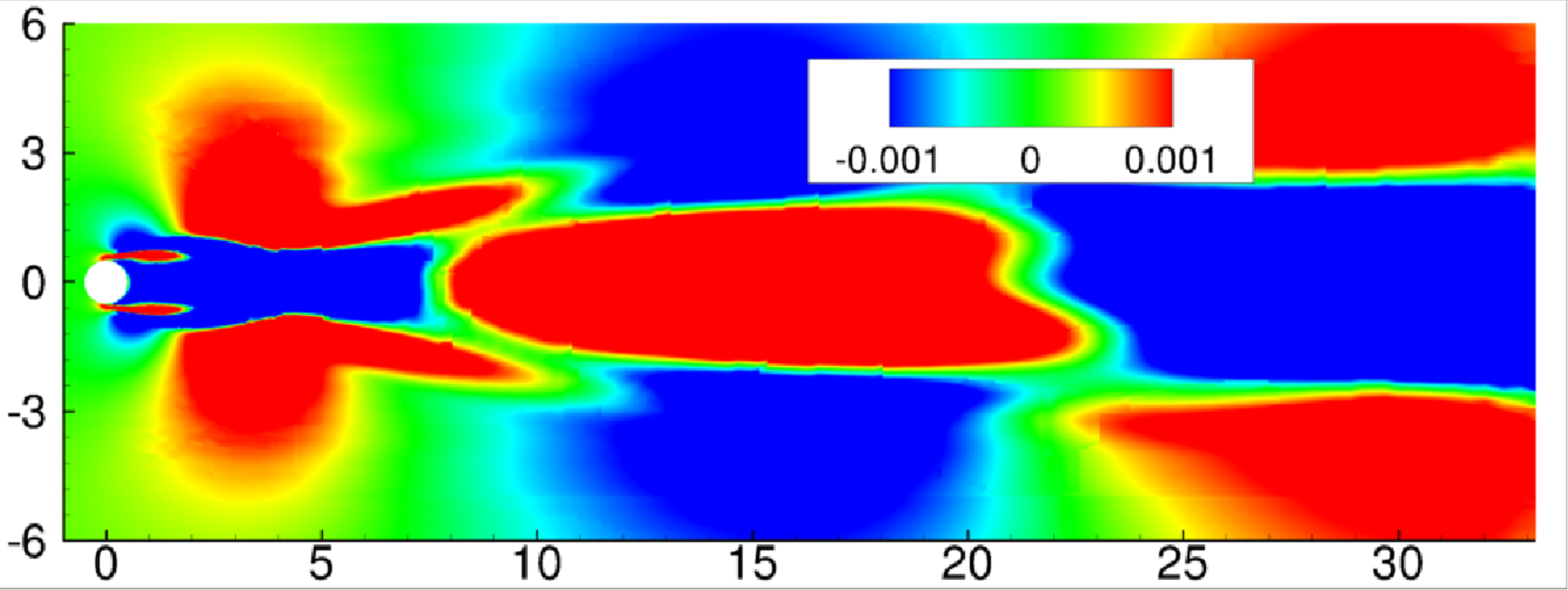}
\put(-315,105){$(b)$}
\put(-150,-10){$x/D$}
\put(-320,57){$y/D$}
\end{minipage}
\caption{Modes corresponding to drag peak frequency at $\sigma=1.0$ $(a)$ and $\sigma=0.7$ $(b)$ for Case A200, colored with streamwise velocity.}
\label{shedding-modes_drag}
\end{figure}
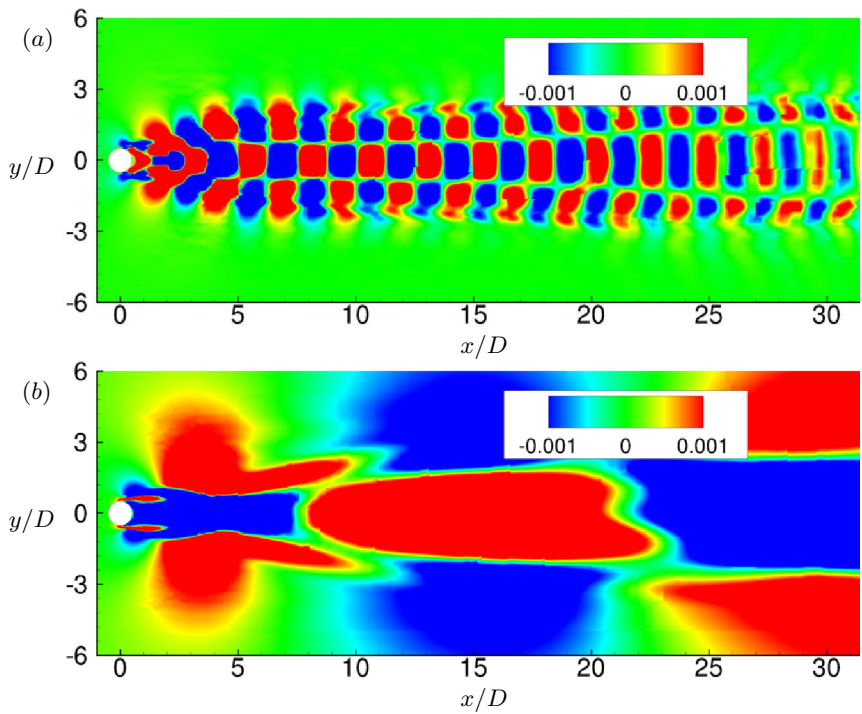

Mode shapes of axial velocity corresponding to the dominant frequency of the drag spectra are significantly altered moving from the cyclic to the transitional regime (figure \ref{shedding-modes_drag}). 
Length scales of the corresponding modes are an order of magnitude larger for the transitional regime, explaining the sharp jump in $St$. In the transitional regime (figure \ref{shedding-modes_drag}$(b)$), the modes are horizontally stretched and their length scales are significantly higher than the distance between subsequent vortex shedding as observed in the cyclic regime (figure \ref{shedding-modes_drag}$(a)$). In addition, in the transitional regime immediately following the cylinder trailing edge, the mode shows large region of negative axial velocity, suggesting the flow reversal due to the condensation front propagation. 

\subsection{Mean flow characteristics}\label{topology}

\subsubsection{Distribution of vapor and \textit{NCG} in the cylinder wake}\label{distri_lowalpha}
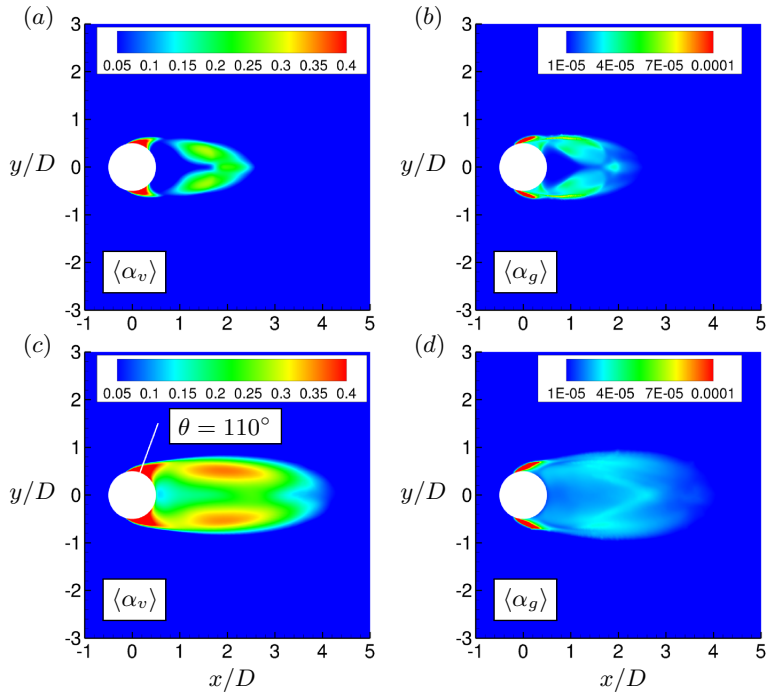
\begin{figure}
\centering
\begin{minipage}{10pc}
\includegraphics[width=10pc,trim={0 0.1cm 0.1cm 0}, clip]{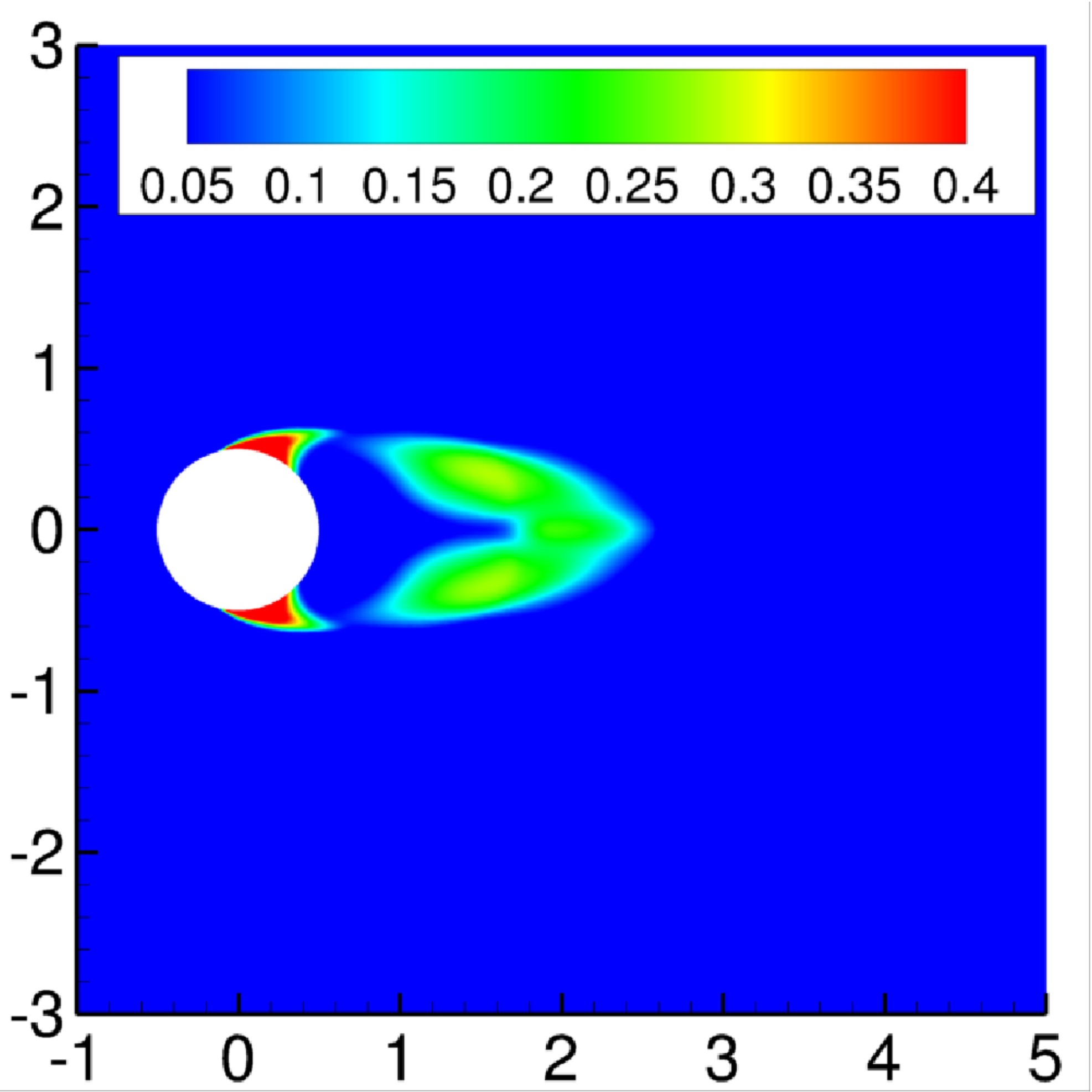}
\put(-135,115){$(a)$}
\put(-140,60){$y/D$}
\put(-105,20){\fcolorbox{black}{white}{\parbox{6mm}{\footnotesize{$\langle \alpha_v \rangle$}}}}
\end{minipage}\hspace{2pc}
\begin{minipage}{10pc}
\includegraphics[width=10pc,trim={0 0.1cm 0.1cm 0}, clip]{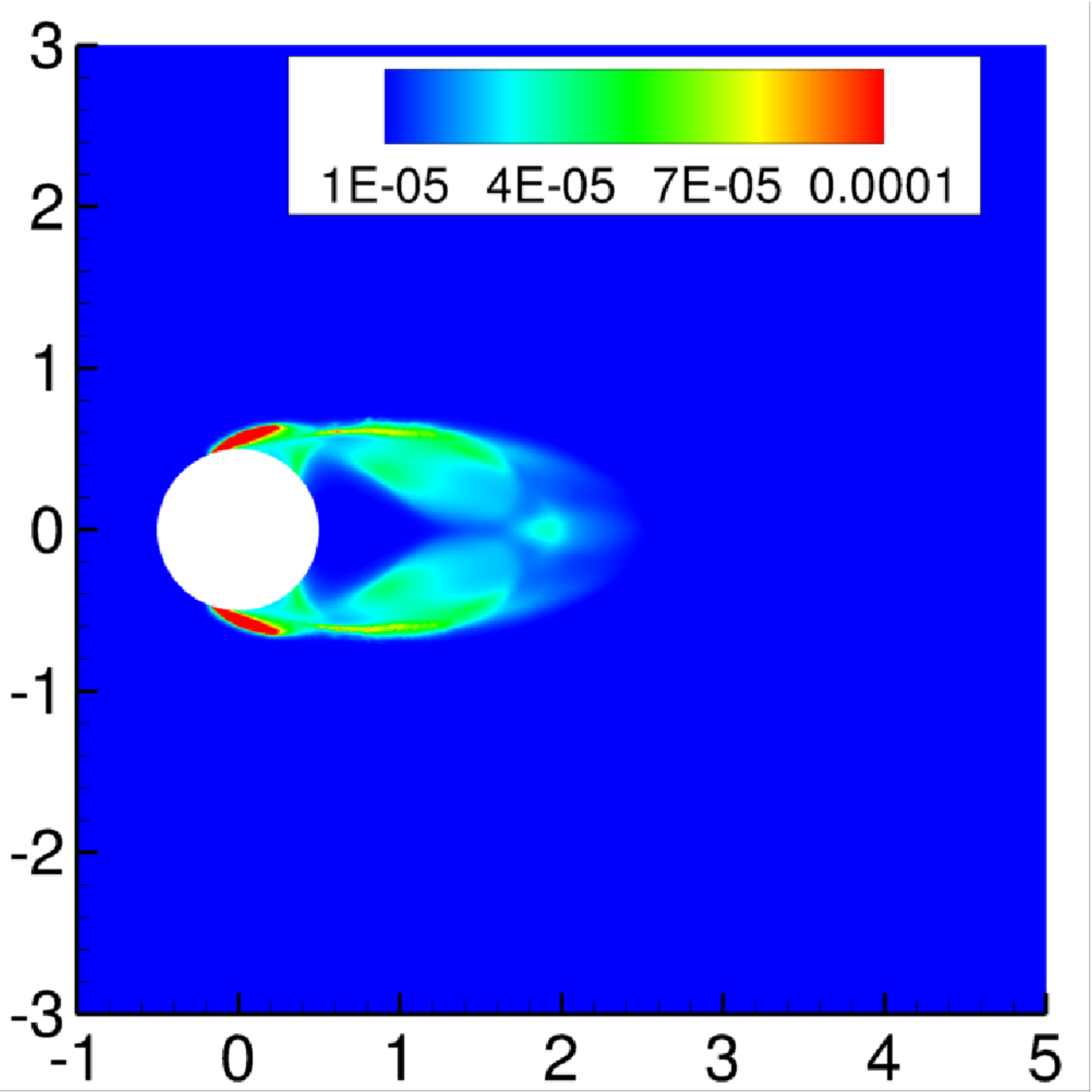}
\put(-135,115){$(b)$}
\put(-140,60){$y/D$}
\put(-105,20){\fcolorbox{black}{white}{\parbox{6mm}{\footnotesize{$\langle \alpha_g \rangle$}}}}
\end{minipage}
\begin{minipage}{10pc}
\includegraphics[width=10pc,trim={0 0.1cm 0.1cm 0}, clip]{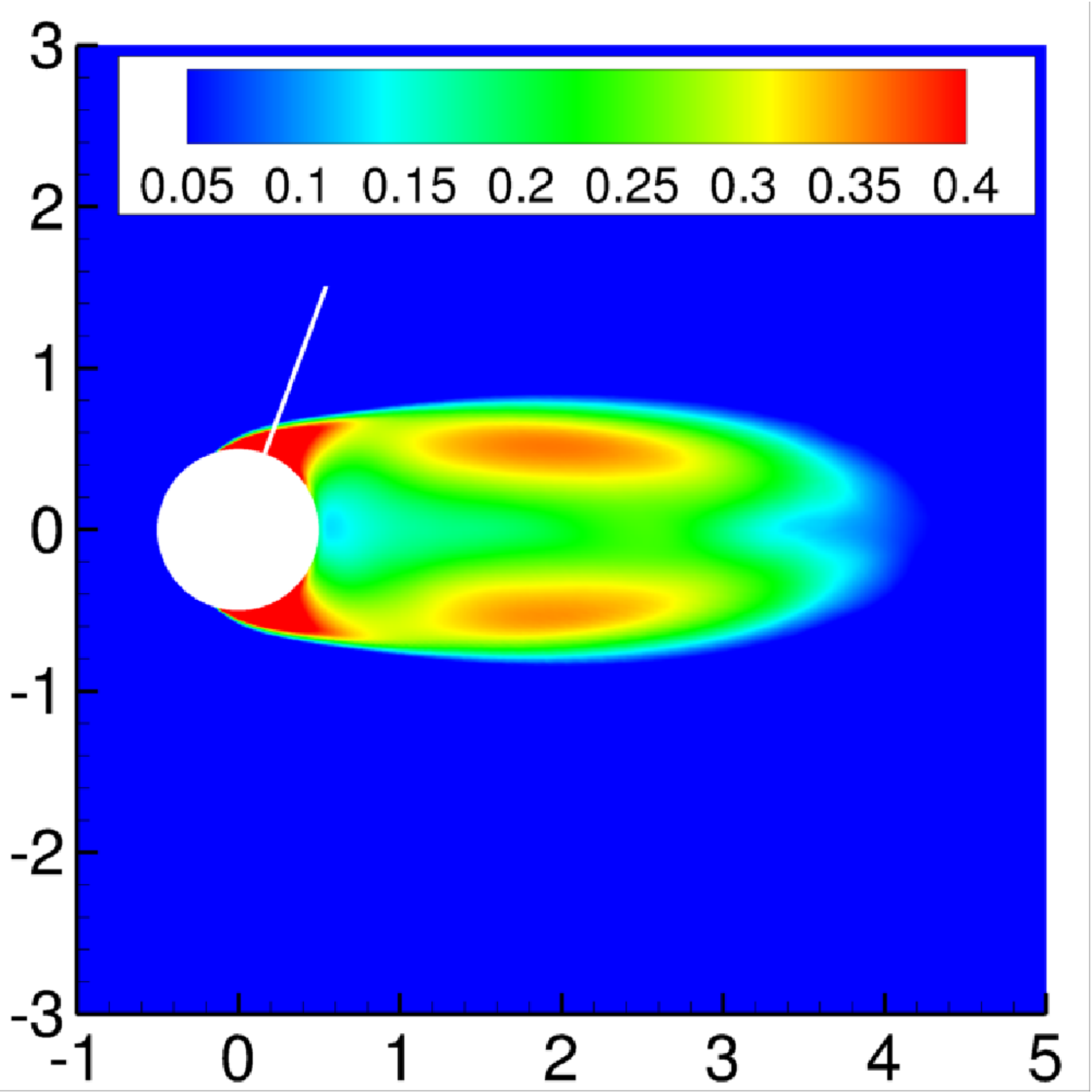}
\put(-135,115){$(c)$}
\put(-140,60){$y/D$}
\put(-65,-10){$x/D$}
\put(-105,20){\fcolorbox{black}{white}{\parbox{6mm}{\footnotesize{$\langle \alpha_v \rangle$}}}}
\put(-80,85){\fcolorbox{black}{white}{\parbox{12.5mm}{\footnotesize{$\theta = 110 \degree$}}}}
\end{minipage}\hspace{2pc}
\begin{minipage}{10pc}
\includegraphics[width=10pc,trim={0 0.1cm 0.1cm 0}, clip]{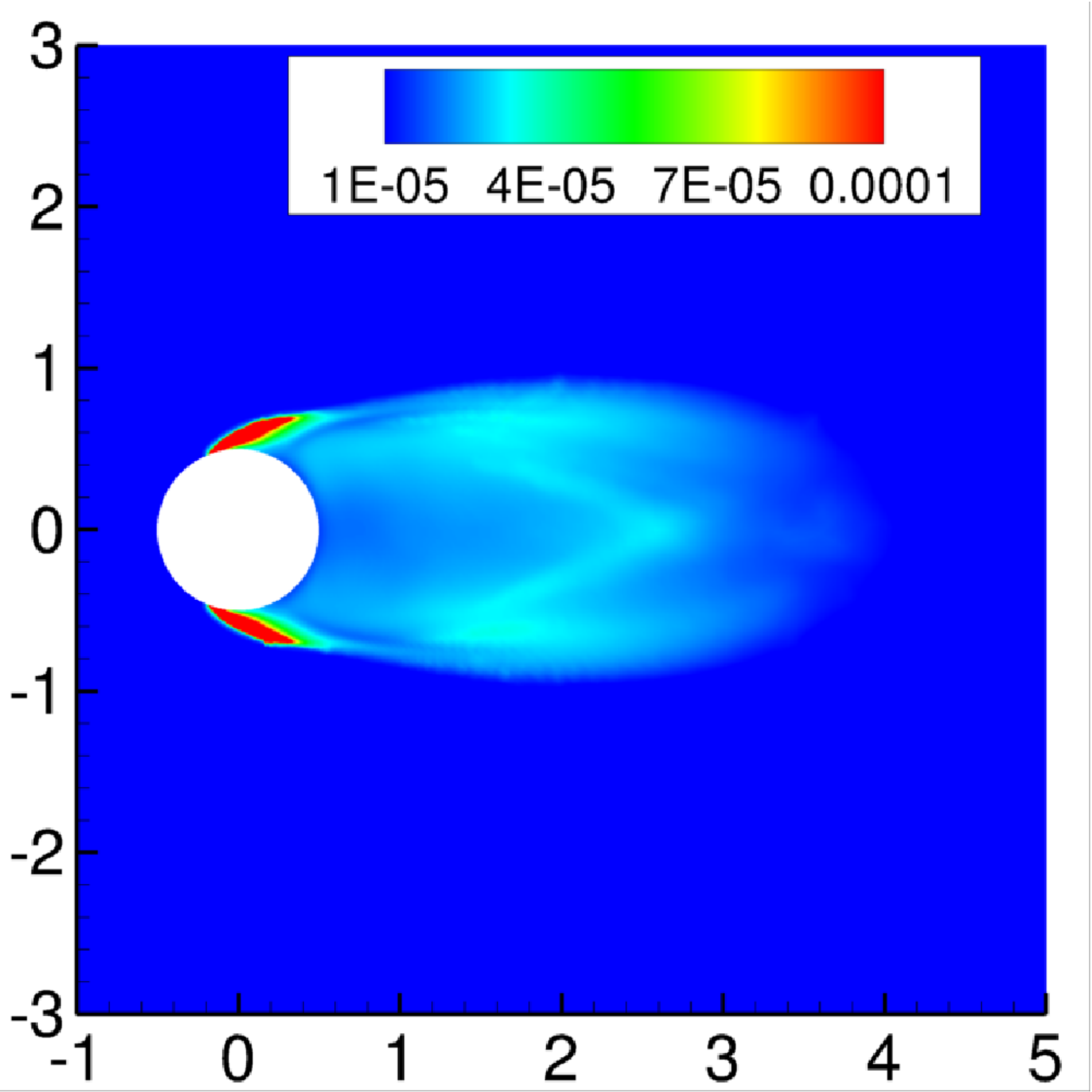}
\put(-135,115){$(d)$}
\put(-140,60){$y/D$}
\put(-65,-10){$x/D$}
\put(-105,20){\fcolorbox{black}{white}{\parbox{6mm}{\footnotesize{$\langle \alpha_g \rangle$}}}}
\end{minipage}
\caption{Time averaged vapor and \textit{NCG} volume fraction contours respectively at $\sigma = 1$ $(a,b)$ and $\sigma =0.7$ $(c,d)$ for the Case A200. White line in $(c)$ indicates azimuthal position of $110 \degree$.}
\label{fig:vfc:allsigma-lowalpha}
\end{figure}

We consider the distribution of mean volume fractions of vapor and \textit{NCG} in the near wake of cylinder for Case A200 at $\sigma = 1$ and $0.7$, respectively in the cyclic and the transitional regime, as shown in figure \ref{fig:vfc:allsigma-lowalpha}. 
In the cyclic regime (figure \ref{fig:vfc:allsigma-lowalpha}$(a,b)$), majority of the vapor is concentrated on the cylinder surface and core of shed vortices from top and bottom. Regions near the cavity trailing edge and in the immediate wake remain cavitation free. \textit{NCG} is concentrated in the incoming shear layer beginning at the cylinder surface into the near wake. Also, note that \textit{NCG} is distributed in the neighbouring regions of the vapor concentration. \textit{NCG} volume fractions are orders of magnitude smaller than vapor as additional \textit{NCG} cannot be produced through phase change and volume fractions are observed only through expansion of existing amount of gas in the freestream. 
In the transitional regime (figure \ref{fig:vfc:allsigma-lowalpha}$(c,d)$), in addition to the cylinder surface and the core of shed vortices, vapor is produced near the cavity trailing edge and in the immediate wake as the local pressure in the immediate wake drops below vapor pressure with reduction in $\sigma$. \textit{NCG} volume fractions are smaller than those observed in the cyclic regime and are concentrated mainly in the incoming shear layer.

\subsubsection{Cavitation inside the boundary layer} 

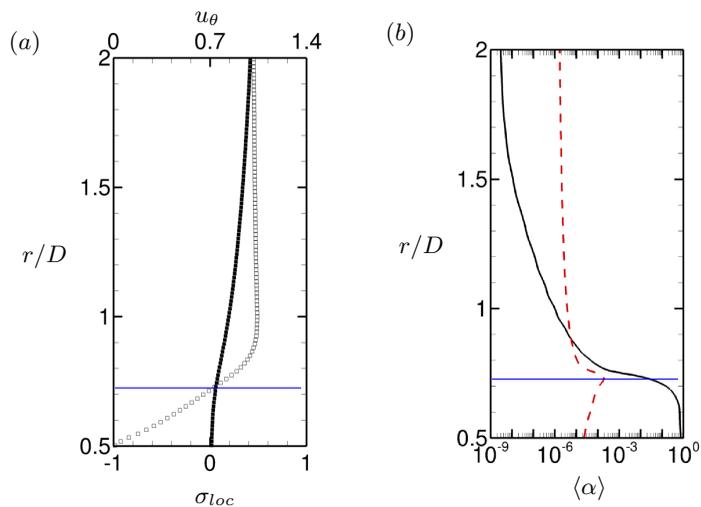
\begin{figure}
\centering
\begin{minipage}{8pc}
\includegraphics[width=8pc, trim={0 0.1cm 0.1cm 0},clip]{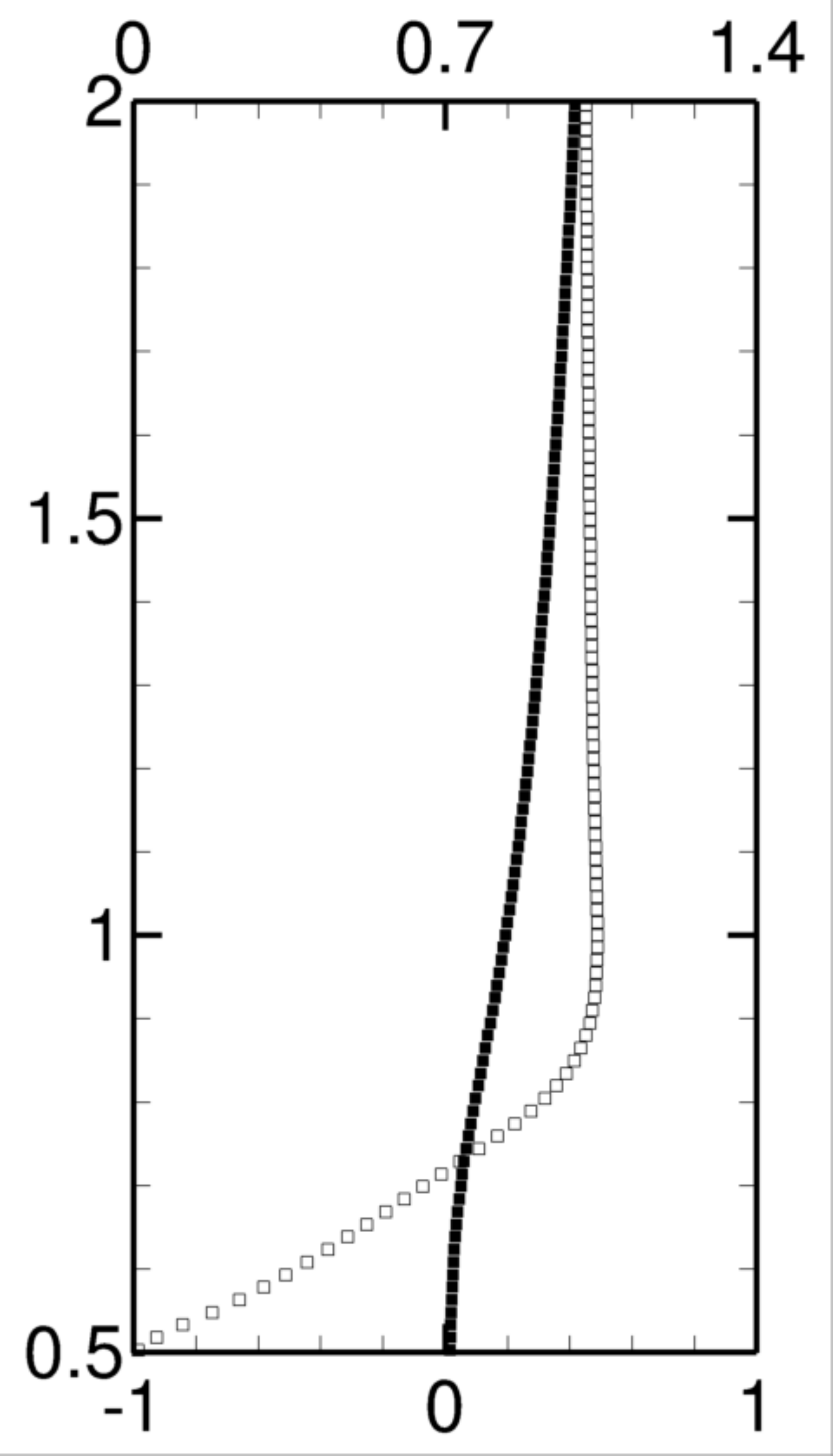}
\put(-120,160){$(a)$}
\put(-115,80){$r/D$}
\put(-50,170){$u_{\theta}$}
\put(-50,-10){$\sigma_{loc}$}
\put(-80,33){$\textcolor{blue}{\line(1,0){70}}$}
\end{minipage}\hspace{15mm}
\begin{minipage}{8pc}
\includegraphics[width=8pc, trim={0 0.1cm 0.1cm 0},clip]{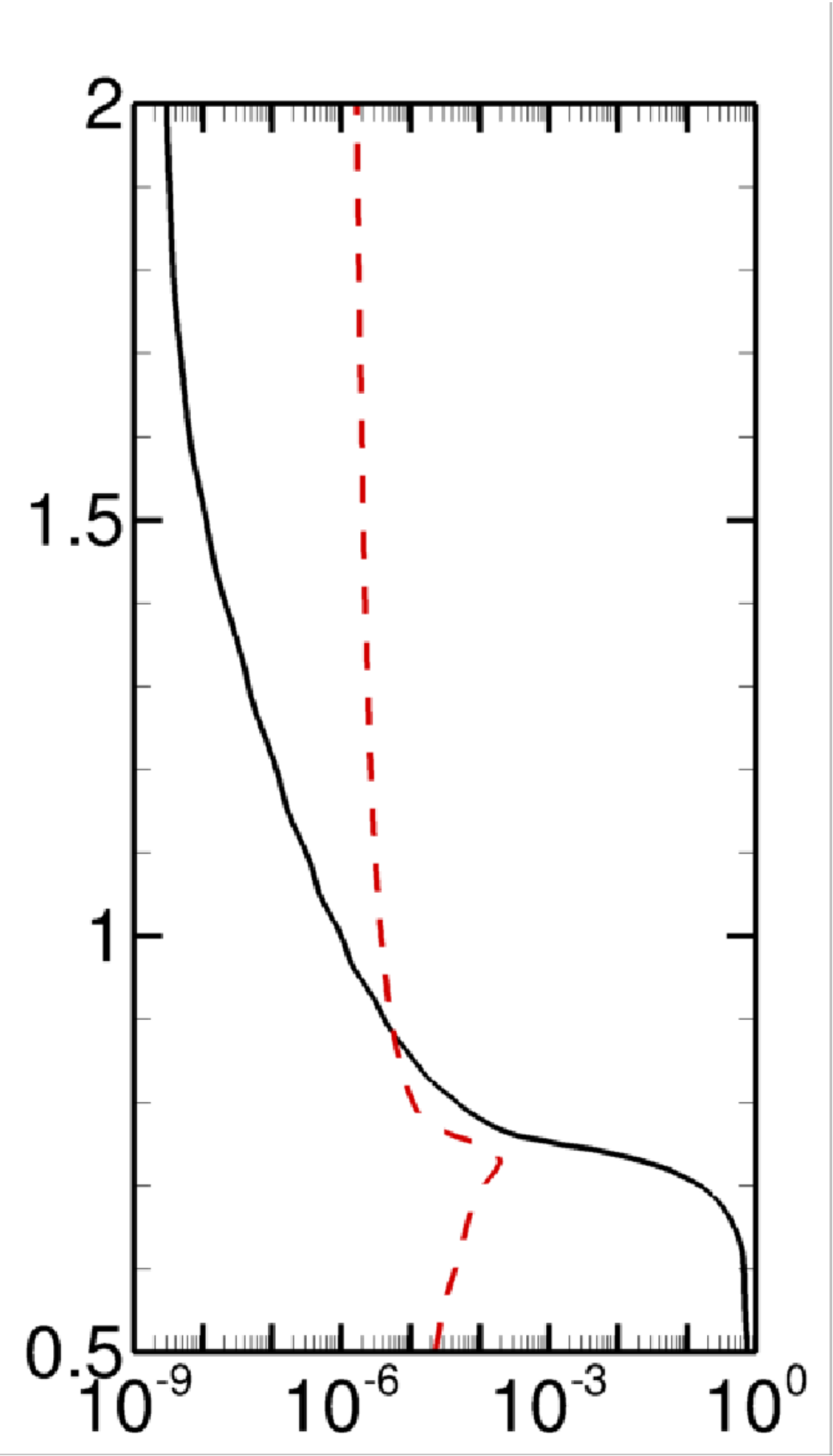}
\put(-120,160){$(b)$}
\put(-115,80){$r/D$}
\put(-50,-10){$\langle \alpha \rangle$}
\put(-80,33){$\textcolor{blue}{\line(1,0){70}}$}
\end{minipage}
\caption{Cavitation inside the boundary layer at $\sigma=0.7$ and for Case A200. Profiles taken at $110 \degree$ from the leading edge. $(a)$ Azimuthal velocity profile ($\square$) with local cavitation number ($\blacksquare$) and $(b)$ mean vapor (\textcolor{black}{$-$}) and \textit{NCG} (\textcolor{red}{$--$}) volume fraction.}  
\label{vfbl}
\end{figure}

In order to distinguish the mass transfer process due to phase change from expansion, we consider the local cavitation number which is defined as $\sigma_{loc} = \frac{p_{loc} - p_v}{0.5\rho_{\infty}u_{\infty}^2}$, where $p_{loc}$ is the local pressure inside a cell. At a given instant if $\sigma_{loc}$ is positive in the region, the observed increase in vapor volume is only due to the expansion or the advection from nearby regions. If it is negative, the resulting increase in the volume of vapor is also accompanied by mass transfer. Consequently, in the regions of negative $\sigma_{loc}$, we expect the vapor to distinguish itself from \textit{NCG}. We choose $\sigma =0.7$ and Case A200 for explanation.  
Figure \ref{vfbl}$(a)$ shows boundary layer profile radially at $110 \degree$ from leading edge (as indicated in figure \ref{fig:vfc:allsigma-lowalpha}$(c)$) of the cylinder along with $\sigma_{loc}$. The region separating positive $\sigma_{loc}$ within the boundary layer is indicated by the solid blue line. Note that vapor and \textit{NCG} volume fractions deviate significantly in this region (figure \ref{vfbl}$(b)$) as vapor is produced due to the mass transfer. The maximum \textit{NCG} volume fraction is observed at $\sigma_{loc}=0$. As one moves radially outward, both vapor and \textit{NCG} gas volume fractions are comparable in the remaining regions within the boundary layer, predominantly due to the expansion and the advection process; finally reaching to the corresponding freestream values. Hence, cavitation as a mass transfer process is only observed in a finite near--wall region within the boundary layer.

\subsubsection{Effect of freestream nuclei}

\begin{figure}
\centering
\begin{minipage}{10pc}
\includegraphics[width=10pc, trim={0 0.1cm 0.1cm 0},clip]{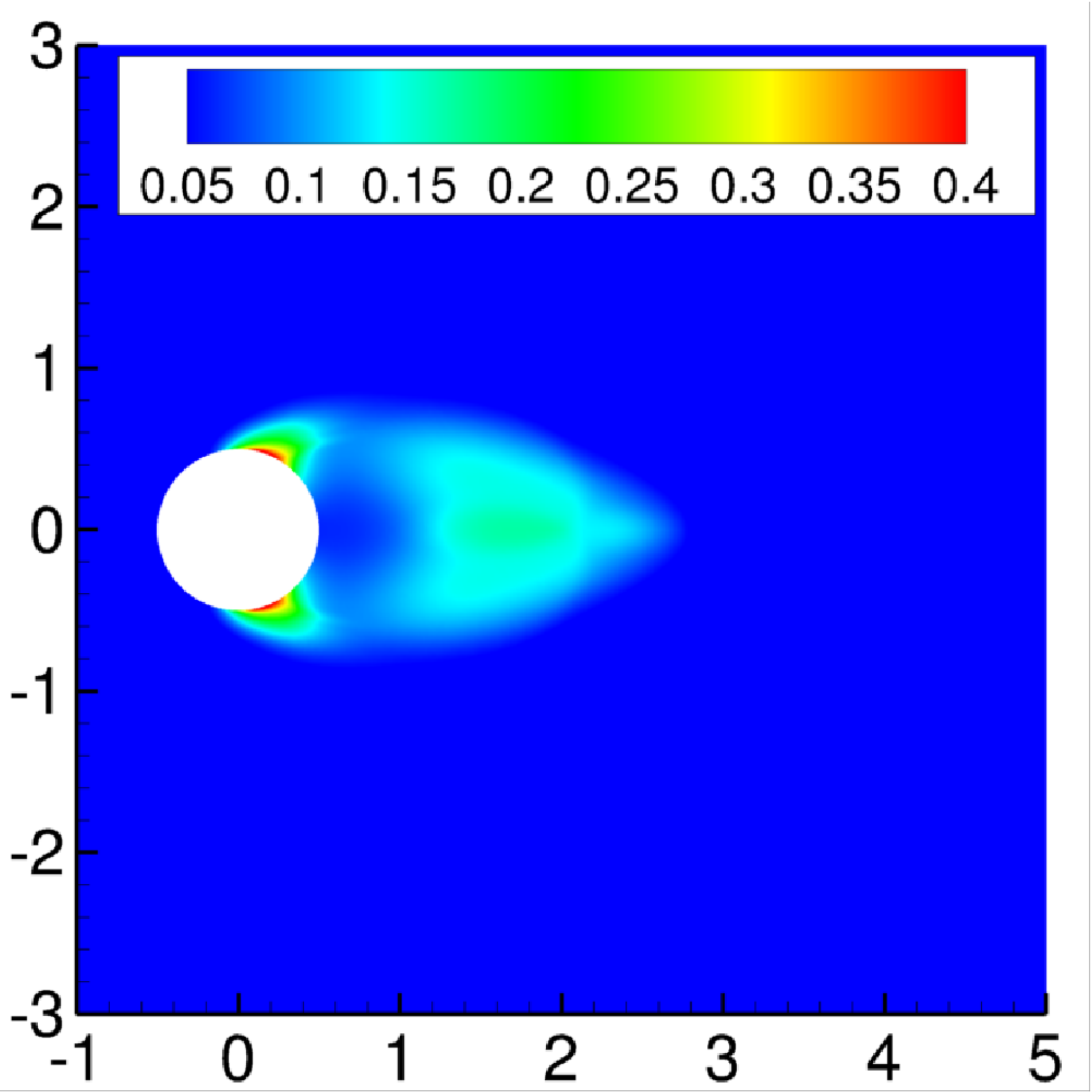}
\put(-140,60){$y/D$}
\put(-135,110){$(a)$}
\put(-105,20){\fcolorbox{black}{white}{\parbox{6mm}{\footnotesize{$\langle \alpha_v \rangle$}}}}
\end{minipage}\hspace{2pc}
\begin{minipage}{10pc}
\includegraphics[width=10pc, trim={0 0.1cm 0.1cm 0},clip]{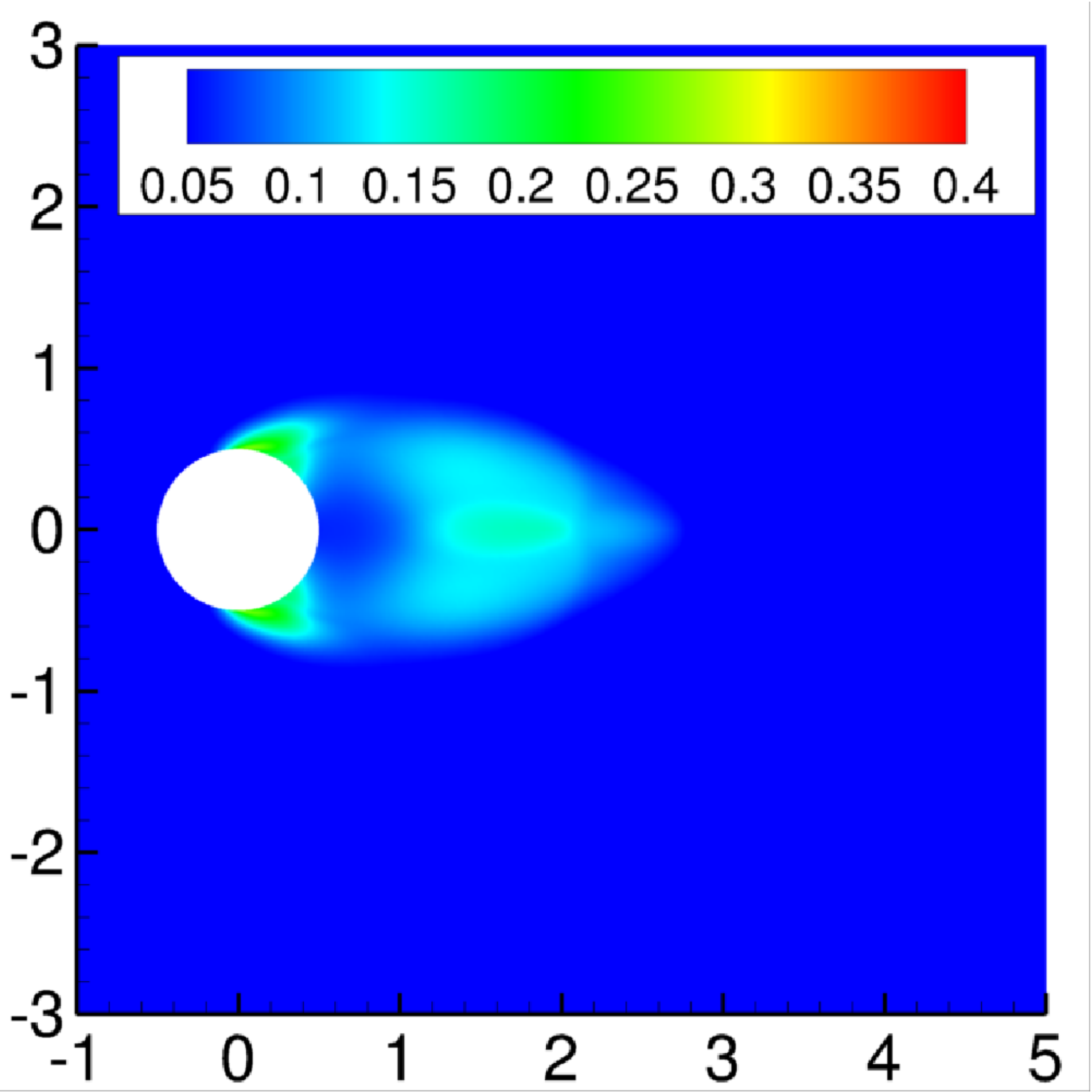}
\put(-135,110){$(b)$}
\put(-140,60){$y/D$}
\put(-105,20){\fcolorbox{black}{white}{\parbox{6mm}{\footnotesize{$\langle \alpha_g \rangle$}}}}
\end{minipage}
\begin{minipage}{10pc}
\includegraphics[width=10pc, trim={0 0.1cm 0.1cm 0},clip]{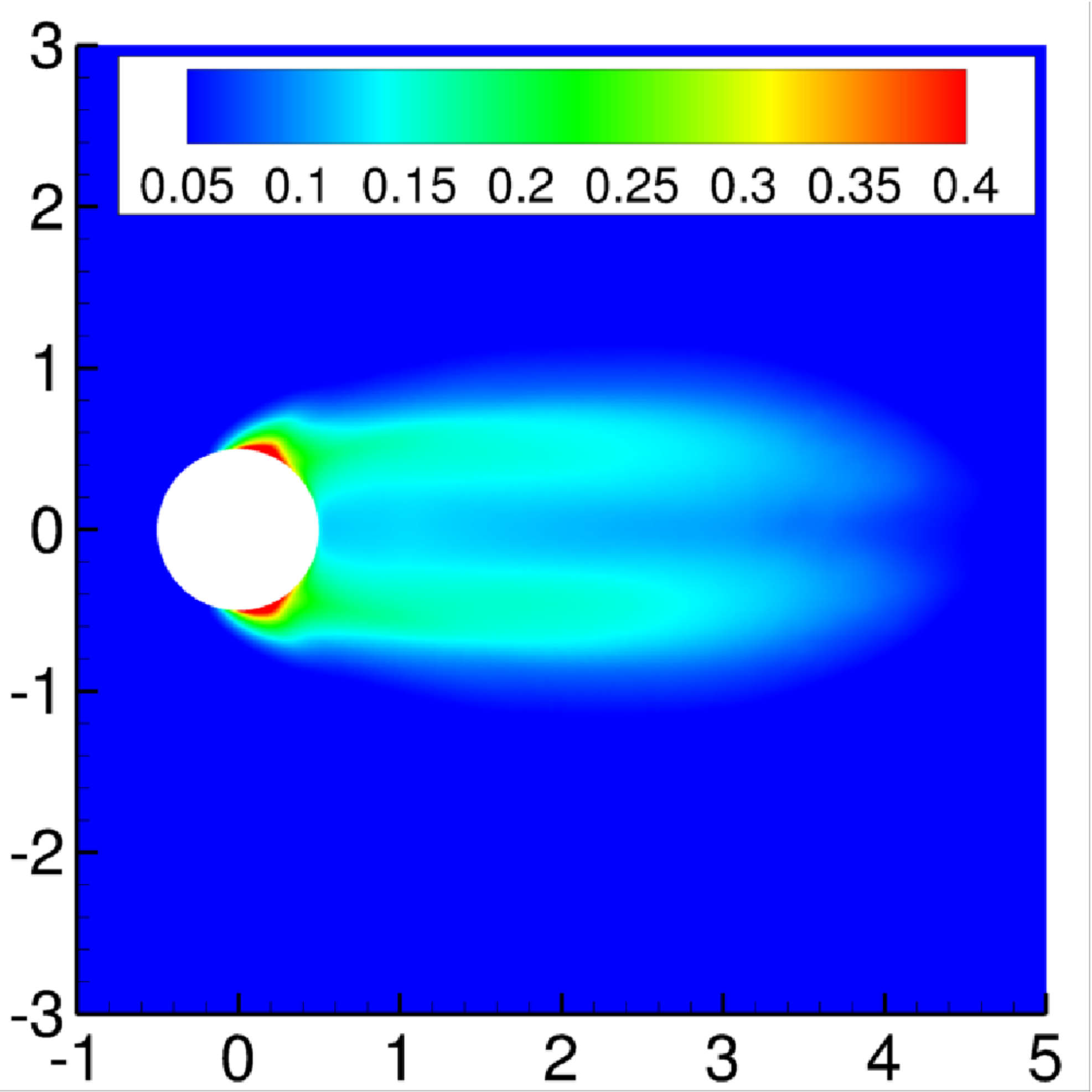}
\put(-135,110){$(c)$}
\put(-140,60){$y/D$}
\put(-65,-10){$x/D$}
\put(-105,20){\fcolorbox{black}{white}{\parbox{6mm}{\footnotesize{$\langle \alpha_v \rangle$}}}}
\end{minipage}\hspace{2pc}
\begin{minipage}{10pc}
\includegraphics[width=10pc, trim={0 0.1cm 0.1cm 0},clip]{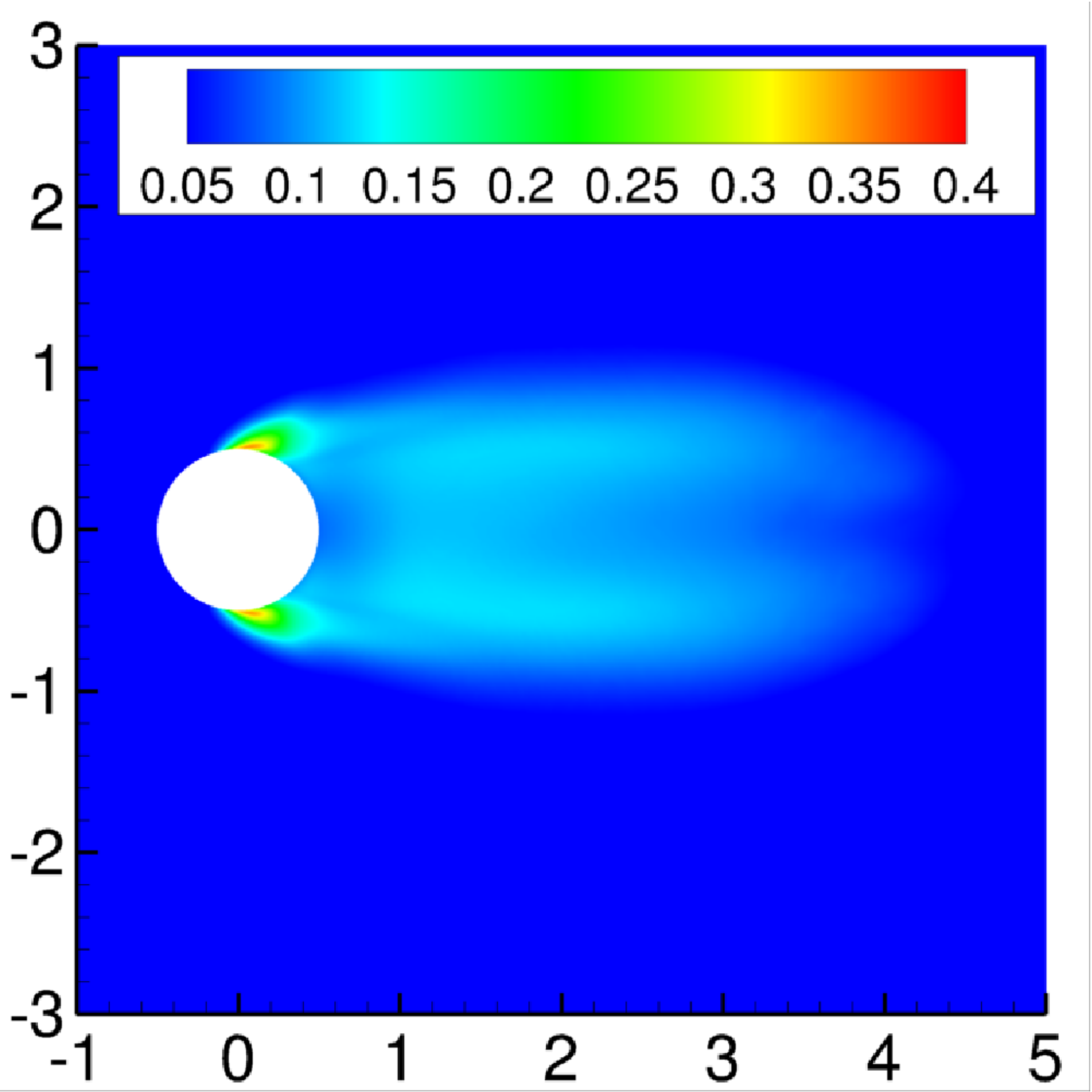}
\put(-135,110){$(d)$}
\put(-140,60){$y/D$}
\put(-65,-10){$x/D$}
\put(-105,20){\fcolorbox{black}{white}{\parbox{6mm}{\footnotesize{$\langle \alpha_g \rangle$}}}}
\end{minipage}
\caption{Time averaged vapor and \textit{NCG} volume fraction contours respectively at $\sigma = 1$ $(a,b)$ and $\sigma =0.7$ $(c,d)$ for Case B.}
\label{fig:vfc:allsigma}
\end{figure}
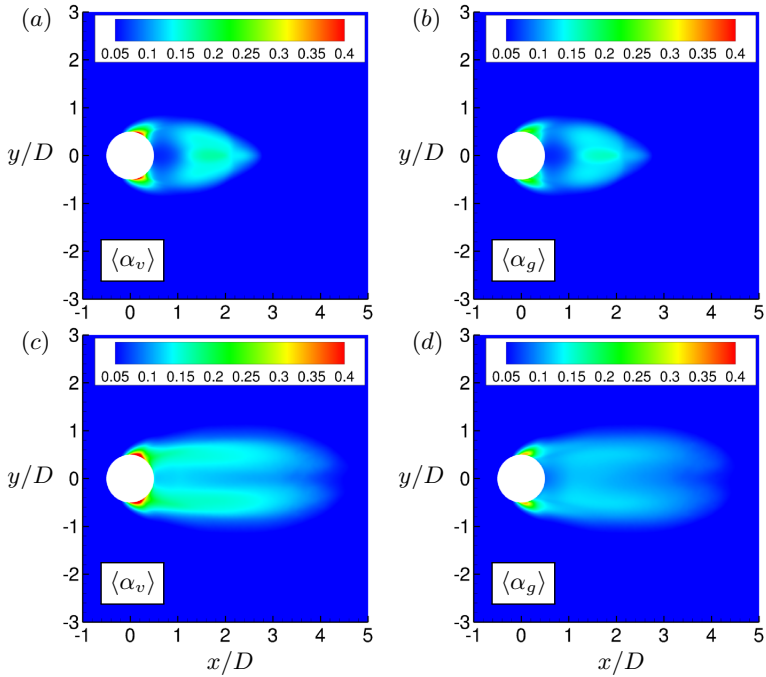

We discuss two important effects of the freestream nuclei content: i) distribution of vapor/\textit{NCG} in the near wake and ii) laminar separation of the boundary--layer. Figure \ref{fig:vfc:allsigma} shows vapor/\textit{NCG} distribution as discussed in the section \ref{distri_lowalpha}, although at a high concentration of vapor/\textit{NCG} (Case B). In the cyclic regime, vapor volume fractions show only minor difference in magnitude and distribution as compared to the low freestream nuclei concentration (Case A200, in figure \ref{fig:vfc:allsigma-lowalpha}). However, the \textit{NCG} volume fraction is orders of magnitude higher as compared to the low freestream nuclei case (figure \ref{fig:vfc:allsigma-lowalpha}$(b)$, figure \ref{fig:vfc:allsigma}$(b)$) and its distribution is almost indistinguishable from vapor at high nuclei concentration (figure \ref{fig:vfc:allsigma}$(a,b)$). As \textit{NCG} does not undergo phase change, its initial concentration in the freestream has a very significant effect on the wake of the cylinder. While due to the significant effect of mass transfer, vapor is not as sensitive as \textit{NCG} to the initial nuclei content. The same is also observed in the transitional regime (figure \ref{fig:vfc:allsigma}$(c,d)$). Note that vapor/gas diffusion can influence the distribution shown in the near wake. Although, we are unable to consider it at the current level of modeling.

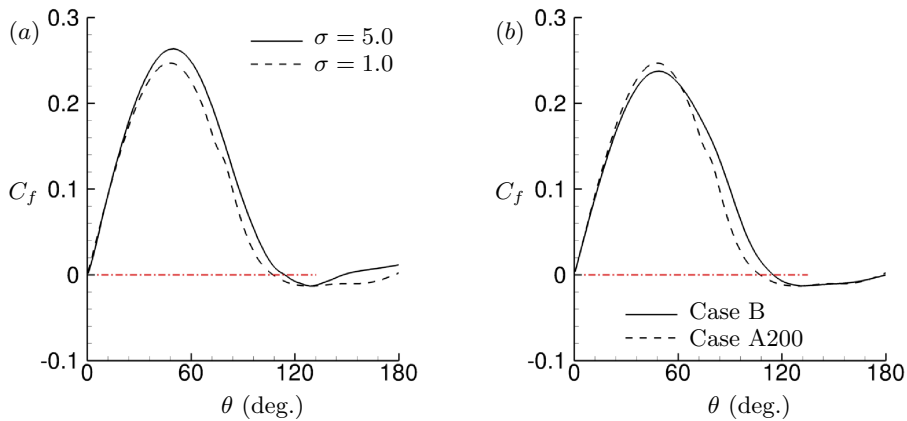
\begin{figure}
\centering
\begin{minipage}{12pc}
\includegraphics[width=12pc, trim={0 0.1cm 0.1cm 0}, clip]{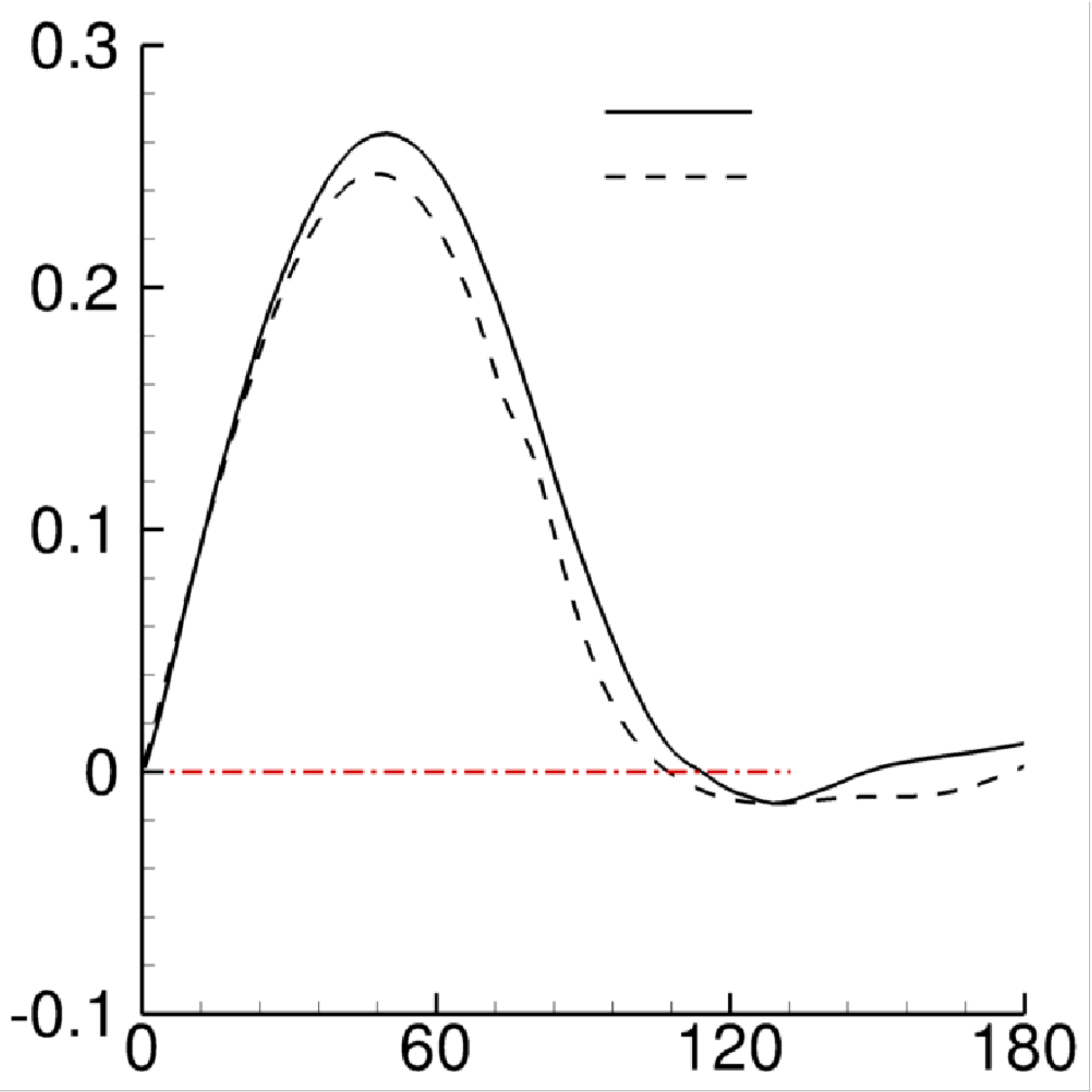}
\put(-75,-10){$\theta$ (deg.)}
\put(-155,70){$C_f$}
\put(-155,130){$(a)$}
\put(-40,128){$\sigma = 5.0$}
\put(-40,118){$\sigma = 1.0$}
\end{minipage}\hspace{3pc}
\begin{minipage}{12pc}
\includegraphics[width=12pc, trim={0 0.1cm 0.1cm 0}, clip]{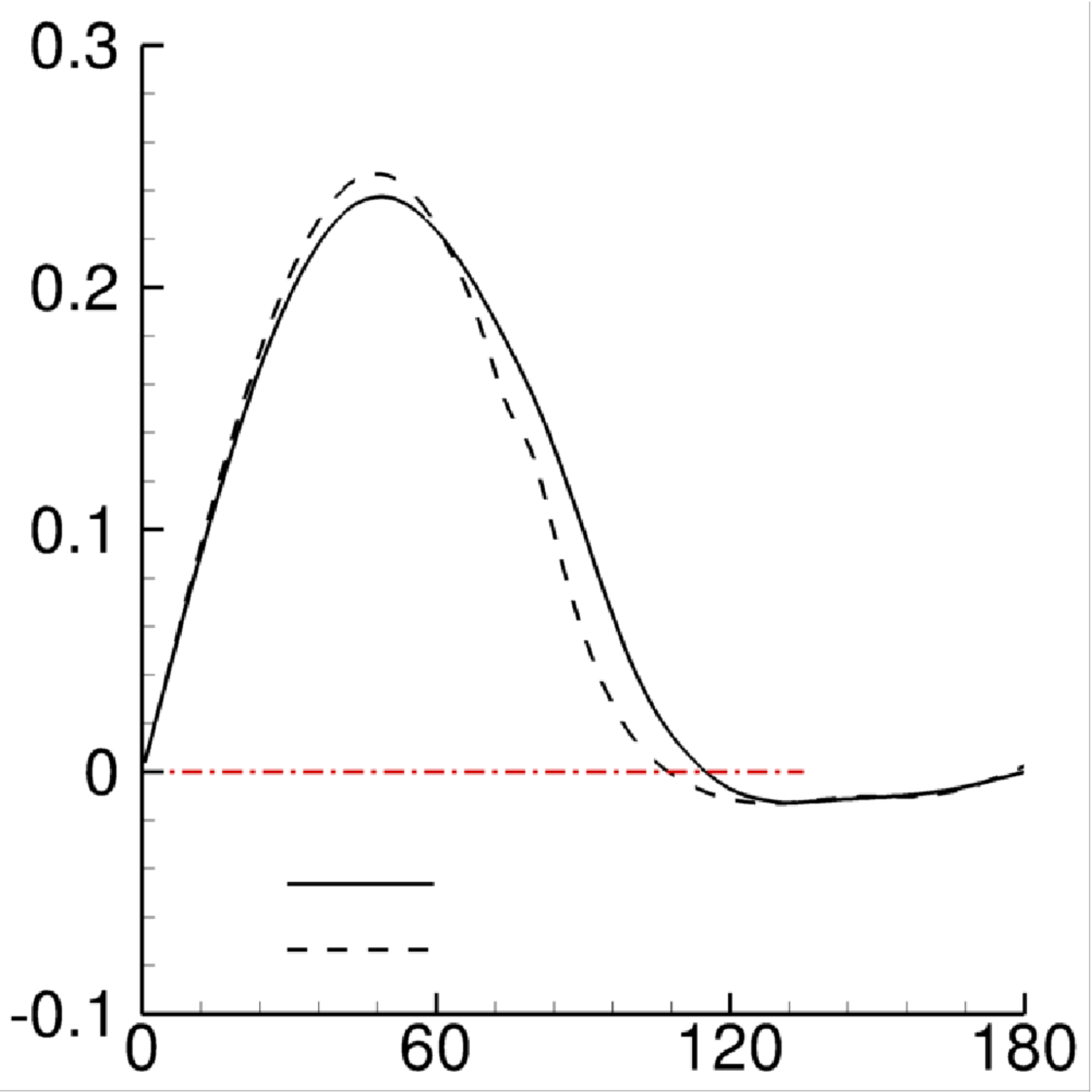}
\put(-75,-10){$\theta$ (deg.)}
\put(-155,70){$C_f$}
\put(-155,130){$(b)$}
\put(-82,25){Case B}
\put(-82,15){Case A200}
\end{minipage}
\caption{Skin friction at different $\sigma$ for Case A200 $(a)$ and at $\sigma = 1.0$ for different cases $(b)$. The dash red line represents $C_f = 0$ and indicates the separation point.}
\label{skin-friction}
\end{figure}

One point of divergence between experiments and simulations using the homogeneous mixture approach involves the location of boundary--layer separation. While experiments show that the boundary--layer separation point moves upstream along the cylinder as the flow cavitates \citep{Arakeri}, the same was not observed numerically by \cite{AswinJFM}. The reason behind this discrepancy, as explained in \cite{AswinJFM}, is the fact that the homogeneous mixture approach predicts the inception point upstream of the boundary--layer separation point. In our simulations with low freestream void fraction (Case A200), the inception point is also observed to be upstream of the separation point (not shown here). Differently from the work in \cite{AswinJFM}, where a high freestream void fraction is employed, the boundary--layer separation point is shifted upstream as the cavitation number is reduced from non--cavitating condition to a cavitating one when the freestream contains small amounts of vapor and gas (Case A200). This is evident from figure \ref{skin-friction}$(a)$ that shows the skin friction along the cylinder surface, with the separation point shifting from $115\degree$ to $106\degree$ as $\sigma$ is reduced from 5.0 to 1.0. 

It is found that the contents of vapor and {\it NCG} in the freestream have significant impact on the separation point, as displayed in figure \ref{skin-friction}$(b)$, with it moving downstream (from $106\degree$ to $116\degree$) as the freestream volume fraction increases. In \cite{AswinJFM}, the authors discussed that the cavitation would start as soon as the local pressure is reduced to values below the vapor pressure. The expansion due to cavitation would then push the separation point downstream. The same behavior is observable for Case A200. Beside this is the fact that before the flow cavitates, both vapor and gas traveling along the cylinder surface expand due to a decrease in pressure. Both the ideal gas expansion and the subsequently expansion due to phase change contribute to pushing the boundary--layer separation further downstream. The reason for Case A200 showing the correct change in the boundary--layer separation point as the flow cavitates in comparison with the high volume fraction cases, however, lies on the fact that as the amounts of vapor and gas at the cylinder surface are substantially reduced, so are their effects on the flow due to ideal gas expansion. This leads to the conclusion that by adding {\it NCG} to a cavitating flow, the separation point would move downstream.

\subsection{Condensation shock}\label{bubbly}

\begin{figure}
\centering
\begin{minipage}{13pc}
\includegraphics[width=13pc, trim={0 0.1cm 0.1cm 0}, clip]{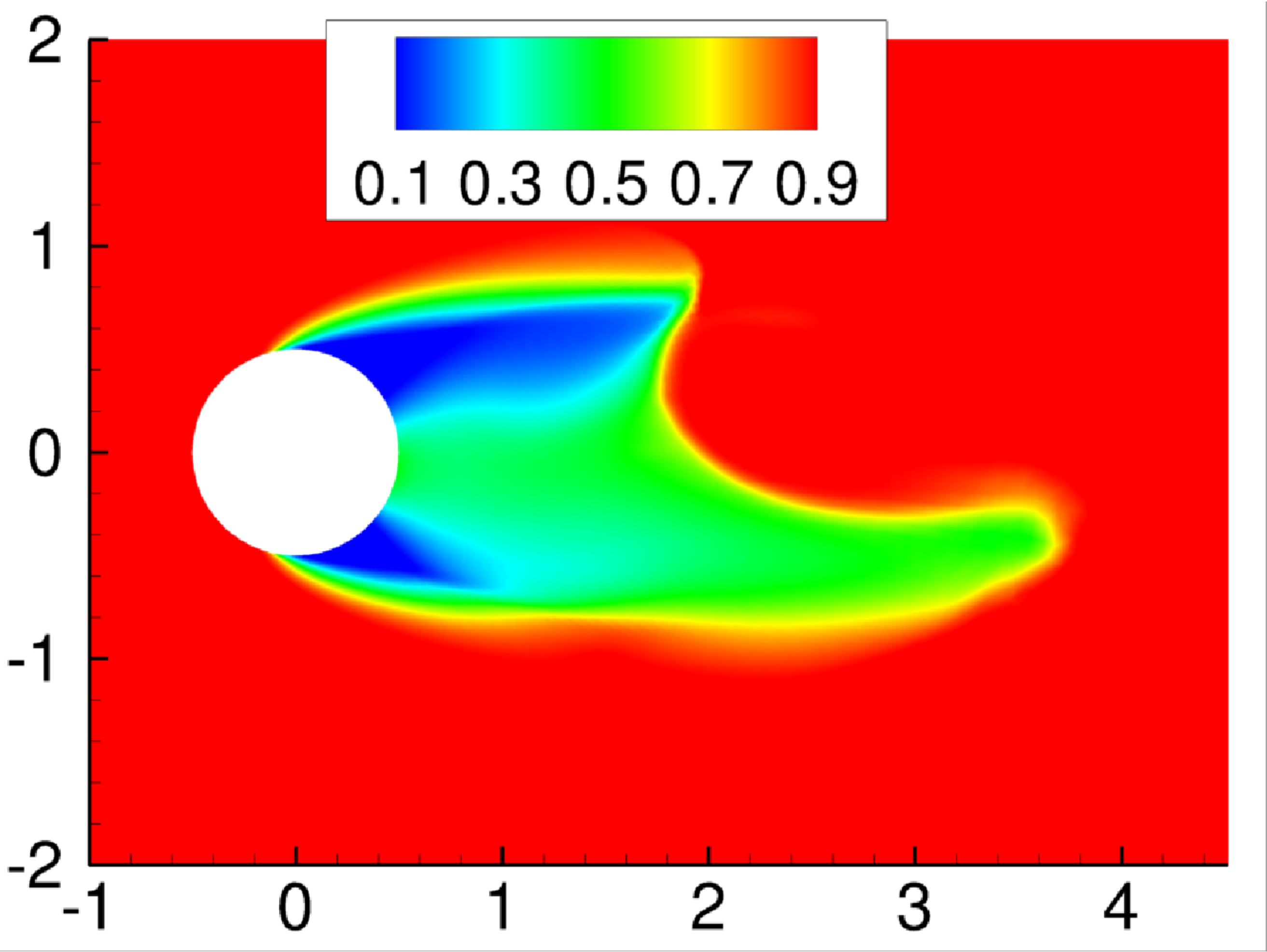}
\put(-177,57){$y/D$}
\put(-175,110){$(a)$}
\put(-50,75){{\textcolor{black}{\vector(-1,-1){15}}}}
\put(-70,85){\fcolorbox{black}{white}{\parbox{20mm}{\footnotesize{Condensation \\ front}}}}
\end{minipage}\hspace{3pc}
\begin{minipage}{13pc}
\includegraphics[width=13pc, trim={0 0.1cm 0.1cm 0}, clip]{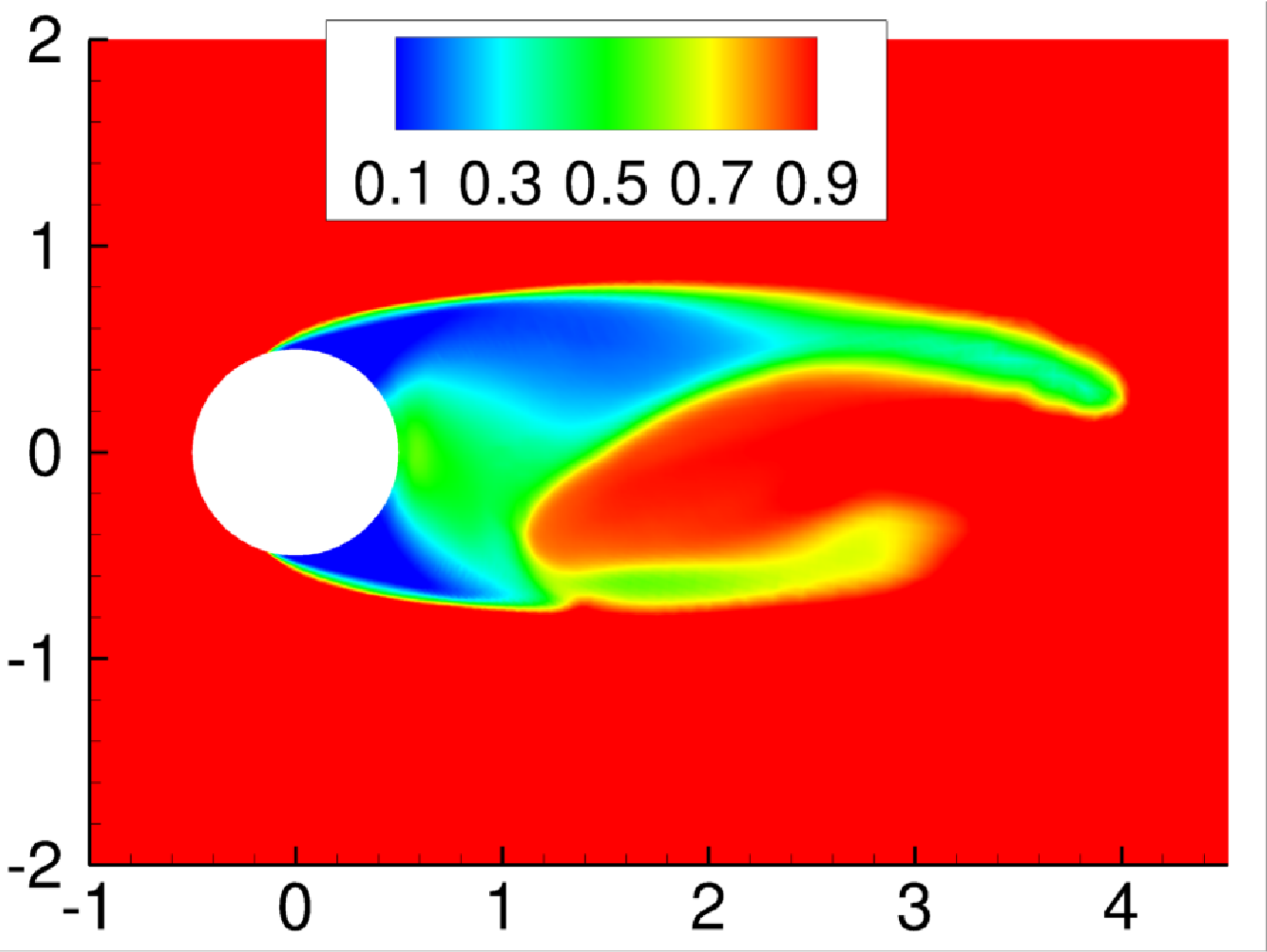}
\put(-177,57){$y/D$}
\put(-175,110){$(b)$}
\put(-63,75){{\textcolor{black}{\vector(-1,-1){15}}}}
\put(-83,80){\fcolorbox{black}{white}{\parbox{15mm}{\footnotesize{$1^{st}$ front}}}}
\end{minipage}\hspace{3pc}
\begin{minipage}{13pc}
\includegraphics[width=13pc, trim={0 0.1cm 0.1cm 0}, clip]{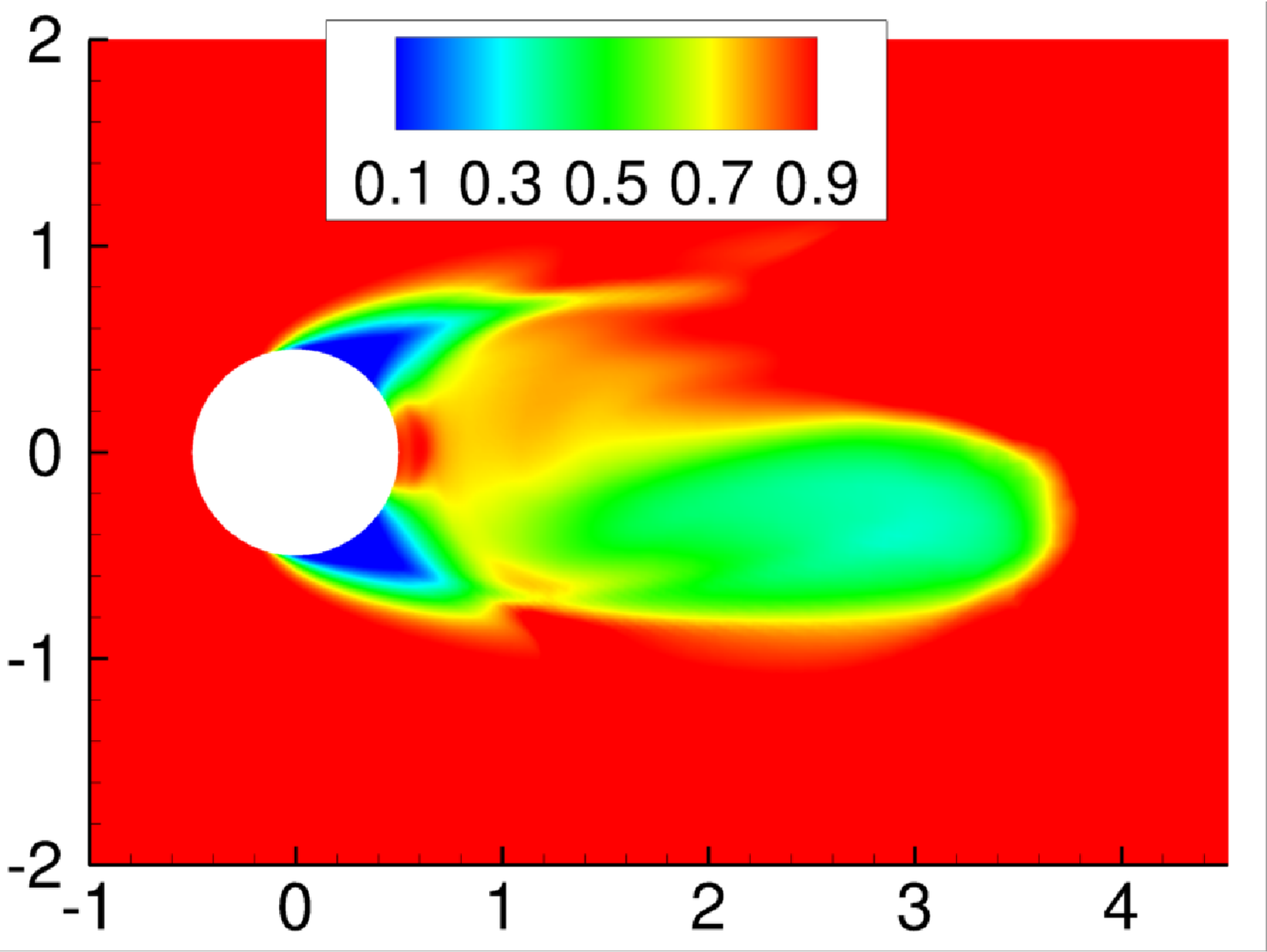}
\put(-85,-10){$x/D$}
\put(-177,57){$y/D$}
\put(-175,110){$(c)$}
\end{minipage}\hspace{3pc}
\begin{minipage}{13pc}
\includegraphics[width=13pc, trim={0 0.1cm 0.1cm 0}, clip]{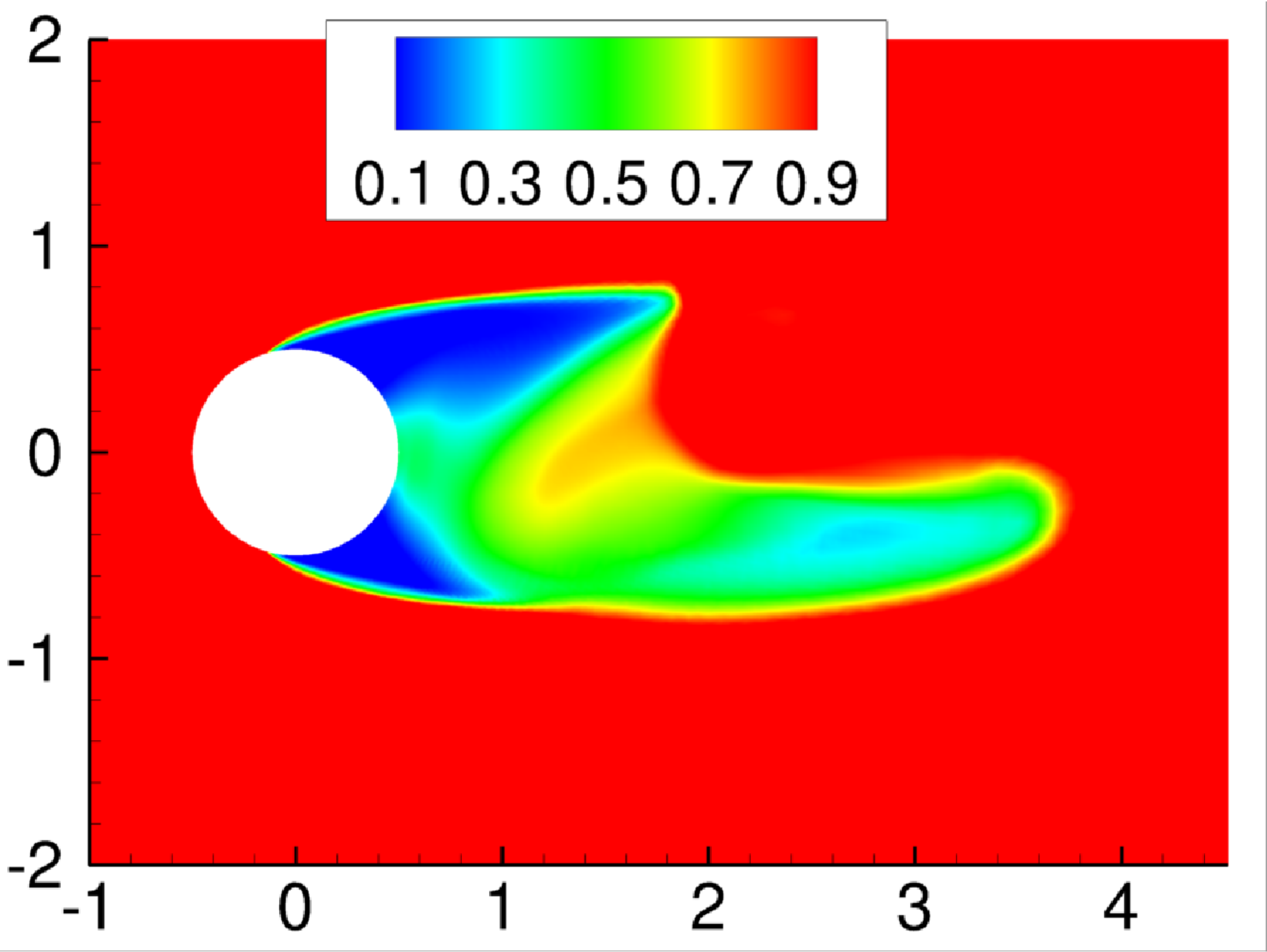}
\put(-85,-10){$x/D$}
\put(-177,57){$y/D$}
\put(-175,110){$(d)$}
\put(-58,50){{\textcolor{black}{\vector(-1,1){15}}}}
\put(-73,45){\fcolorbox{black}{white}{\parbox{15mm}{\footnotesize{$2^{nd}$ front}}}}
\put(-105,35){{\textcolor{black}{\vector(0,1){25}}}}
\put(-125,27){\fcolorbox{black}{white}{\parbox{35mm}{\footnotesize{cavity remains attached}}}}
\end{minipage}\hspace{3pc}
\caption{Density contours showing the propagation of a condensation front in the absence (Case C) $(a,c)$ and presence (Case A200) $(b,d)$ of \textit{NCG}.}
\label{shock-prop-sigma-0.7}
\end{figure}
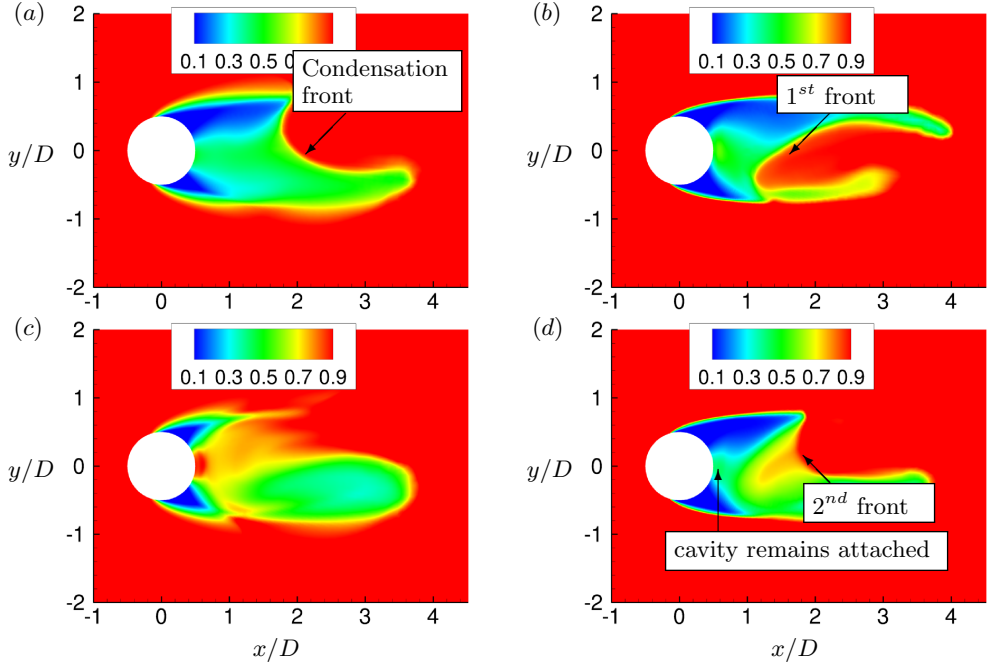

In this section, we consider the low frequency cavity shedding in the transitional regime using $\sigma = 0.7$ to illustrate the shedding cycle in the presence (Case A200) and absence (Case C) of \textit{NCG}. 
Figure \ref{shock-prop-sigma-0.7} shows density contours taken at two different instances during the cycle for Case C (figure \ref{shock-prop-sigma-0.7}$(a,c)$) and Case A200 (figure \ref{shock-prop-sigma-0.7}$(b,d)$). The condensation front is visualized by the density discontinuity in the cavity closure region as indicated by the arrows in figure \ref{shock-prop-sigma-0.7}$(a)$.
As the front propagates upstream when \textit{NCG} is absent, it condenses the vapor along the way, finally detaching the cavity completely as it impinges on the trailing edge of the cylinder as shown in figure \ref{shock-prop-sigma-0.7}$(c)$. In the presence of \textit{NCG} however, figure \ref{shock-prop-sigma-0.7}$(d)$ shows that the cavity remains attached after the cylinder is struck by the first propagating front. A second front is formed approaching the cylinder trailing edge. This indicates that when \textit{NCG} is present, the pressure recovery in the back of the cylinder after the passage of the first front is not enough to condense the vapor and to compress large amounts of \textit{NCG} in order to lead to cavity detachment. This indicates that a weaker condensation front impinges the cylinder surface in the presence of \textit{NCG}. Likewise, \cite{Vennig} in the experimental investigation on a flow over a hydrofoil observed that when the flow was rich in nuclei, a first shock wave only partially condensed the cavity prior to the passage of second shock wave leading to full spanwise detachment. 

\begin{figure}
\centering
\begin{minipage}{11pc}
\hspace{10mm}
\includegraphics[height=70mm, trim={0 0.1cm 0.1cm 0}, clip]{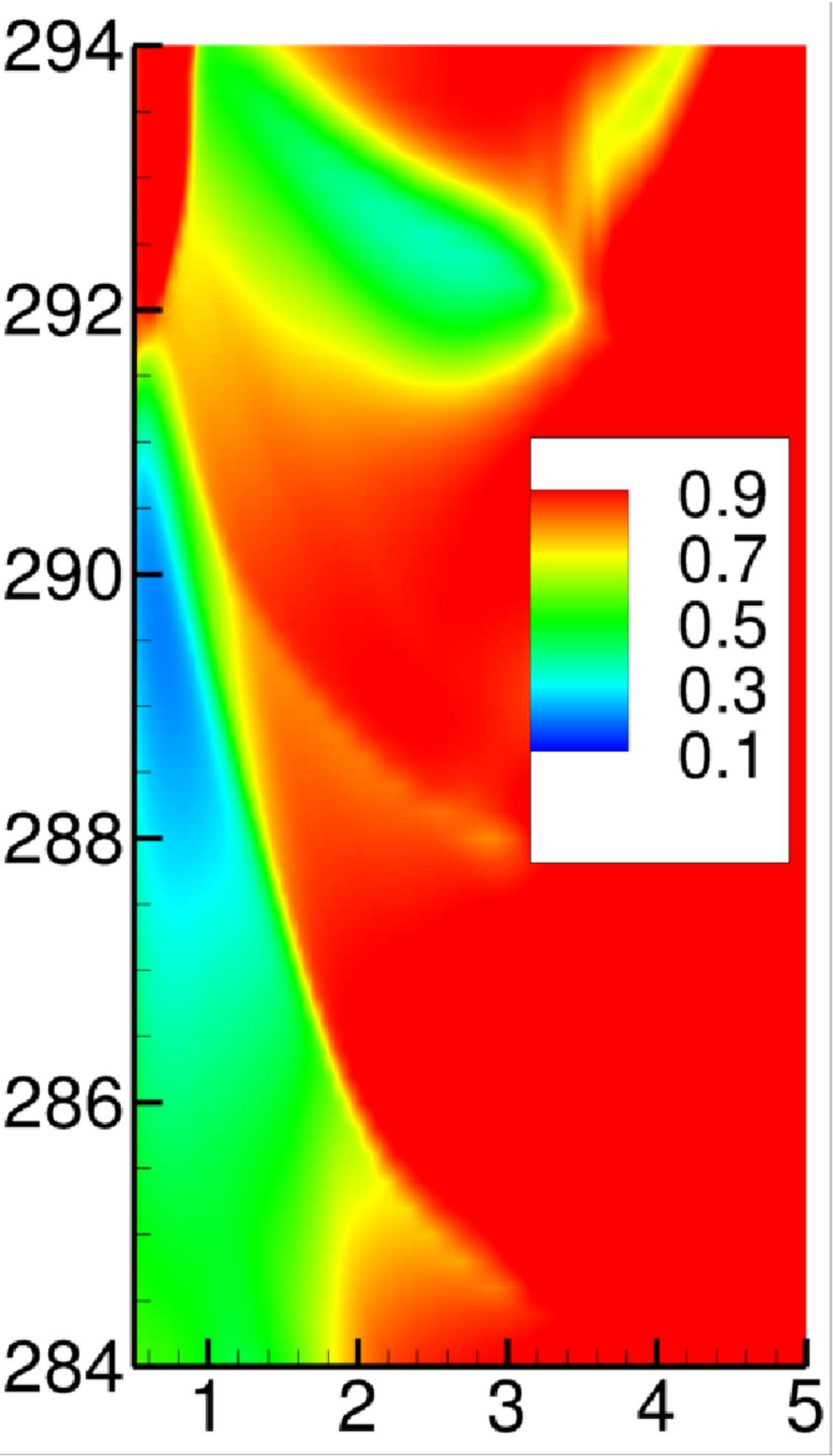}
\put(-60,-10){$x/D$}
\put(-135,100){$tU/D$}
\put(-135,185){$(a)$}
\put(-45,40){{\textcolor{black}{\vector(-2,1){25}}}}
\put(-45,35){\fcolorbox{black}{white}{\parbox{20mm}{\footnotesize{Condensation \\ front}}}}
\put(-68,150){{\textcolor{black}{\vector(-2,1){25}}}}
\put(-68,150){\fcolorbox{black}{white}{\parbox{20mm}{\footnotesize{Total \\ condensation}}}}
\put(-80,63){$\bullet$}
\put(-70,63){$\bullet$}
\put(-60,63){{\textcolor{black}{\textbf{$\overline{t}=1$}}}}
\put(-85,81){$\bullet$}
\put(-75,81){$\bullet$}
\put(-65,81){{\textcolor{black}{\textbf{$\overline{t}=2$}}}}
\put(-87,99){$\bullet$}
\put(-77,99){$\bullet$}
\put(-67,99){{\textcolor{black}{\textbf{$\overline{t}=3$}}}}
\put(-93,117){$\bullet$}
\put(-80,117){$\bullet$}
\put(-70,117){{\textcolor{black}{\textbf{$\overline{t}=4$}}}}
\end{minipage}\hspace{5pc}
\begin{minipage}{11pc}
\includegraphics[height=70mm, trim={0 0.1cm 0.1cm 0}, clip]{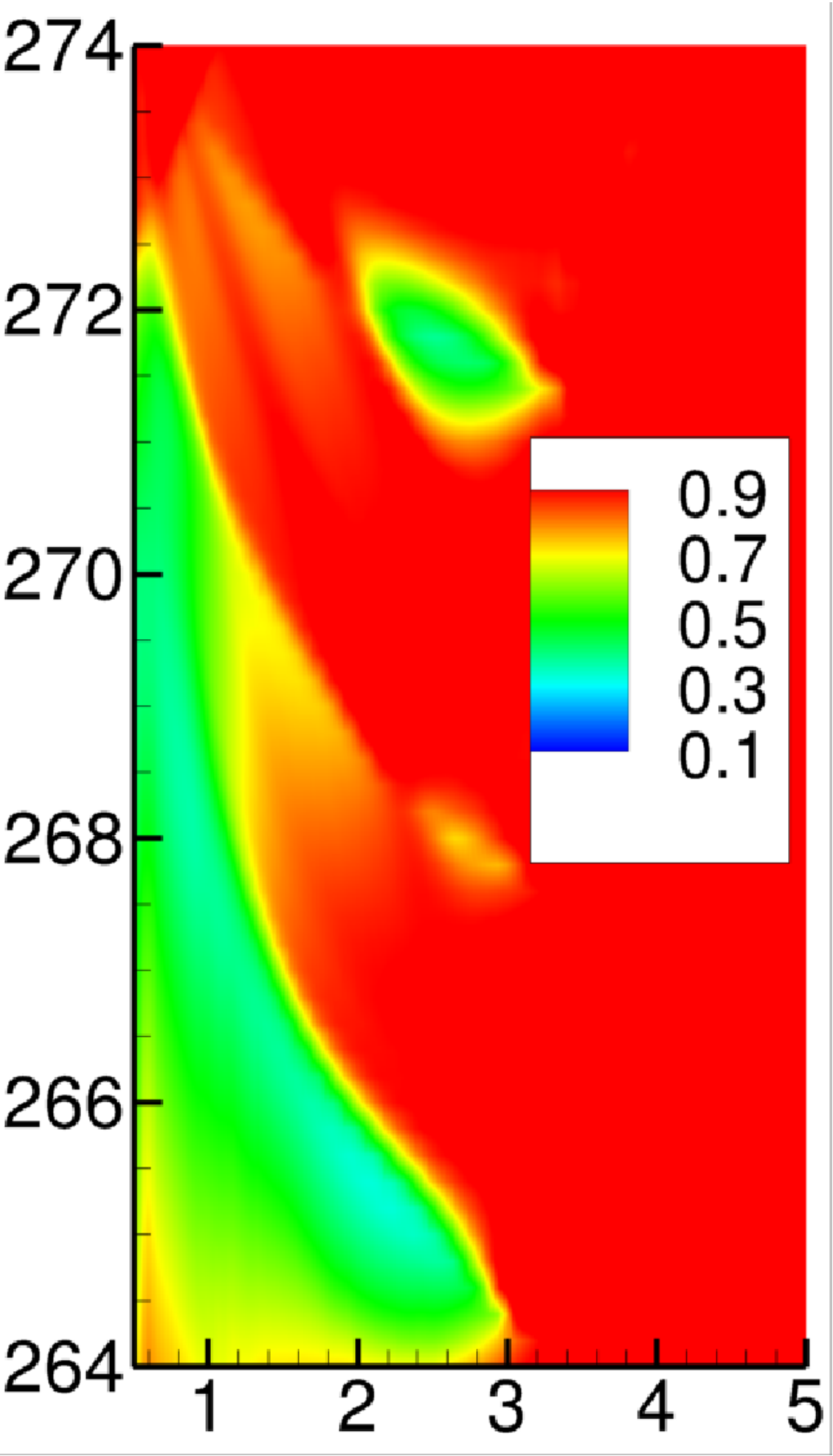}
\put(-60,-10){$x/D$}
\put(-135,95){$tU/D$}
\put(-135,185){$(b)$}
\put(-30,60){{\textcolor{black}{\vector(-1,0){40}}}}
\put(-50,150){{\textcolor{black}{\vector(-1,0){40}}}}
\put(-65,150){\fcolorbox{black}{white}{\parbox{13mm}{\footnotesize{$2^{nd}$ front}}}}
\put(-35,60){\fcolorbox{black}{white}{\parbox{12mm}{\footnotesize{$1^{st}$ front}}}}
\put(-71,45){$\bullet$}
\put(-60,45){$\bullet$}
\put(-50,45){{\textcolor{black}{\textbf{$\overline{t}=1$}}}}
\put(-80,63){$\bullet$}
\put(-70,63){$\bullet$}
\put(-60,63){{\textcolor{black}{\textbf{$\overline{t}=2$}}}}
\put(-85,81){$\bullet$}
\put(-75,81){$\bullet$}
\put(-65,81){{\textcolor{black}{\textbf{$\overline{t}=3$}}}}
\put(-87,99){$\bullet$}
\put(-77,99){$\bullet$}
\put(-67,99){{\textcolor{black}{\textbf{$\overline{t}=4$}}}}
\end{minipage}\hspace{2pc}
\caption{$x-t$ plot of density along the wake centerline for $\sigma=0.7$ when $(a)$ {\it NCG} is absent (Case C) and $(b)$ present (Case A200).}
\label{xt-rho-sigma-0.7}
\end{figure}

In order to quantify the behavior we construct an $x-t$ diagram by taking data along the wake centerline (starting from trailing edge of the cylinder to a $5D$ distance along the wake) and stacking them in time. The $x-t$ diagram is shown in figure \ref{xt-rho-sigma-0.7}$(a)$ and $(b)$ for Case C and Case A200, respectively. 
The density discontinuity moving towards the cylinder when advancing in time indicates the condensation front. The slope of this discontinuity represents the inverse of the speed of the propagating front. It is evident here that in the presence of \textit{NCG}, two fronts propagate before the complete detachment of the cavity as shown in figure \ref{xt-rho-sigma-0.7}$(b)$. Curvature in the density discontinuity indicates that the speed of the condensation front changes as it travels towards the cylinder in the presence of \textit{NCG} as compared to the almost straight line for the case without gas. Therefore, we use Rankine--Hugoniot jump conditions derived in Appendix \ref{RH} at different time instances ($\overline{t}$), as given in figure \ref{xt-rho-sigma-0.7}, to obtain the speed of the front. Left and right states in the equation (\ref{shock_speed}) are obtained from quantities across the condensation front in the $x-t$ plot at the given time instants and are indicated by the bullet points. The speed of sound ahead of the front is obtained using equation (\ref{soseq}). 

\subsubsection{Shock Mach numbers}\label{shock_mach}

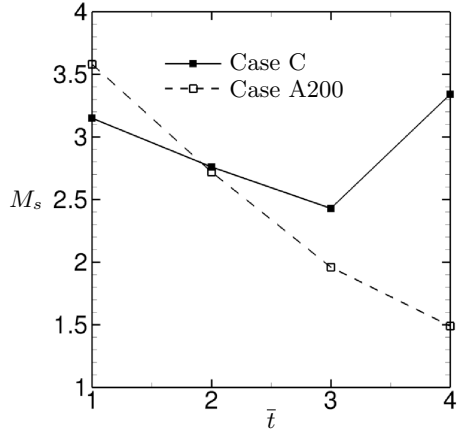
\begin{figure}
\centering
\includegraphics[width=13pc, trim={0 0.1cm 0.1cm 0}, clip]{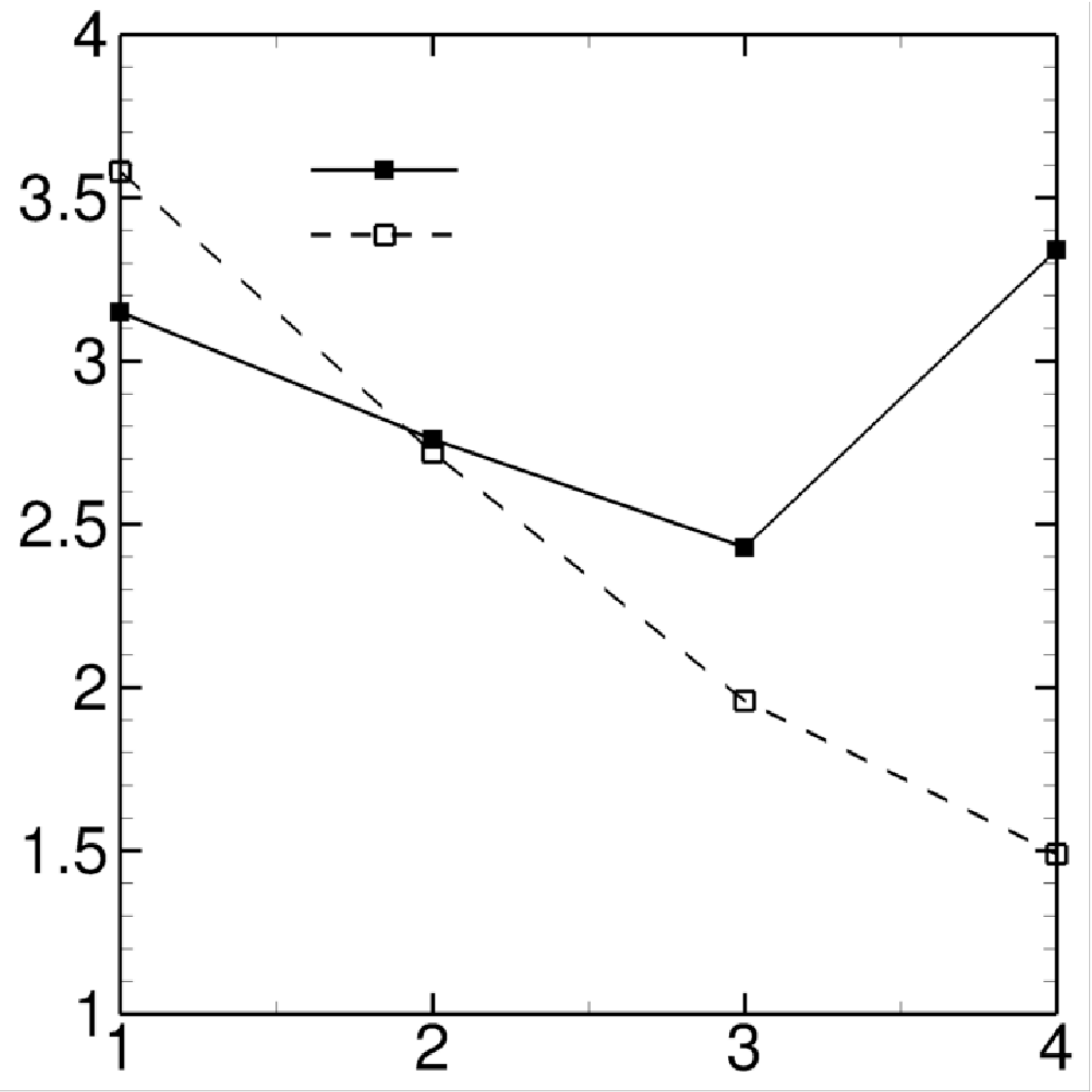}
\put(-170,78){$M_s$}
\put(-73,-5){$\overline{t}$}
\put(-87,129){Case C}
\put(-87,119){Case A200}
\caption{Mach number of condensation front for $\sigma=0.7$.}
\label{shock-Mach-sigma-0.7}
\end{figure}

The speed of condensation front along with the sound speed allows us to comment on the Mach number at which the front propagates. Figure \ref{shock-Mach-sigma-0.7} shows the computed Mach number for the condensation front when {\it NCG} is absent (Case C) and present (Case A200) for the time instances mentioned in figure \ref{xt-rho-sigma-0.7}. Note that the computed Mach numbers refer to the first condensation front that impinges on the cylinder. Interestingly, all the Mach numbers in the figure \ref{shock-Mach-sigma-0.7} are greater than 1, indicating that the front is indeed supersonic; it is henceforth referred to as condensation shock. An important distinction is that in the presence of \textit{NCG}, the Mach number at which the condensation shock impinges the cylinder is much smaller than in the case without the gas, despite both having nearly same Mach numbers when they are formed. It is also evident that the shock Mach number monotonically reduces as it approaches cylinder in the presence of \textit{NCG}. This reduction in shock Mach number is associated with a decrease in pressure jump across the discontinuity in the presence of gas, which will be discussed in section \ref{weakening}.   

\subsubsection{Cavity Mach numbers} 

\begin{figure}
\centering
\begin{minipage}{13pc}
\includegraphics[width=13pc, trim={0 0.1cm 0.1cm 0}, clip]{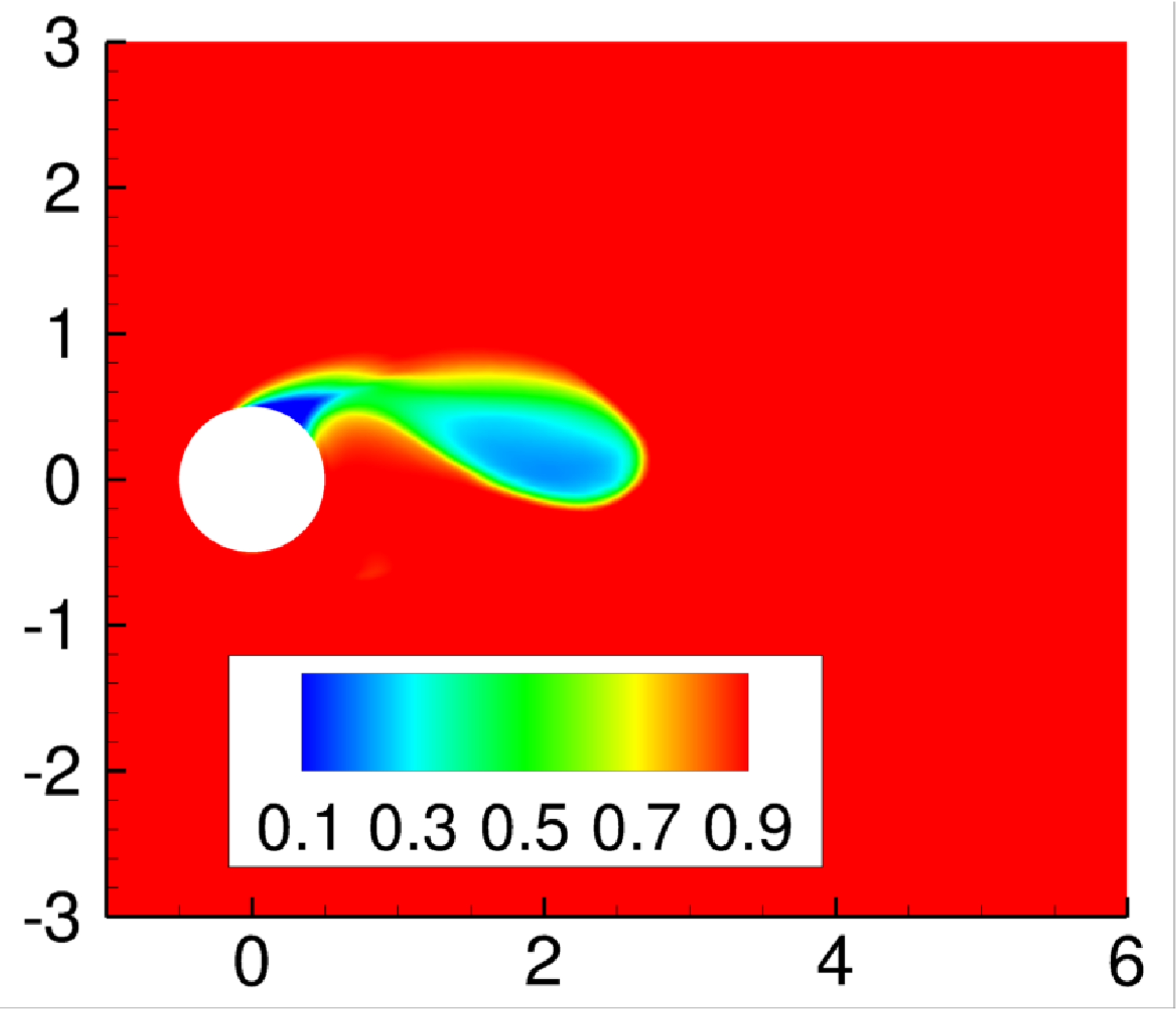}
\put(-85,-10){$x/D$}
\put(-178,80){$y/D$}
\put(-170,135){$(a)$}
\end{minipage}\hspace{2pc}
\begin{minipage}{13pc}
\includegraphics[width=13pc, trim={0 0.1cm 0.1cm 0}, clip]{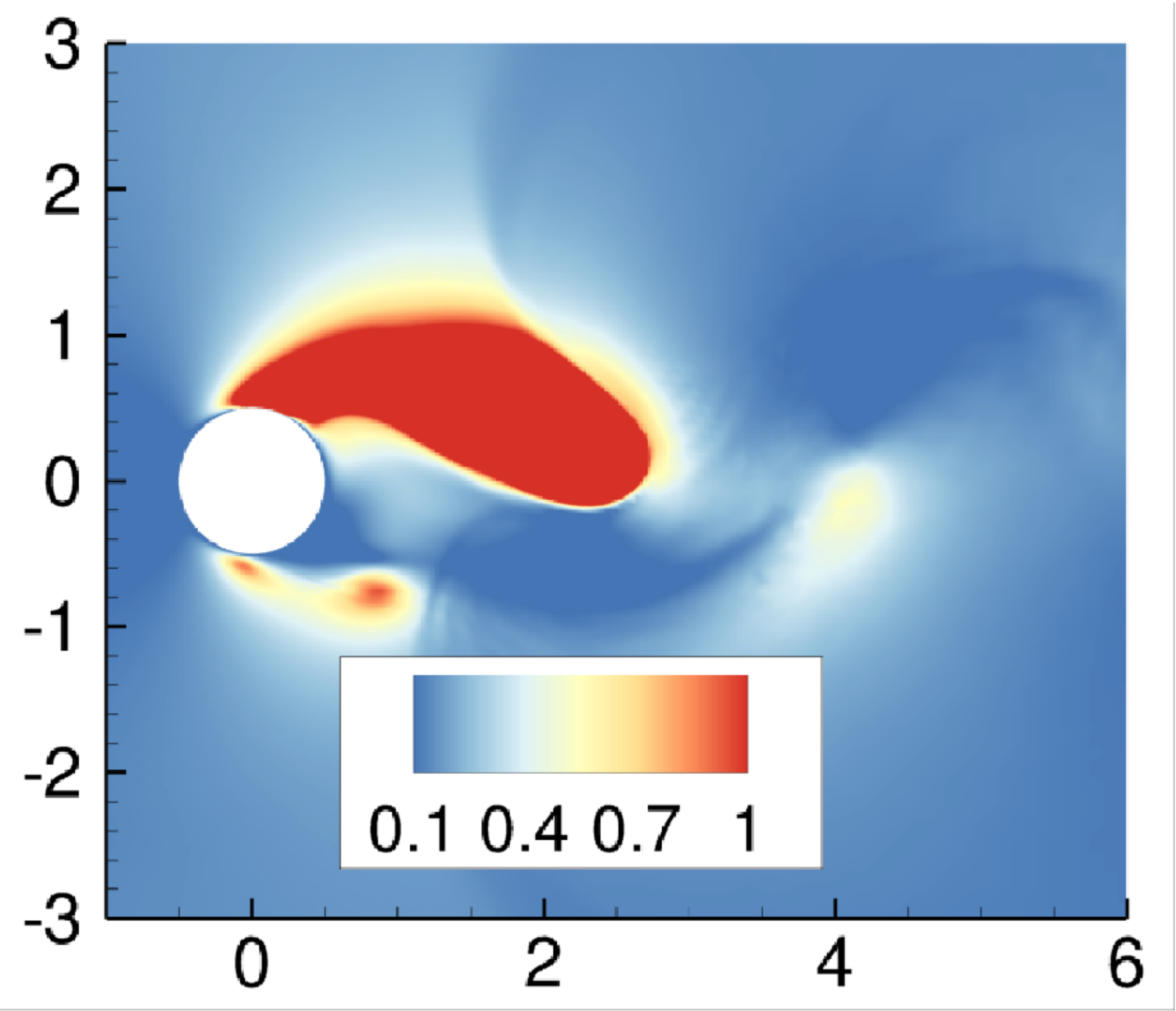}
\put(-85,-10){$x/D$}
\put(-178,80){$y/D$}
\put(-170,135){$(b)$}
\put(-70,110){{\textcolor{white}{\vector(-1,0){20}}}}
\put(-77,51){{\textcolor{white}{\vector(-1,1){10}}}}
\put(-55,55){{\textcolor{black}{\vector(-1,1){15}}}}
\put(-55,55){\fcolorbox{black}{white}{\parbox{15mm}{\footnotesize{Supersonic cavity}}}}
\end{minipage}
\caption{Instantaneous density $(a)$ and Mach number $(b)$ contour of the attached cavity for $\sigma=0.85$.}
\label{mach_local_sigma-085}
\end{figure}
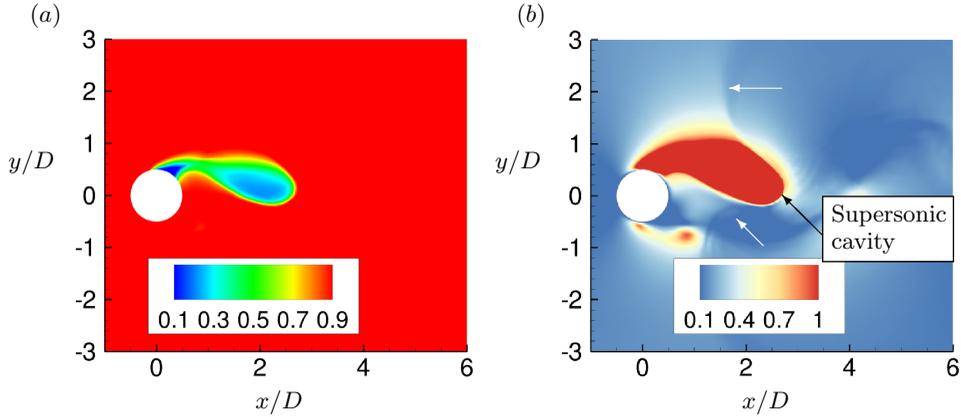

\begin{figure}
\centering
\begin{minipage}{13pc}
\includegraphics[width=13pc, trim={0 0.1cm 0.1cm 0}, clip]{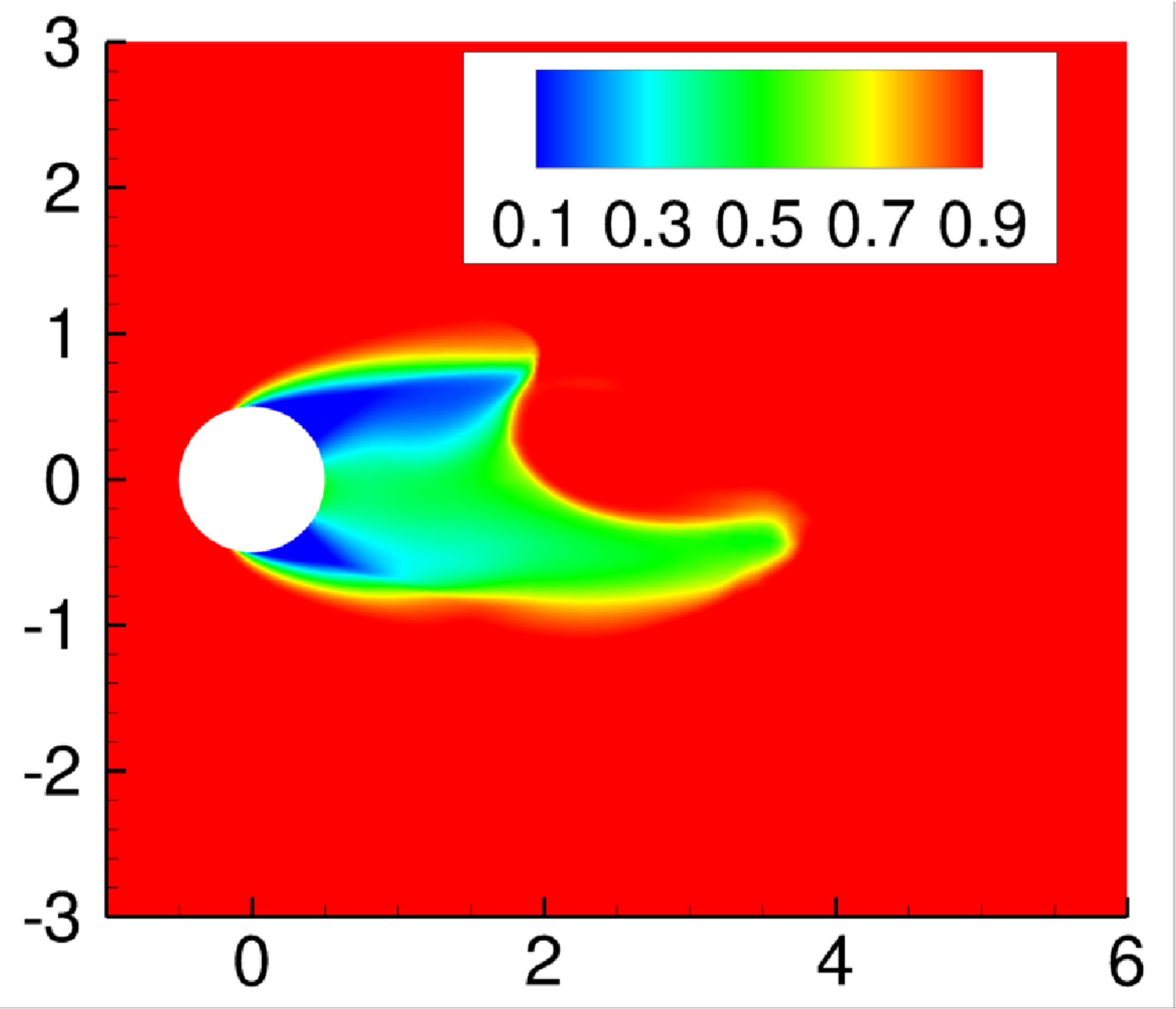}
\put(-85,-10){$x/D$}
\put(-178,67){$y/D$}
\put(-170,135){$(a)$}
\end{minipage}\hspace{2pc}
\begin{minipage}{13pc}
\includegraphics[width=13pc, trim={0 0.1cm 0.1cm 0}, clip]{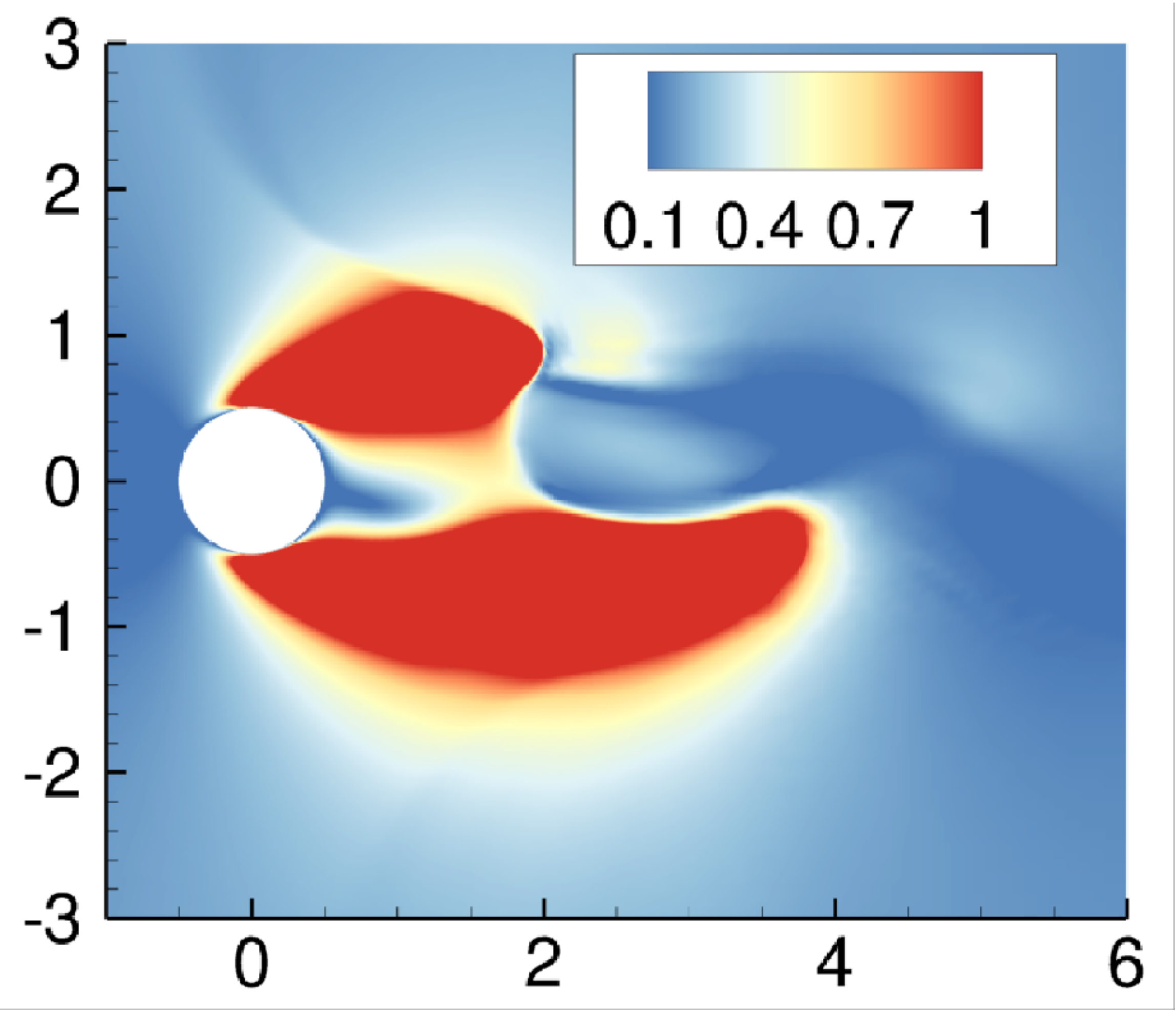}
\put(-85,-10){$x/D$}
\put(-178,67){$y/D$}
\put(-170,135){$(b)$}
\put(-73,73){\vector(-1,0){20}}
\put(-75,50){\vector(-1,1){20}}
\put(-80,40){\fcolorbox{black}{white}{\parbox{15mm}{\footnotesize{Subsonic region}}}}
\put(-97,110){{\textcolor{white}{\vector(-1,0){20}}}}
\end{minipage}
\caption{Instantaneous density $(a)$ and Mach number $(b)$ contour of the attached cavity for $\sigma=0.7$.}
\label{mach_local_sigma-07}
\end{figure}
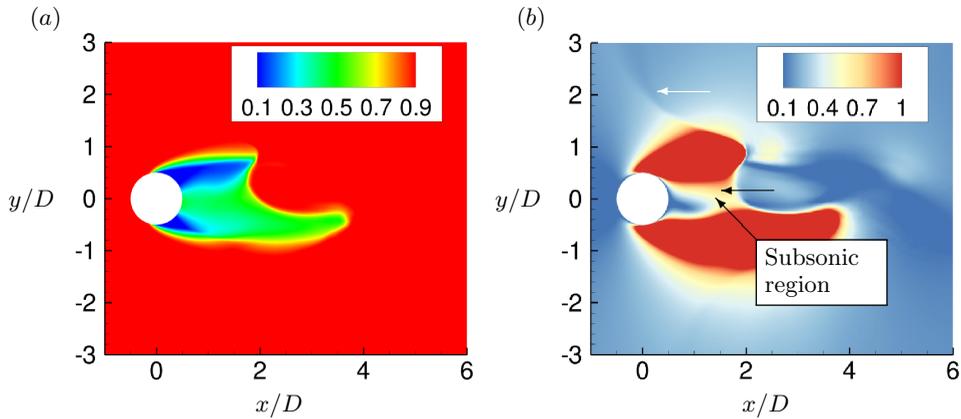

In the condensation shock regime, the shock wave induced by the collapse of a previously shed cavity, propagates upstream through the growing cavity initiating the condensation process \citep{jahangir}. Since the shock cannot propagate through the cavity if it is supersonic, it is important to consider the Mach numbers inside the cavity for the range of cavitation numbers studied, in order to assess the onset of the condensation shock. Here we consider two $\sigma$ for Case C; $\sigma=0.85$ (figure \ref{mach_local_sigma-085}) in the cyclic regime and $\sigma=0.7$ (figure \ref{mach_local_sigma-07}) in the transitional regime. Note that the cavity Mach numbers are different from the shock Mach numbers computed in the earlier section. For $\sigma=0.85$, instantaneous density contours showing the cavity region (figure \ref{mach_local_sigma-085}$(a)$) and corresponding Mach number contours (figure \ref{mach_local_sigma-085}$(b)$) indicate that the cavity is supersonic. It is noticeable that the pressure wave generated due to the collapse of the previously shed cavity (indicated by the white arrow), does not propagate through the cavity as shown in figure \ref{mach_local_sigma-085}$(b)$, and instead travels through the surrounding subsonic liquid region. In contrast, at $\sigma=0.7$, the cavity region surrounding the cylinder trailing edge and near wake is subsonic as shown in figure \ref{mach_local_sigma-07}$(a)$. The propagation of the shock wave through the cavity is indicated by the black arrow.

\subsubsection{Effect of \textit{NCG} on condensation shock}\label{weakening}

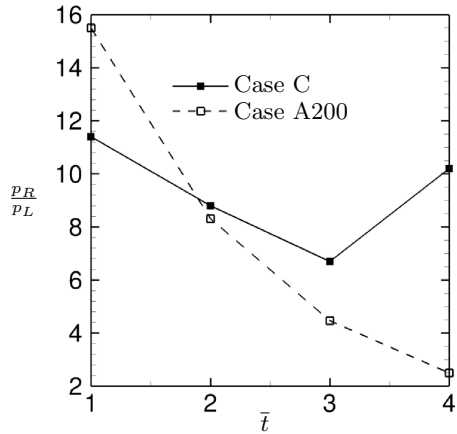
\begin{figure}
\centering
\includegraphics[width=13pc, trim={0 0.1cm 0.1cm 0}, clip]{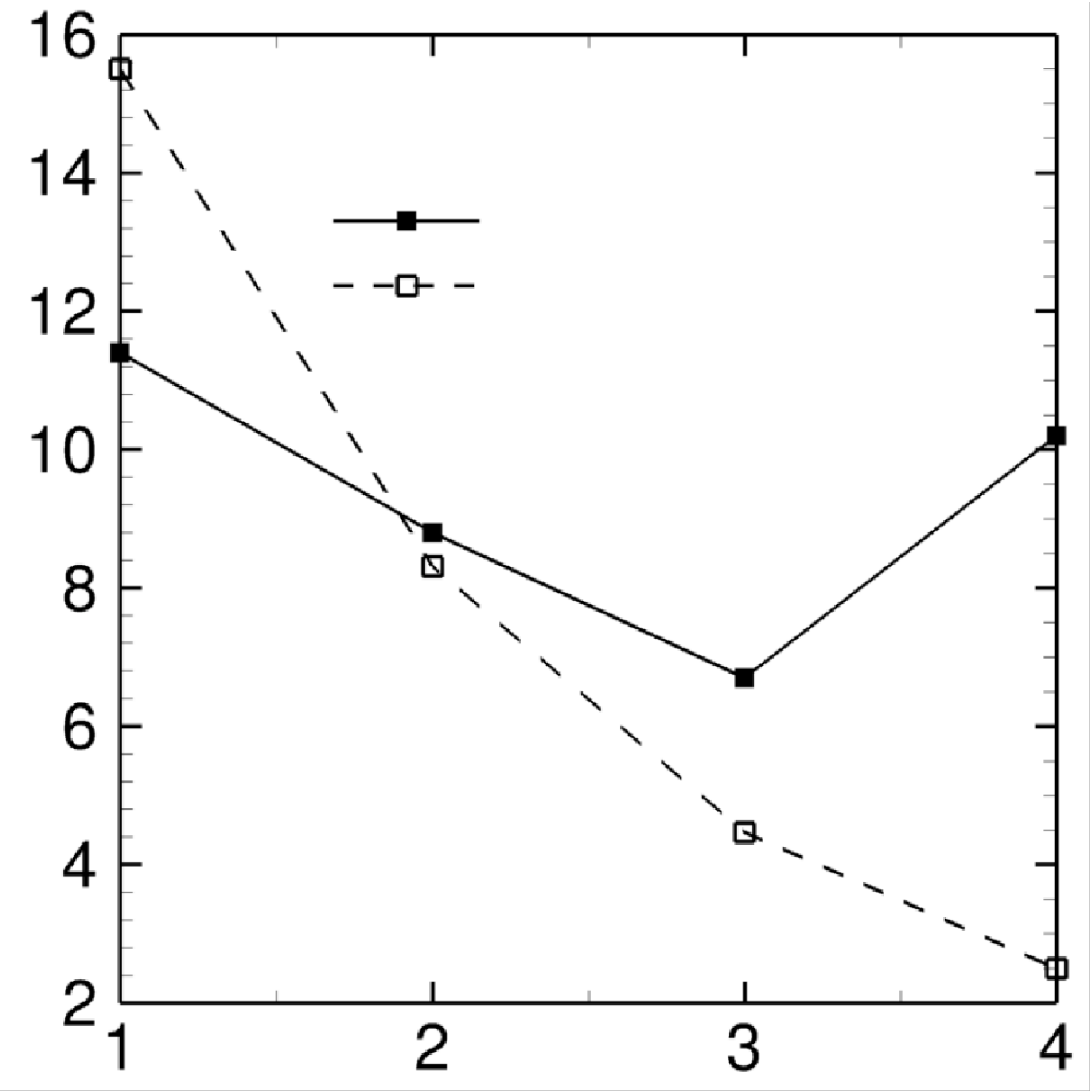}
\put(-170,80){$\frac{p_R}{p_L}$}
\put(-75,-5){$\overline{t}$}
\put(-85,122){Case C}
\put(-85,112){Case A200}
\caption{Pressure ratio across a condensation shock for $\sigma=0.7$.}
\label{press_ratio-sigma-0.7}
\end{figure}

Section \ref{shock_mach} demonstrated that in the presence of \textit{NCG}, the Mach number of the condensation shock decreases as it approaches the cylinder. Consequently, two condensation shock waves were necessary for complete condensation and detachment of the cavity. This section explains this weakening of the condensation shock as it travels in the presence of \textit{NCG}. The strength of the condensation shock is characterized by the pressure ratio and is plotted in figure \ref{press_ratio-sigma-0.7} for the instants of time mentioned in figure \ref{xt-rho-sigma-0.7}. It is evident that the pressure ratio monotonically decreases as the shock approaches the cylinder surface in the presence of \textit{NCG} (Case A200). The initial pressure ratio of 16 reduces to 2.5 near the cylinder trailing edge. Variation in the pressure ratio when \textit{NCG} is absent (Case C) is expected due to the finite rate of condensation; if the condensation rate is not strong enough to condense large portions of the vapor as the shock moves, some amount of uncondensed vapor remain after the passage of the shock decreasing its strength. In the cases considered here, the pressure rise observed across the condensation shock is much lower as compared to the pressure jump across the typical shock wave generated due to cavity collapse. 
 
\begin{figure}
\centering
\begin{minipage}{30pc}
\includegraphics[width=15pc, trim={0.1cm 0.1cm 0.1cm 0}, clip]{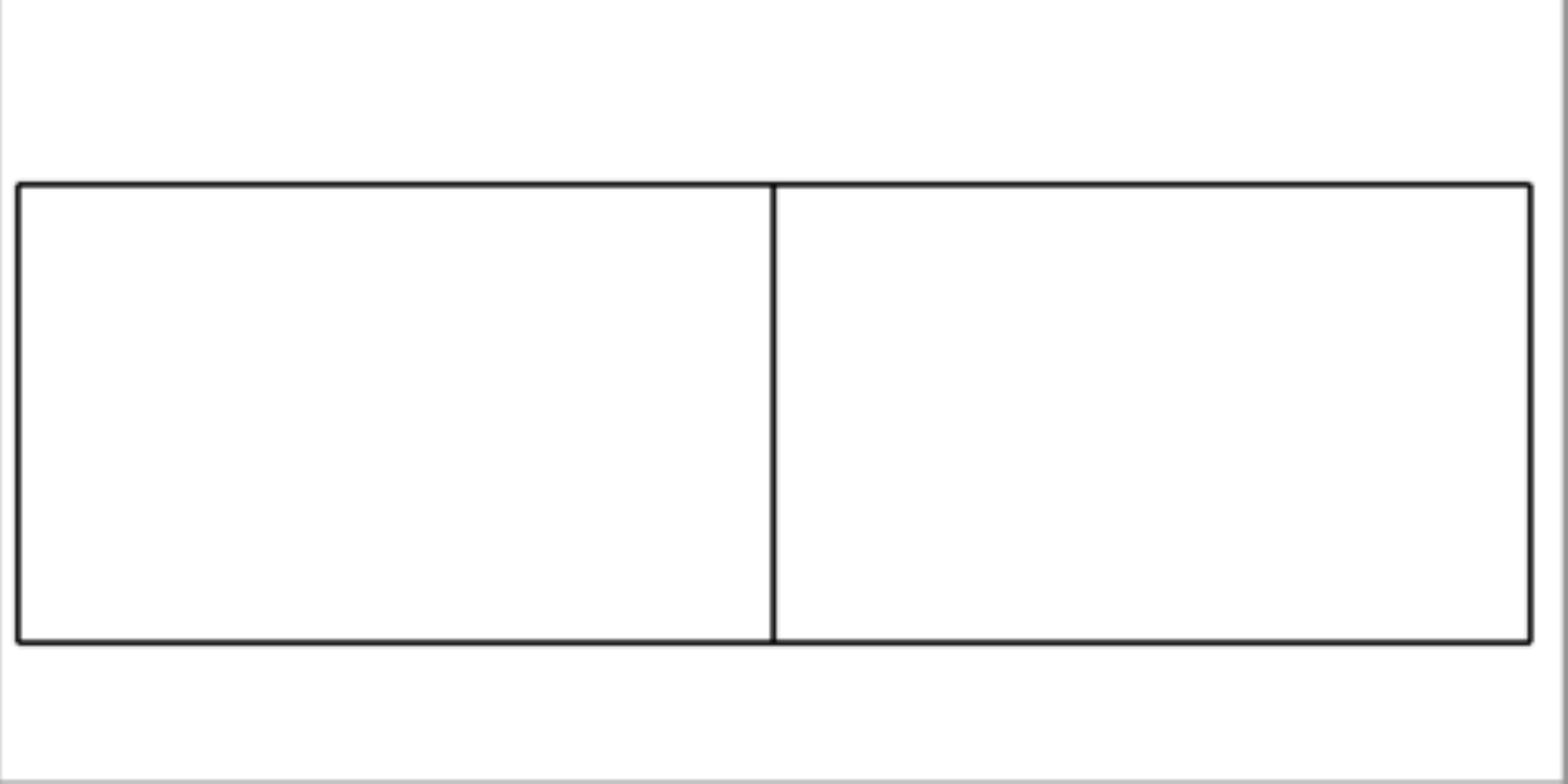}
\put(-180,95){$(a)$}
\put(-95,95){$t$}
\put(-95,75){$x$}
\put(-140,45){$L$}
\put(-50,45){$R$}
\put(-160,30){vapor or} 
\put(-160,20){vapor + \textit{NCG}}
\put(-60,30){water}
\put(50,45){$\rho_L$ $<$ $\rho_R$}
\put(50,35){$p_L$ $<$ $p_R$}
\end{minipage}\hspace{5pc}
\begin{minipage}{30pc}
\includegraphics[width=15pc, trim={0.1cm 0.1cm 0.1cm 0}, clip]{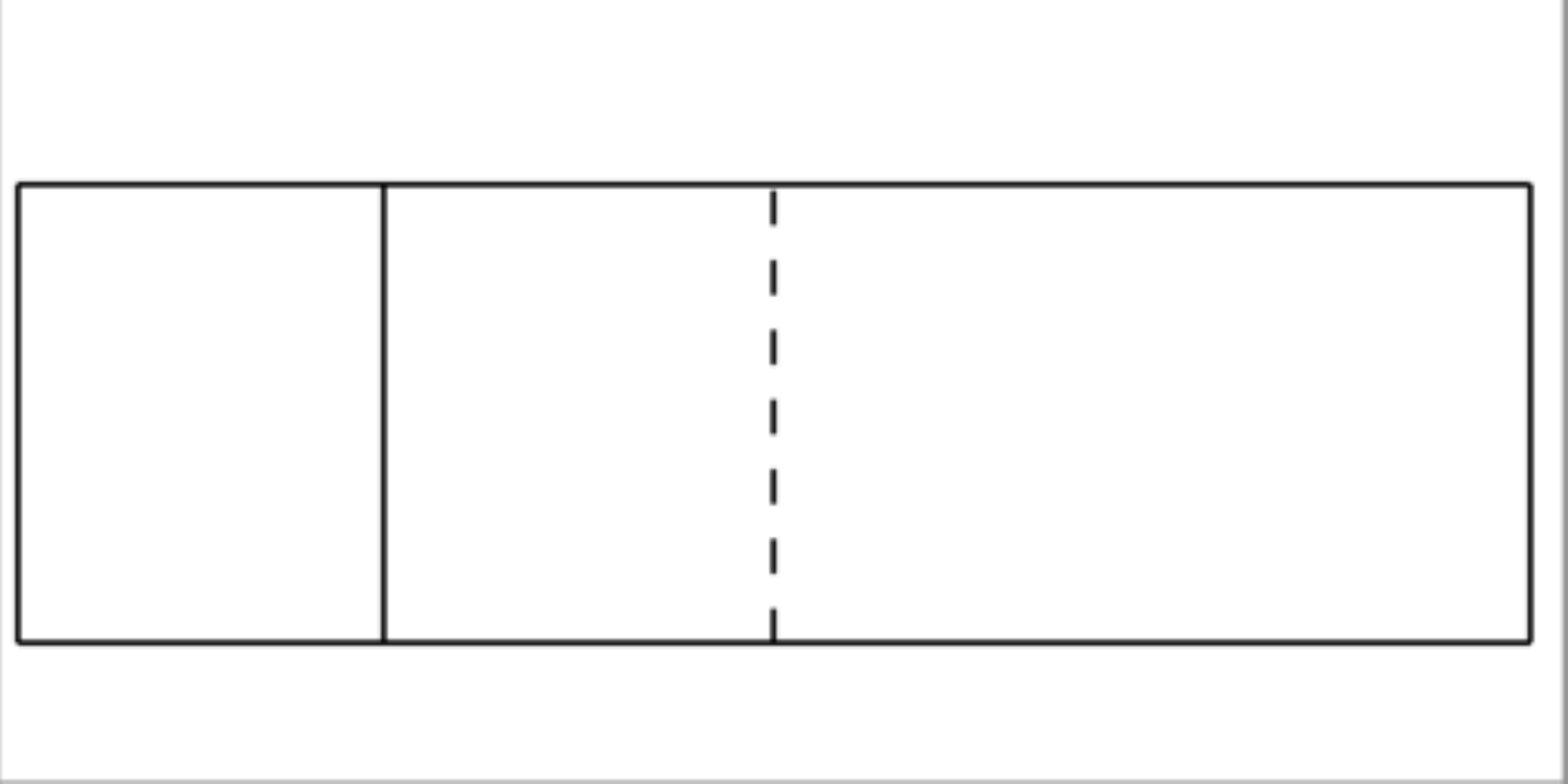}
\put(-180,95){$(b)$}
\put(-101,95){$t + \Delta t$}
\put(-95,75){$x$}
\put(-150,75){$x + \Delta x$}
\put(-120,40){$R^*$}
\put(-50,40){$R$}
\put(-160,40){$L$}
\put(30,95){\textcolor{red}{\textit{NCG} absent}}
\put(20,85){vapor condenses}
\put(50,75){$\downarrow$}
\put(20,65){$R^*$ is mostly water}
\put(35,55){$\rho_{R^*}$ $\approx$ $\rho_R$}
\put(35,45){$p_{R^*}$ $\approx$ $p_R$}
\put(50,35){$\downarrow$}
\put(20,25){shock still strong}
\put(130,95){\textcolor{red}{\textit{NCG} present}}
\put(120,85){vapor condenses}
\put(150,75){$\downarrow$}
\put(120,65){$R^*$ still has \textit{NCG}}
\put(135,55){$\rho_{R^*}$ $<$ $\rho_R$}
\put(135,45){$p_{R^*}$ $<$ $p_R$}
\put(150,35){$\downarrow$}
\put(120,25){shock weakened}
\end{minipage}
\caption{Diagram for the example of a left--moving shock.}
\label{shock-expl}
\end{figure}

Reduction in the pressure ratio due to the presence of gaseous phase behind the shock is explained by considering a typical setting of condensation shock moving from water to the cavity consisted of vapor/\textit{NCG} mixture as shown in figure \ref{shock-expl}$(a)$. Density and pressure behind the shock ($\rho_R$ and $p_R$) are higher than that ahead of the shock ($\rho_L$ and $p_L$). At time $t+\Delta t$, the condensation shock moves through the cavity of vapor/gas mixture to $x+ \Delta x$ creating region $R*$ behind the shock as shown in figure \ref{shock-expl}$(b)$. Now, the jump conditions are determined across the region $L$ and $R*$. Assuming that the vapor is completely condensed as the shock travels through the cavity; if the cavity consisted only of vapor, the region $R*$ behind the shock remains nearly the same as $R$ and is primarily water. In contrast to that, if the cavity also had \textit{NCG}, the region $R*$ would have the mixture of water and \textit{NCG}. Consequently, the overall density is reduced as compared to the original density of water ($\rho_{R^*} < \rho_{R}$), and using the jump conditions derived in (\ref{dens_ratio}) it can be shown that the pressure jump accordingly is also reduced ($p_{R^*}/p_{L} < p_{R}/p_{L}$). Hence, the condensation shock weakens as it propagates through the cavity that contains \textit{NCG}. Subsequently, we consider the conditions at which condensation shock cease to exist as it weakens with the reduction in pressure ratio. 

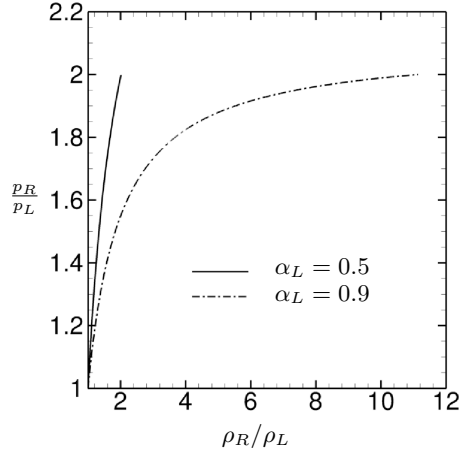
\begin{figure}
\centering
\begin{minipage}{11pc}
\includegraphics[width=13pc, trim={0 0.1cm 0.1cm 0}, clip]{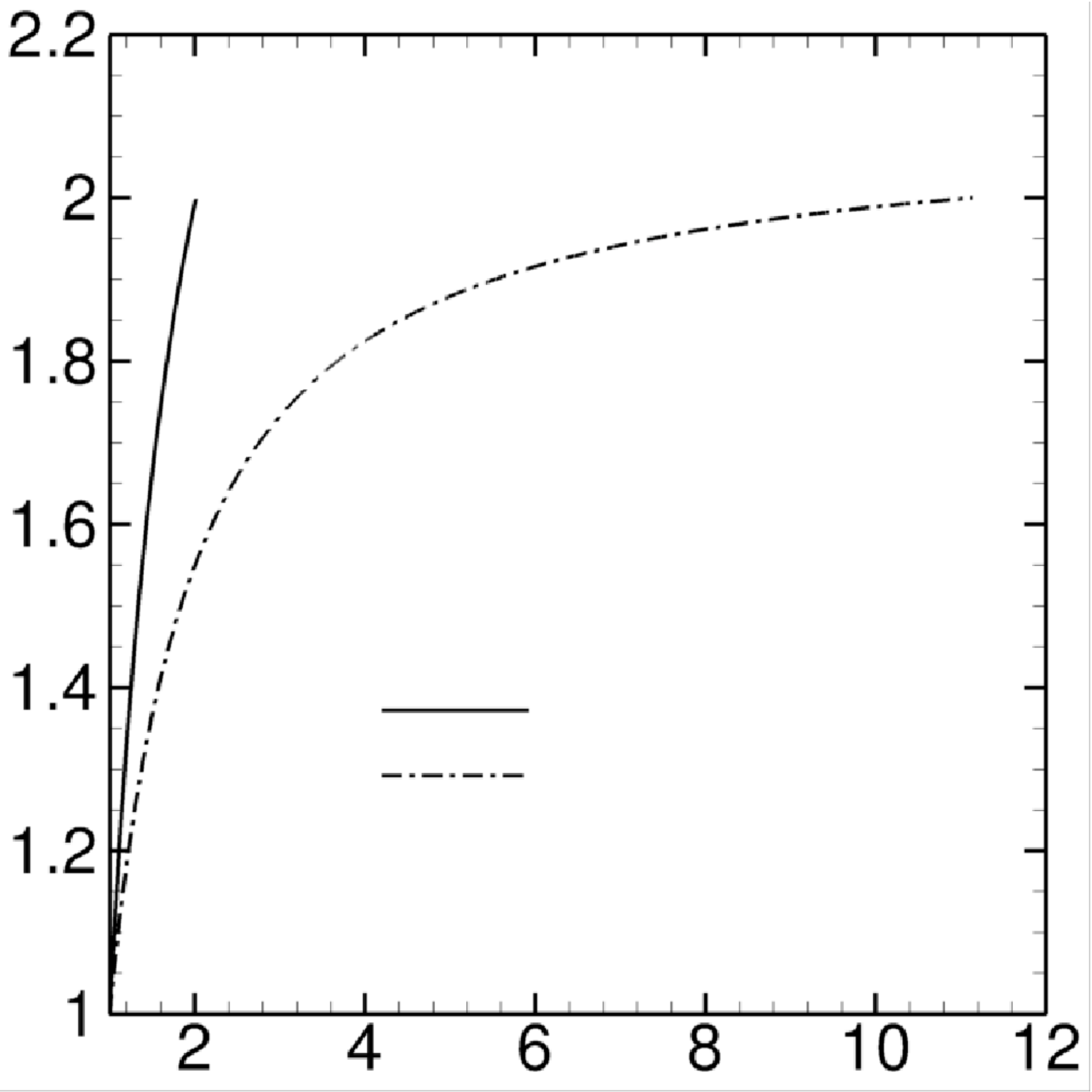}
\put(-90,-10){$\rho_{R}/\rho_{L}$}
\put(-170,80){$\frac{p_R}{p_L}$}
\put(-70,53){$\alpha_L=0.5$}
\put(-70,43){$\alpha_L=0.9$}
\end{minipage}\hspace{2pc}
\caption{Pressure jump across condensation shock at sonic conditions for different density ratios and gaseous phase void fractions ahead of the shock}
\label{parametric-pressratio}
\end{figure}

For single--phase flows, the pressure ratio across the shock is determined uniquely using the shock Mach number \citep{Toro}. However, this is not the case for condensation shocks since multiple phases can be present on either side of the shock. Equation (\ref{shock_speed}) derived in Appendix \ref{RH} shows that the shock speed and consequently the shock Mach number are not uniquely related to the pressure ratio, it also depends on vapor/\textit{NCG} mass fractions ahead and behind the shock and the resulting density difference. In contrast, pressure and density ratio are equal to 1 at sonic conditions for single--phase flows. As an example, figure \ref{parametric-pressratio} shows pressure jump plotted at sonic conditions across a condensation shock using the current system of equations and the jump conditions described by equation (\ref{dens_ratio}) for different amounts of total void fraction ahead of the front ($\alpha_L$). It is clear that in case of condensation shocks, the pressure jump is not unique at a given shock Mach number.

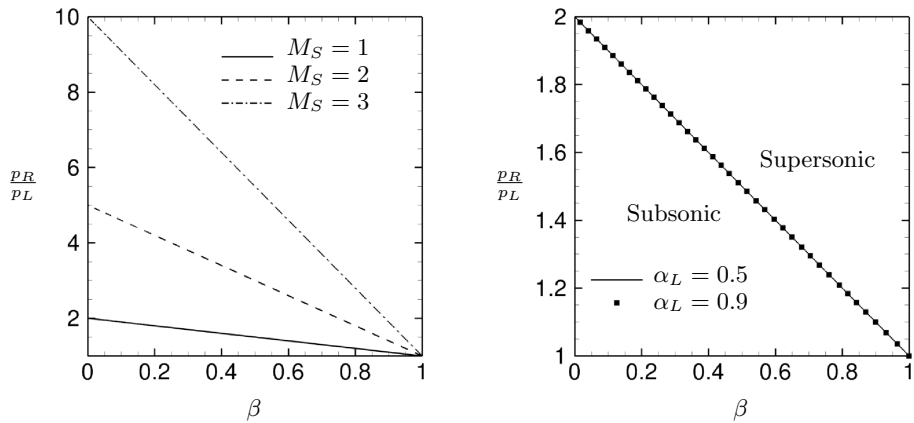
\begin{figure}
\centering 
\begin{minipage}{12pc}
\includegraphics[width=12pc, trim={0 0.1cm 0.1cm 0}, clip]{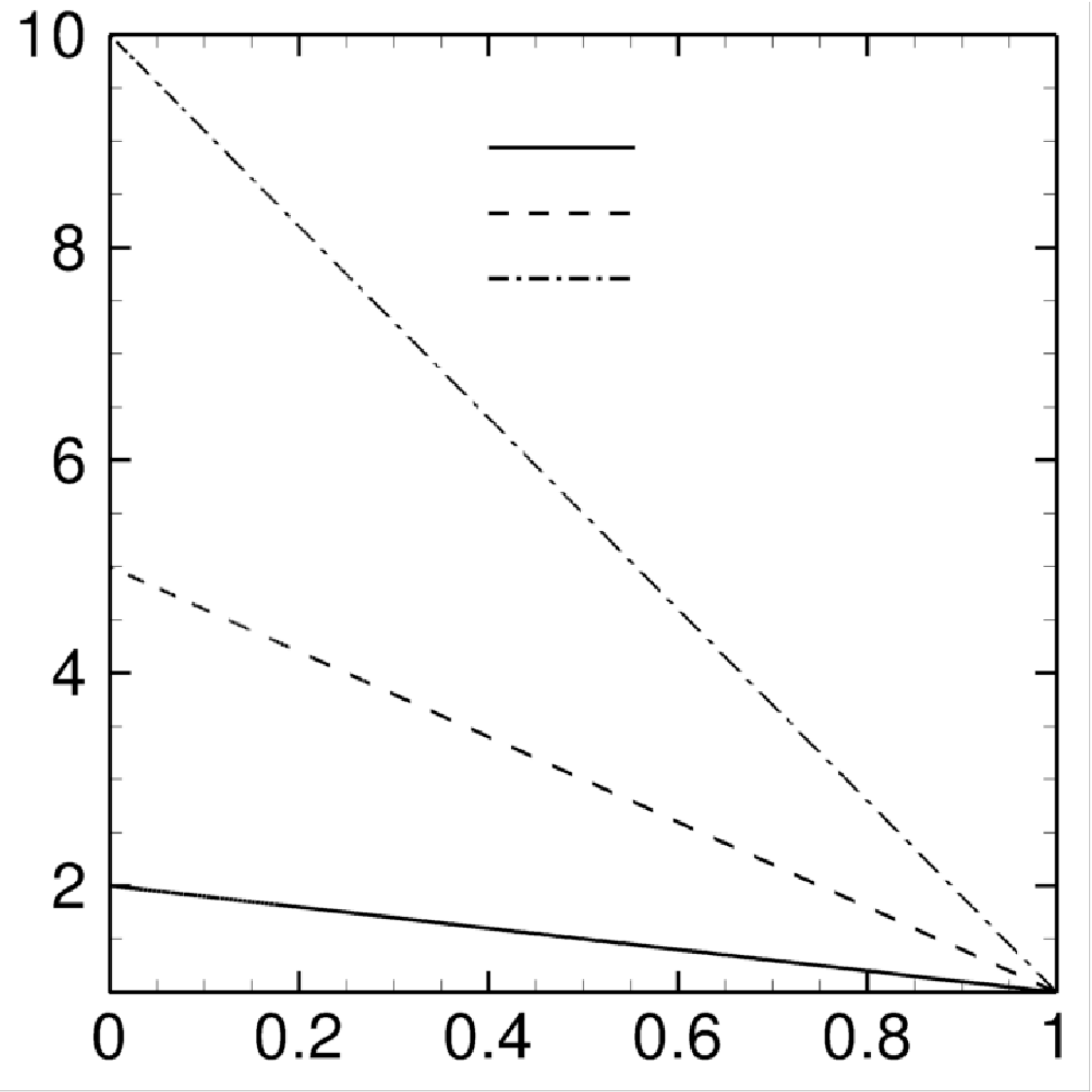}
\put(-70,-10){$\beta$}
\put(-160,75){$\frac{p_R}{p_L}$}
\put(-55,125){$M_S=1$}
\put(-55,115){$M_S=2$}
\put(-55,105){$M_S=3$}
\end{minipage}\hspace{3pc}
\begin{minipage}{12pc}
\includegraphics[width=12pc, trim={0 0.1cm 0.1cm 0}, clip]{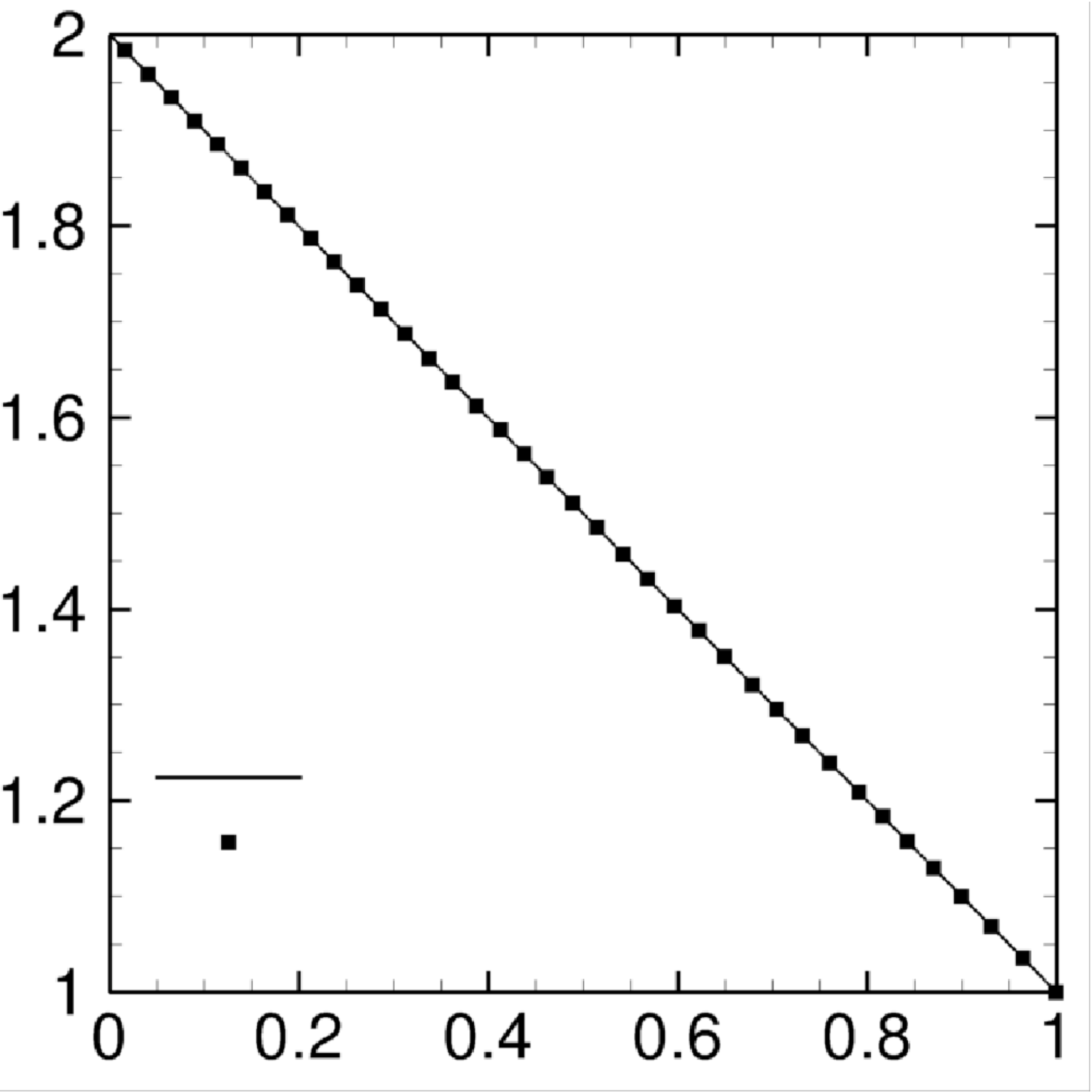}
\put(-70,-10){$\beta$}
\put(-160,75){$\frac{p_R}{p_L}$}
\put(-100,40){$\alpha_L=0.5$}
\put(-110,63){Subsonic}
\put(-60,83){Supersonic}
\put(-100,30){$\alpha_L=0.9$}
\end{minipage}
\caption{Pressure ratio against different values of $\beta$ at different $M_S$ $(a)$. Pressure ratio necessary for condensation shock $(b)$.}
\label{beta_pratio}
\end{figure}

By combining the ratios of density and gaseous phase volume fraction into a single parameter ($\beta=\frac{\alpha_{R}/\alpha_{L}}{\rho_{R}/\rho_{L}}$) and plotting the resultant pressure ratio for different shock Mach number, figure \ref{beta_pratio}$(a)$ is obtained and equation (\ref{pressratio_M_s}) can be derived from it with a linear fit, assuming $u_L$ to be 0 for simplicity.

\begin{equation}
\label{pressratio_M_s}
\begin{aligned}
p_{R}/p_{L} = - M_S^2 \beta + M_S^2 + 1. 
\end{aligned}
\end{equation}

At sonic conditions, the pressure ratio is simplified to
\begin{equation}
\label{pressratio_M_s_1}
\begin{aligned}
p_{R}/p_{L} = 2 - \beta,
\end{aligned}
\end{equation}
and is shown in detail in figure \ref{beta_pratio}$(b)$. From figure \ref{beta_pratio}$(b)$, it is evident that the results of figure \ref{parametric-pressratio} can collapse when plotted using equation (\ref{pressratio_M_s_1}). Figure \ref{beta_pratio}$(b)$ allows us to consider a parameter space for which the condensation front is supersonic. Note that in the single--phase limit ($\alpha_R = \alpha_L = 1$), equation (\ref{eos}) reduces to the ideal gas equation of state and equation (\ref{pressratio_M_s_1}) suggests that $p_{R}/p_{L}=1$ and $\rho_R/\rho_L =1$ has to be identically satisfied at sonic conditions. $\beta = 0$ when the phase is completely liquid behind the shock. This suggest that the pressure ratio has to be greater than 2 for the condensation front propagating from the liquid ($\beta = 0$) into a gaseous cavity to be supersonic. It is important to note, however, that if the phase behind the shock is not completely liquid ($\beta > 0$), the pressure ratio for the occurrence of supersonic Mach number can be less than 2. In the current simulations, although the pressure ratio across the condensation front reduces as it propagates through the cavity containing \textit{NCG}, it remains greater than a factor of 2, indicating that the condensation front remains supersonic as it travels towards the cylinder. Since the state ahead of the shock is the attached cavity, we can conclude that as long as the pressure ratio across the condensation shock does not drop to values below 2, the condensation shock will remain supersonic regardless of the amounts of gaseous phase inside the cavity.

\subsection{LES of cavitating flow at $Re=3900$}\label{LES}

LES of turbulent cavitating flow over a cylinder is performed at $Re=3900$ in both the cyclic and the transitional regimes, respectively at $\sigma=1.0$ and $\sigma =0.7$, using low freestream nuclei concentration (Case A3900). As noted in the Appendix \ref{app_tjump}, the temperature changes in the flow are negligible. Hence, we perform LES calculations using isothermal formulation. 

\subsubsection{Cyclic regime}

\begin{figure}
\centering
\begin{minipage}{10pc}
\includegraphics[width=10pc,trim={0 0.1cm 0.1cm 0}, clip]{Figures/low_alpha/vf_bar_CaseA.pdf}
\put(-135,110){$(a)$}
\put(-140,60){$y/D$}
\put(-105,20){\fcolorbox{black}{white}{\parbox{6mm}{\footnotesize{$\langle \alpha_v \rangle$}}}}
\end{minipage}\hspace{2pc}
\begin{minipage}{10pc}
\includegraphics[width=10pc,trim={0 0.1cm 0.1cm 0}, clip]{Figures/low_alpha/vf_ncg_bar_CaseA.pdf}
\put(-140,60){$y/D$}
\put(-135,110){$(b)$}
\put(-105,20){\fcolorbox{black}{white}{\parbox{6mm}{\footnotesize{$\langle \alpha_g \rangle$}}}}
\end{minipage} 
\begin{minipage}{10pc}
\includegraphics[width=10pc, trim={0 0.1cm 0.1cm 0}, clip]{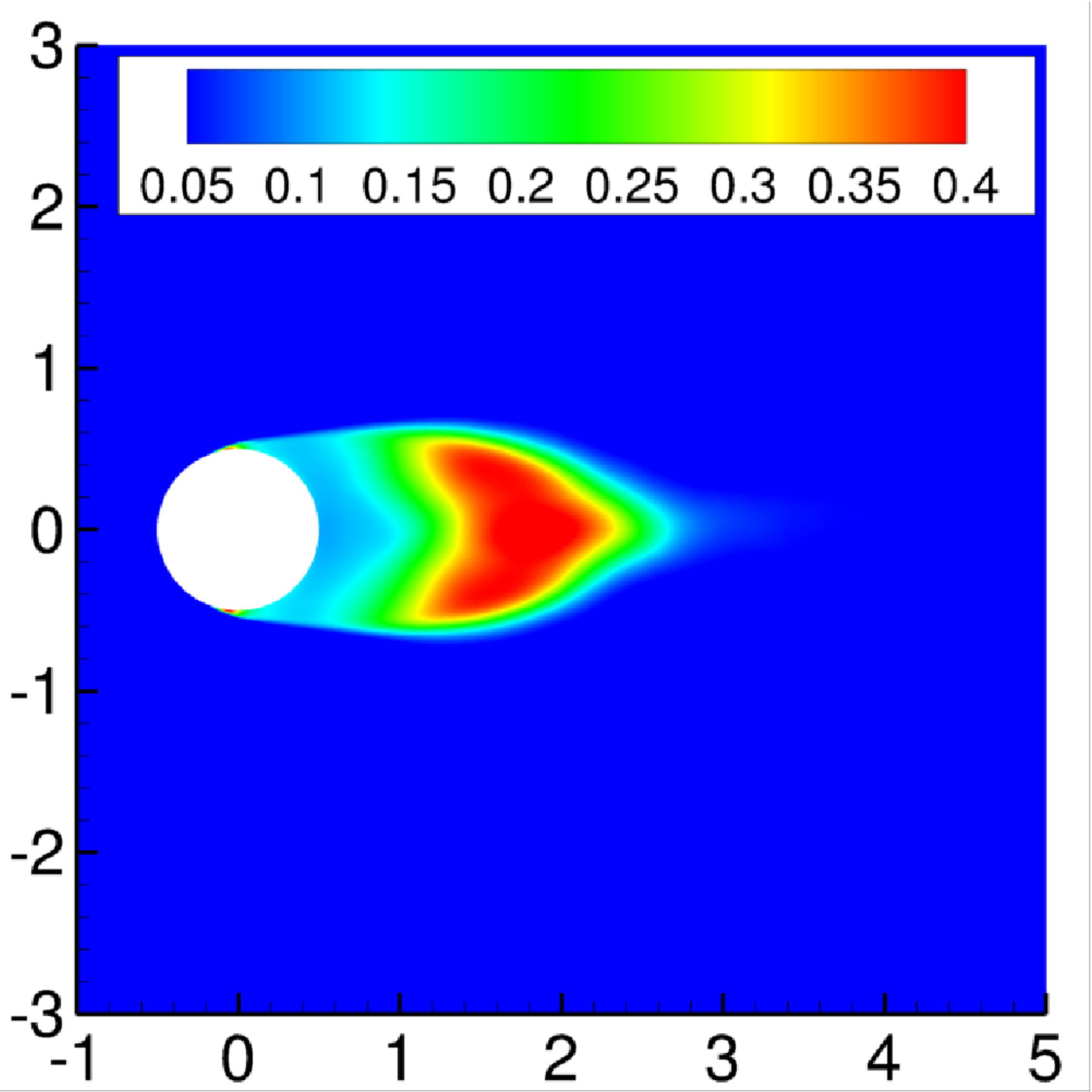}
\put(-65,-8){$x/D$}
\put(-140,60){$y/D$}
\put(-135,110){$(c)$}
\put(-105,20){\fcolorbox{black}{white}{\parbox{6mm}{\footnotesize{$\langle \alpha_v \rangle$}}}}
\end{minipage}\hspace{2pc}
\begin{minipage}{10pc}
\includegraphics[width=10pc, trim={0 0.1cm 0.1cm 0}, clip]{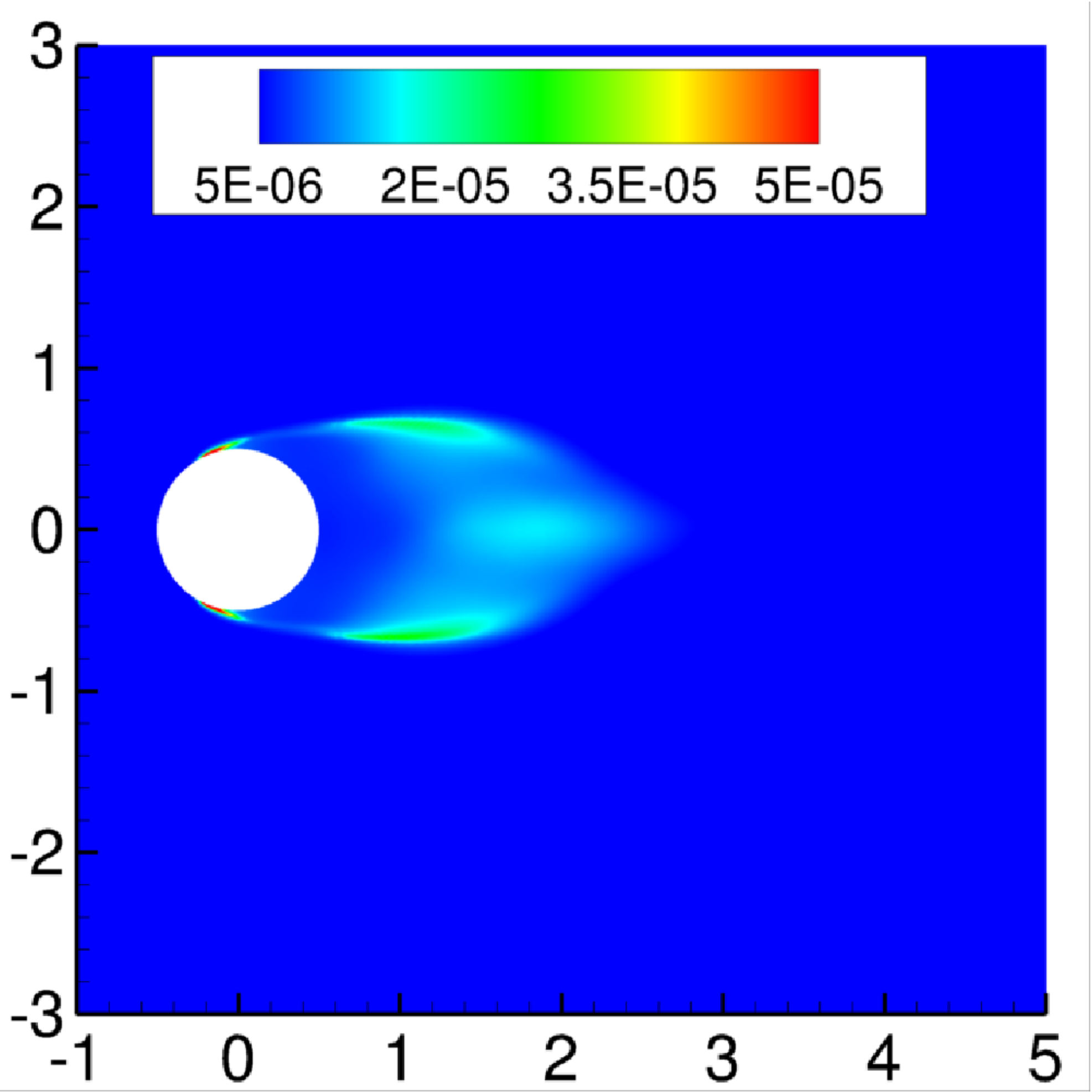}
\put(-65,-8){$x/D$}
\put(-140,60){$y/D$}
\put(-135,110){$(d)$}
\put(-105,20){\fcolorbox{black}{white}{\parbox{6mm}{\footnotesize{$\langle \alpha_g \rangle$}}}}
\end{minipage}
\caption{Mean volume fractions at $Re=200$ ($(a)$ vapor $(b)$ {\it NCG}) and $Re=3900$ ($(c)$ vapor $(d)$ {\it NCG}).}
\label{meanflow_Re3900}
\end{figure}

We consider mean vapor and {\it NCG} volume fraction at $Re=3900$ and compare it to $Re=200$ (figure \ref{meanflow_Re3900}). At $Re=3900$, only a thin layer of vapor is observed at the cylinder surface as compared to the $Re=200$ (figure \ref{meanflow_Re3900}$(a,c)$). Also, at higher $Re$, more vapor is observed in the region of near wake. At $Re=3900$, the incoming boundary layer is turbulent and confined close to the cylinder surface. This results in higher vorticity inside the  K\'arm\'an vortices leading to larger pressure drop and more vapor production in the wake. The mean volume fraction of {\it NCG} at $Re=3900$, however, is observed to be one order of magnitude lower than at $Re=200$ (figure \ref{meanflow_Re3900}$(b,d)$).

\begin{figure}
\centering
\includegraphics[width=13pc, trim={0 0.1cm 0.1cm 0}, clip]{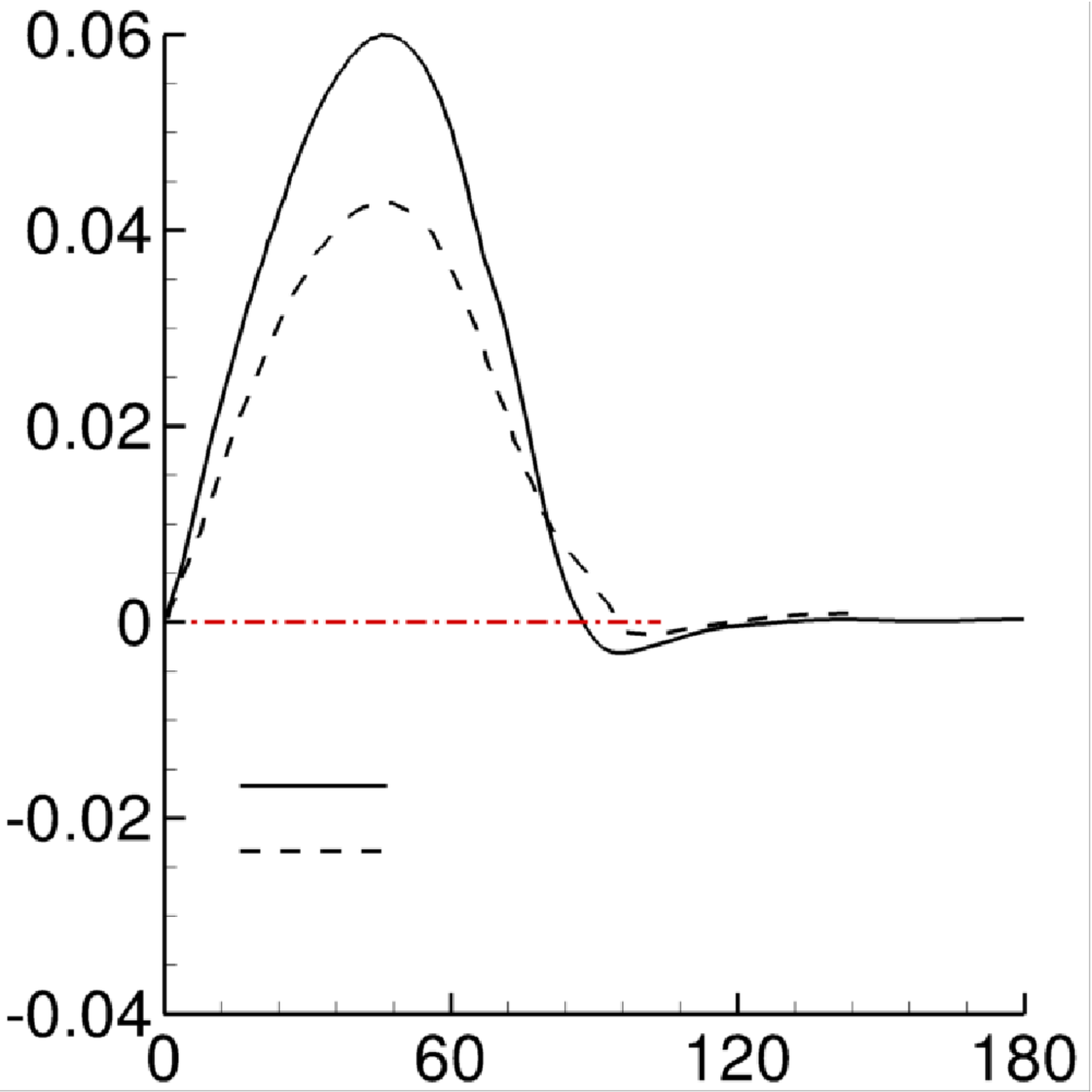}
\put(-82,-8){$\theta$ (deg.)}
\put(-170,77){$C_f$}
\put(-95,42){$\sigma=1.0$, present work}
\put(-95,32){$\sigma=1.0$, \cite{AswinJFM}}
\caption{Skin friction coefficient at $Re=3900$. Red dashed line identifies $C_f=0$}
\label{Cf_Re3900}
\end{figure}

The skin friction coefficient is compared to the result from \cite{AswinJFM} in figure \ref{Cf_Re3900}. They studied the cyclic cavitation regime at $\sigma=1.0$ using freestream nuclei concentration of 0.01 for vapor (Case C of the current simulations). As noted for the laminar separation at $Re=200$, we observe that at $Re=3900$ the boundary--layer separation point moves upstream as compared to \cite{AswinJFM} (figure \ref{Cf_Re3900}). The difference between the separation points is approximately $10 \degree$, the same observed for $Re=200$. These results confirm the role of freestream total void fraction on the location of boundary--layer separation.

\subsubsection{Transitional regime}

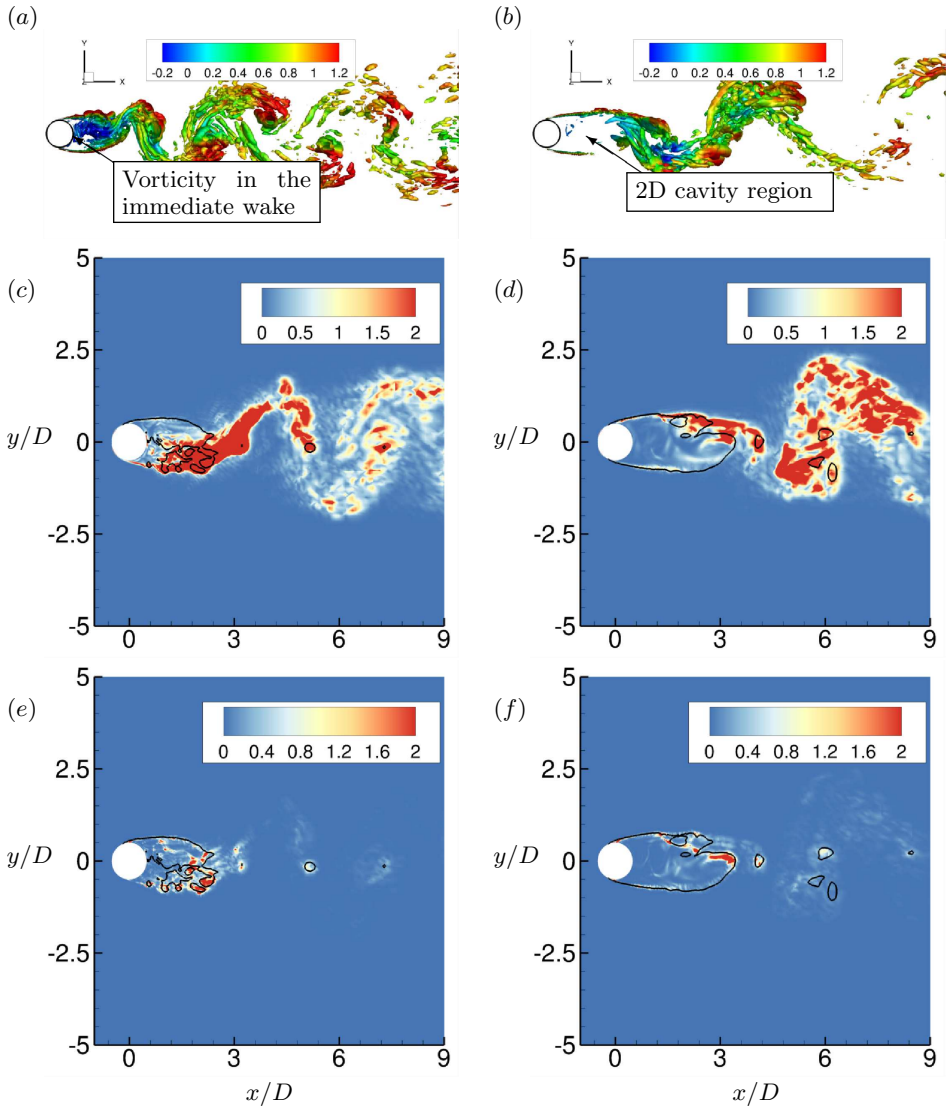
\begin{figure}
\centering
\begin{minipage}{13pc}
\includegraphics[width=13pc, trim={0.1cm 0.1cm 0.1cm 0}, clip]{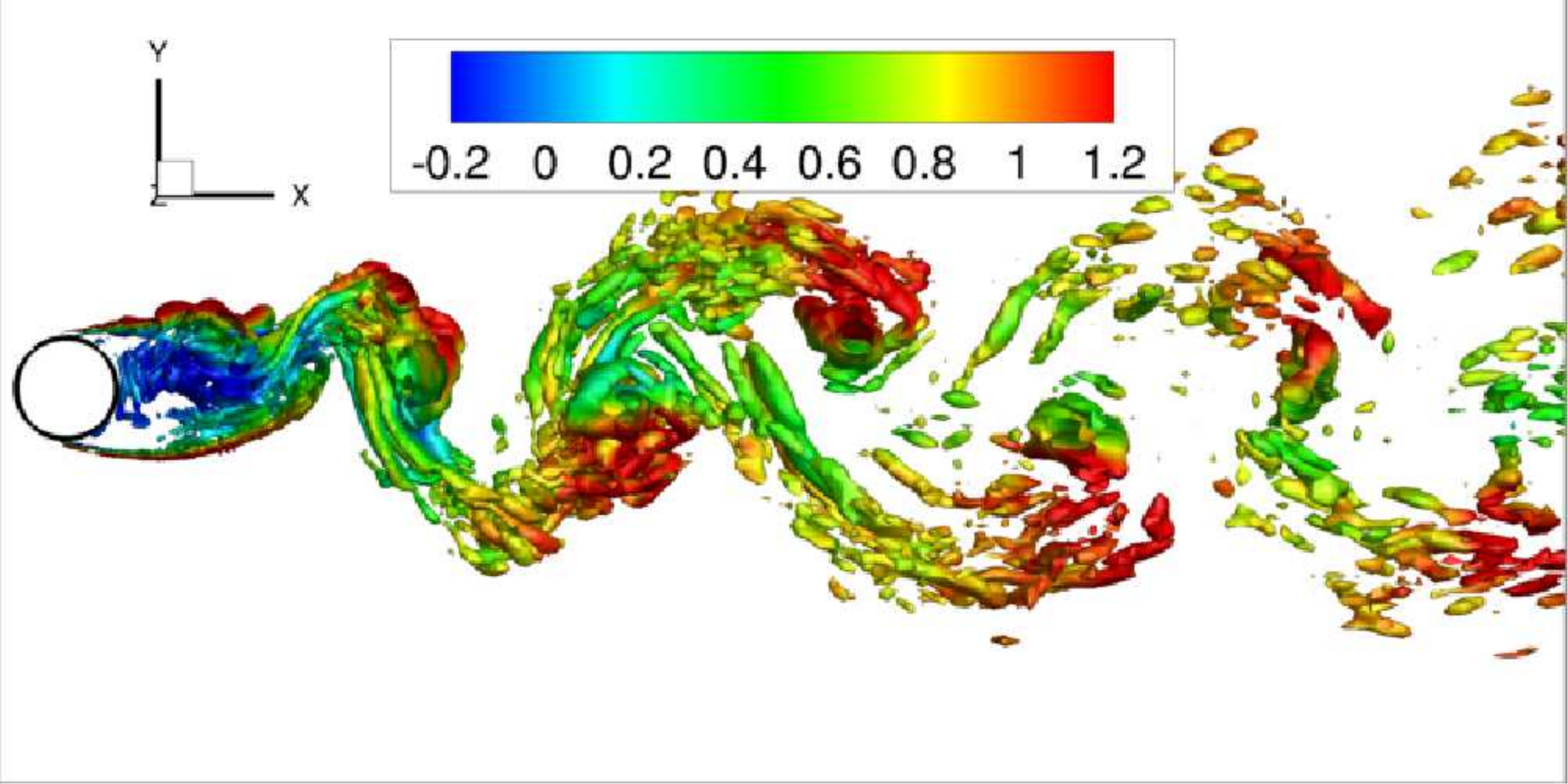}
\put(-110,15){{\textcolor{black}{\vector(-3,2){35}}}}
\put(-130,15){\fcolorbox{black}{white}{\parbox{25mm}{\footnotesize{Vorticity in the immediate wake}}}}
\put(-170,80){$(a)$}
\end{minipage}\hspace{2pc}
\begin{minipage}{13pc}
\includegraphics[width=13pc, trim={0.1cm 0.1cm 0.1cm 0}, clip]{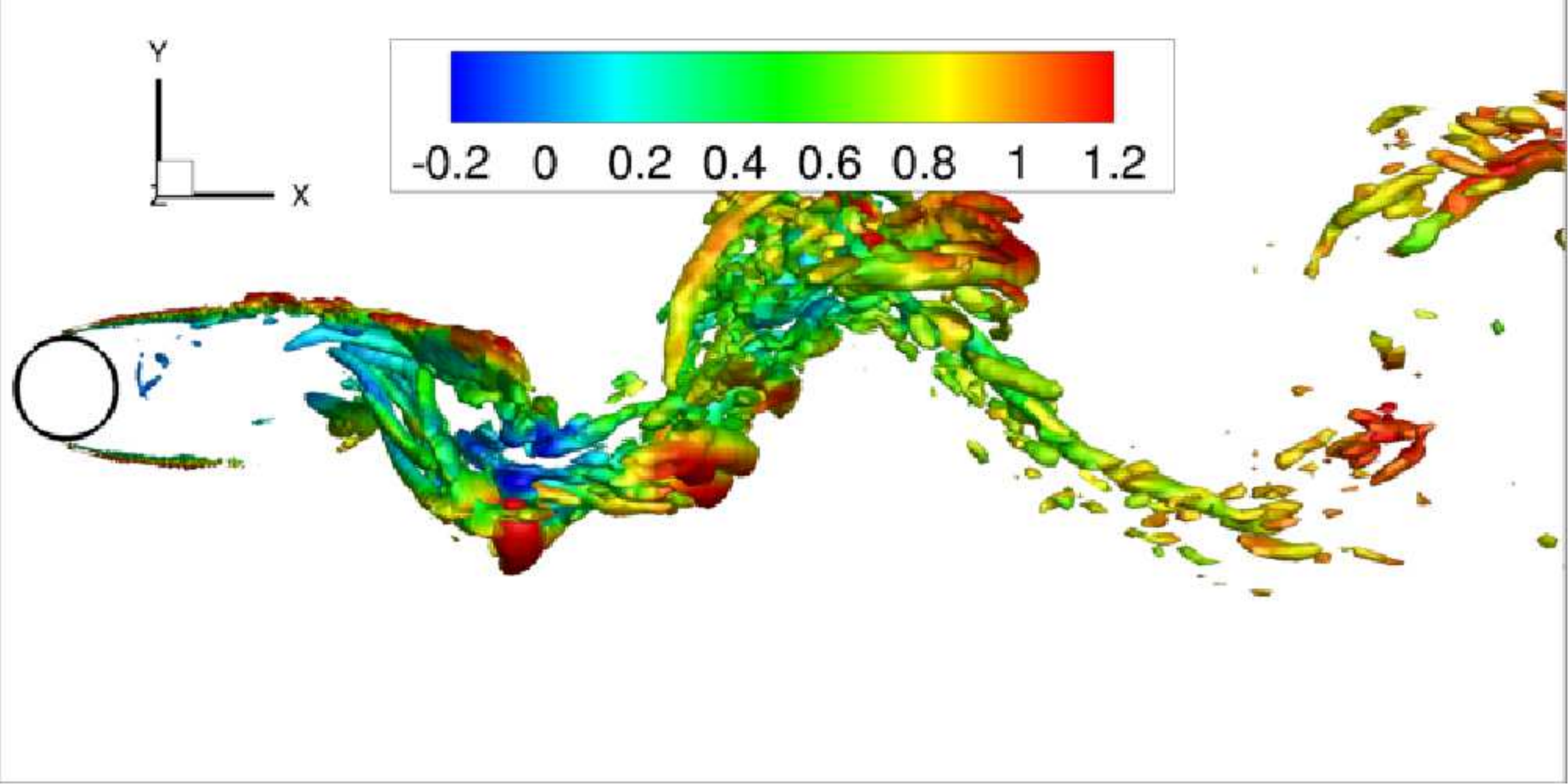}
\put(-100,15){{\textcolor{black}{\vector(-3,2){35}}}}
\put(-120,15){\fcolorbox{black}{white}{\parbox{25mm}{\footnotesize{2D cavity region}}}}
\put(-170,80){$(b)$}
\end{minipage}\hspace{2pc}
\begin{minipage}{13pc}
\includegraphics[width=13pc, trim={0 0.1cm 0.1cm 0}, clip]{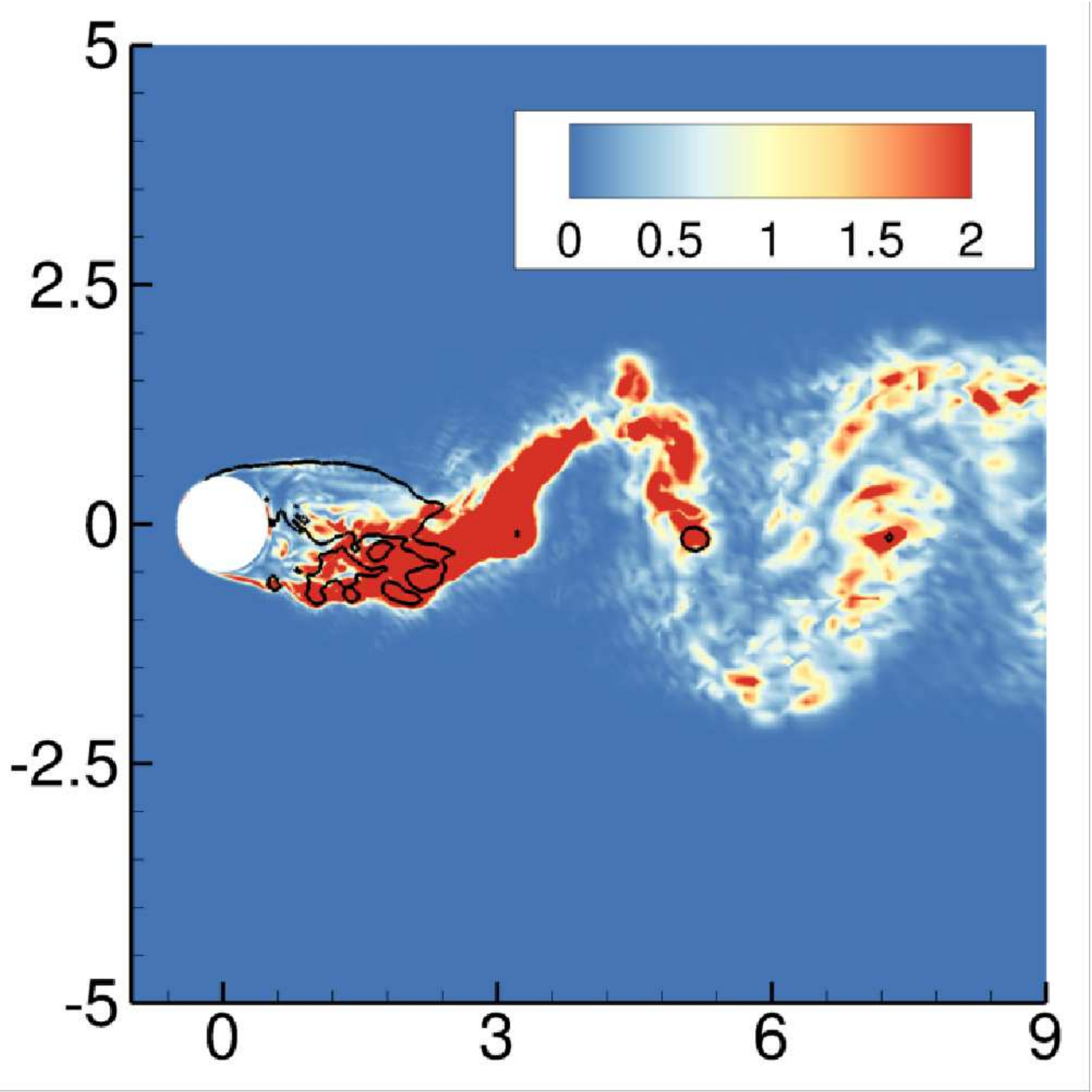}
\put(-170,80){$y/D$}
\put(-170,135){$(c)$}
\end{minipage}\hspace{2pc}
\begin{minipage}{13pc}
\includegraphics[width=13pc, trim={0 0.1cm 0.1cm 0}, clip]{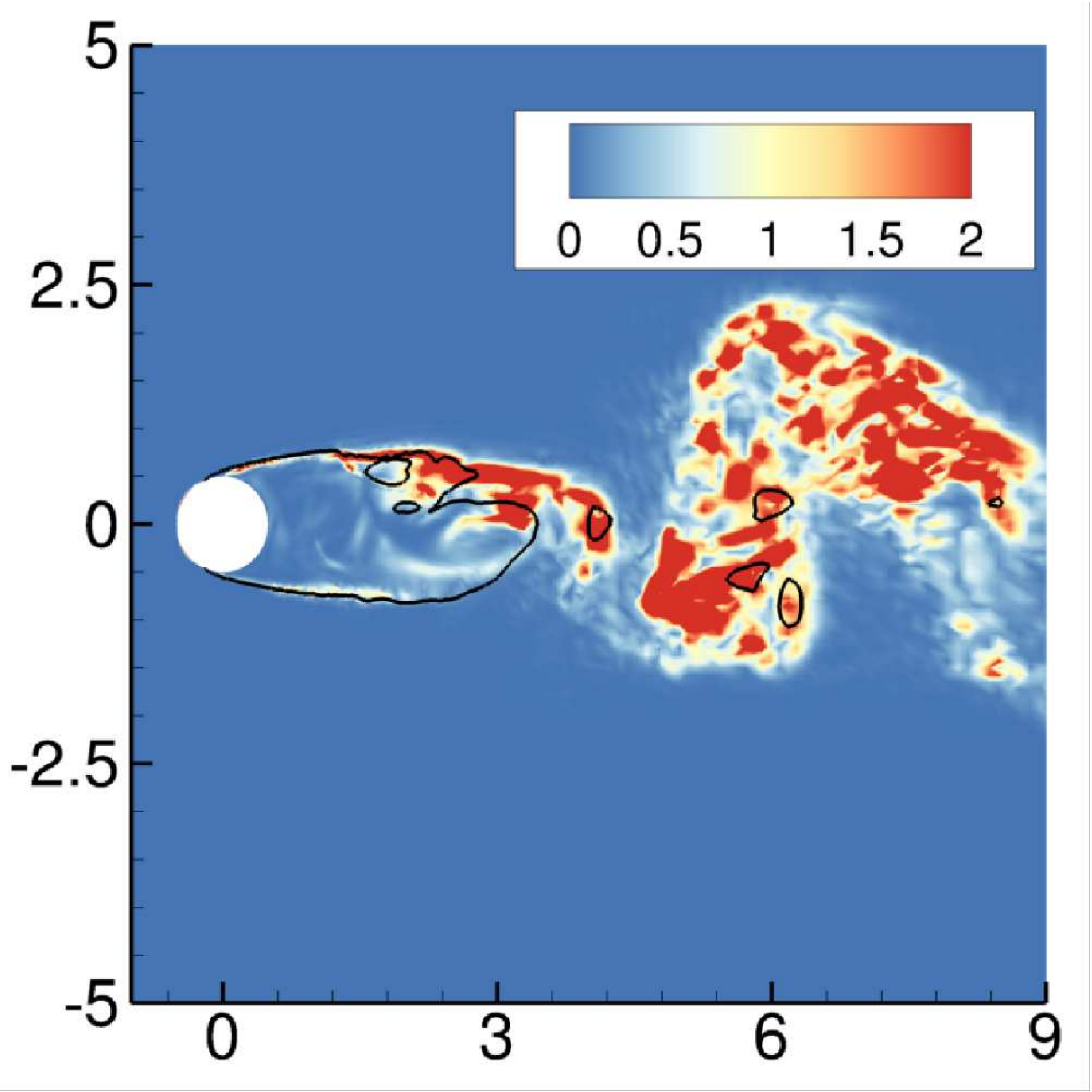}
\put(-170,135){$(d)$}
\put(-170,80){$y/D$}
\end{minipage}\hspace{2pc}
\begin{minipage}{13pc}
\includegraphics[width=13pc, trim={0 0.1cm 0.1cm 0}, clip]{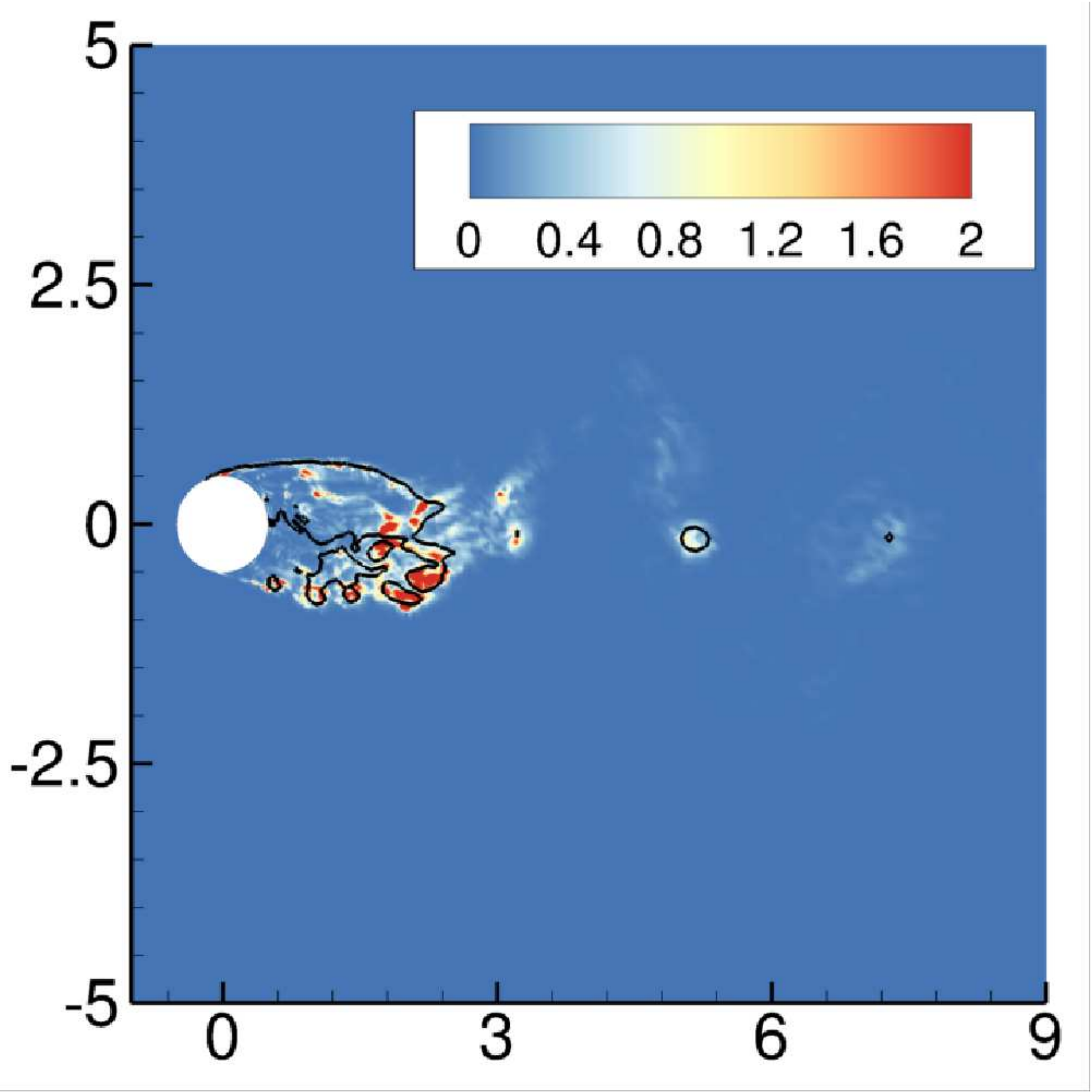}
\put(-80,-9){$x/D$}
\put(-170,80){$y/D$}
\put(-170,135){$(e)$}
\end{minipage}\hspace{2pc}
\begin{minipage}{13pc}
\includegraphics[width=13pc, trim={0 0.1cm 0.1cm 0}, clip]{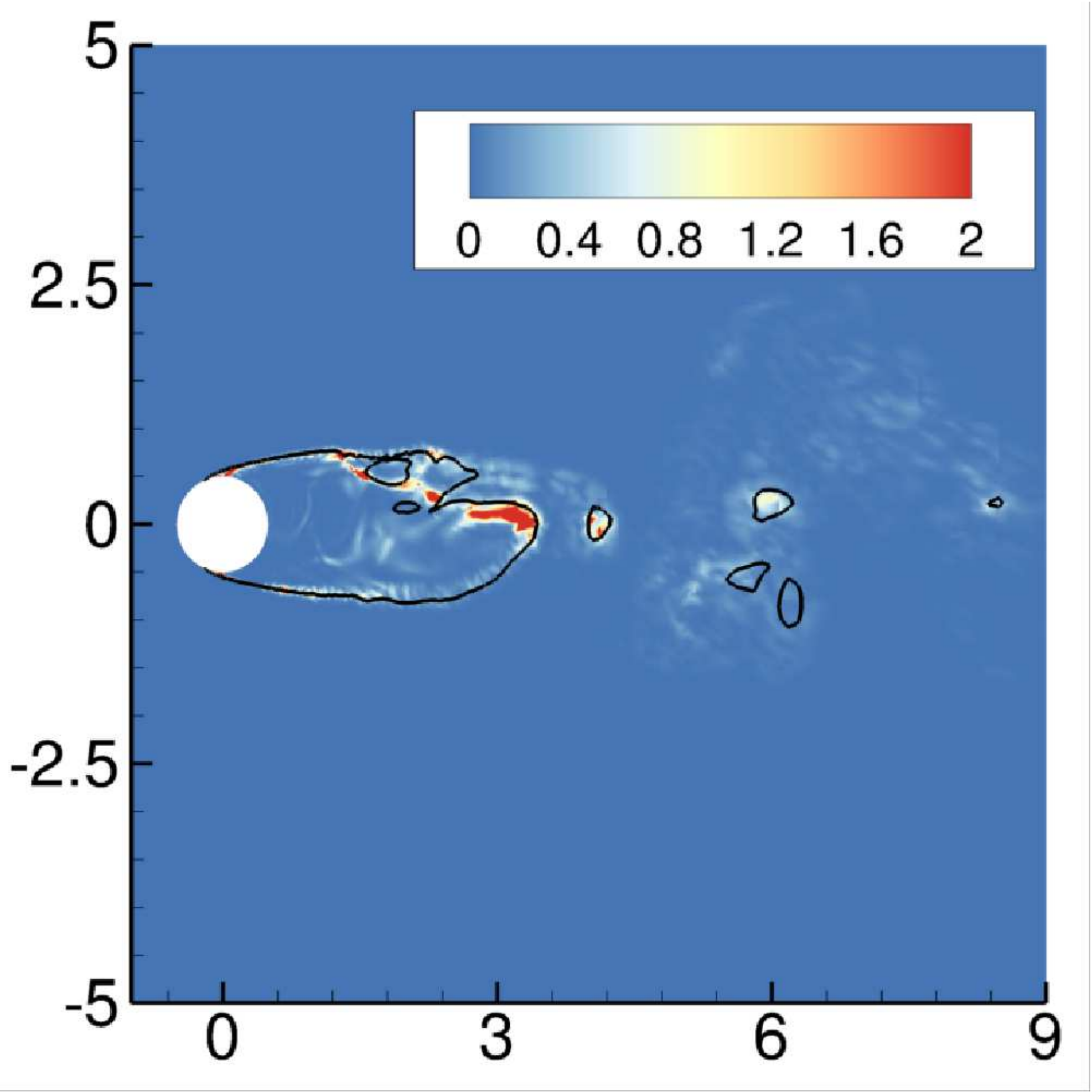}
\put(-80,-9){$x/D$}
\put(-170,80){$y/D$}
\put(-170,135){$(f)$}
\end{minipage}
\caption{Vortex transport for cyclic regime $(a,c,e)$ and transitional regime $(b,d,f)$. Q criterion colored by streamwise velocity $(a,b)$, vorticity stretching/tilting $(c,d)$ and baroclinic torque $(e,f)$. Black lines indicate isosurface of total void fraction of $0.1$ and represent the cavity interface.}
\label{vort_RHS}
\end{figure}

We noted that in the transitional regime, flow cavitates over the entire aft body of the cylinder continuing into the regions in the immediate wake. Consequently, the grown cavity is observed to be nearly two--dimensional with negligible vorticity within the cavity (figure \ref{vort_RHS}$(b)$). Figure \ref{vort_RHS}$(a,b)$ shows a comparison to the cyclic cavitation. In the cyclic regime, significant vorticity is observed in the immediate wake of the cylinder (figure \ref{vort_RHS}$(a)$). Vortex stretching/tilting plotted in figure \ref{vort_RHS}$(c,d)$ confirms this distinction. In the transitional regime, a stable region of incoming shear layer is visible on the either side of the cylinder and majority of stretching/tilting is observed following the cavity closure. Consequently, in the transitional regime, periodic shedding and breakdown of K\'arm\'an vortices is significantly altered. In addition, in the cyclic regime due to the three dimensionality of the flow, significant vorticity production is observed in the near wake by mis--aligned density gradients in the cavity region and the pressure gradients in the pressure waves generated by cavity collapse. In the transitional regime, vorticity production within the 2D cavity is negligible, while vorticity production is observed in the cavity closure at the onset of three dimensionality in the flow. 

\begin{figure}
\centering
\begin{minipage}{12pc}
\includegraphics[width=12pc, trim={0 0.1cm 0.1cm 0}, clip]{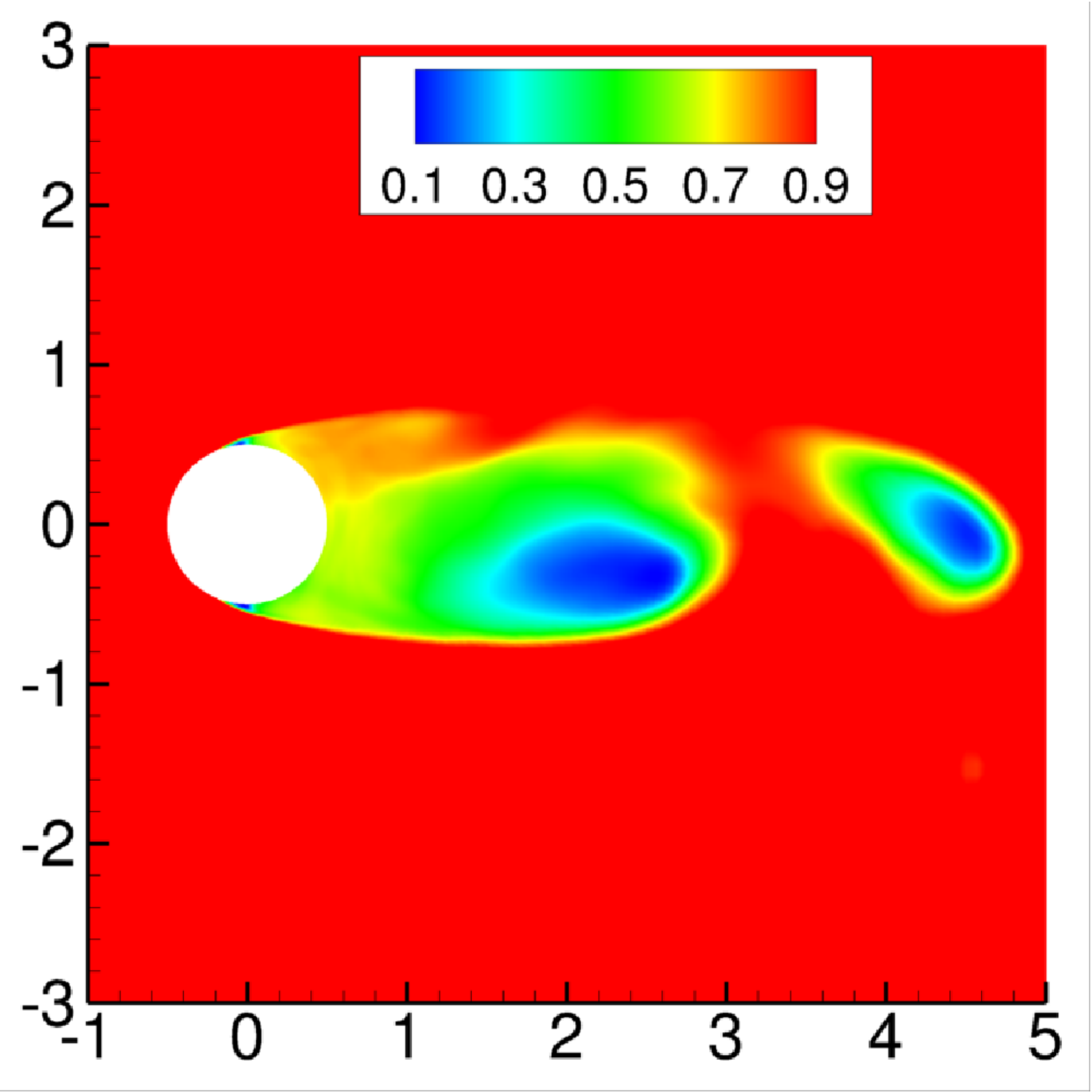}
\put(-77,-8){$x/D$}
\put(-165,70){$y/D$}
\put(-165,140){$(a)$}
\put(-78,100){{\textcolor{black}{\vector(0,-1){15}}}}
\put(-98,100){\fcolorbox{black}{white}{\parbox{20mm}{\footnotesize{$1^{st}$ front}}}}
\end{minipage}\hspace{2pc}
\begin{minipage}{12pc}
\includegraphics[width=12pc, trim={0 0.1cm 0.1cm 0}, clip]{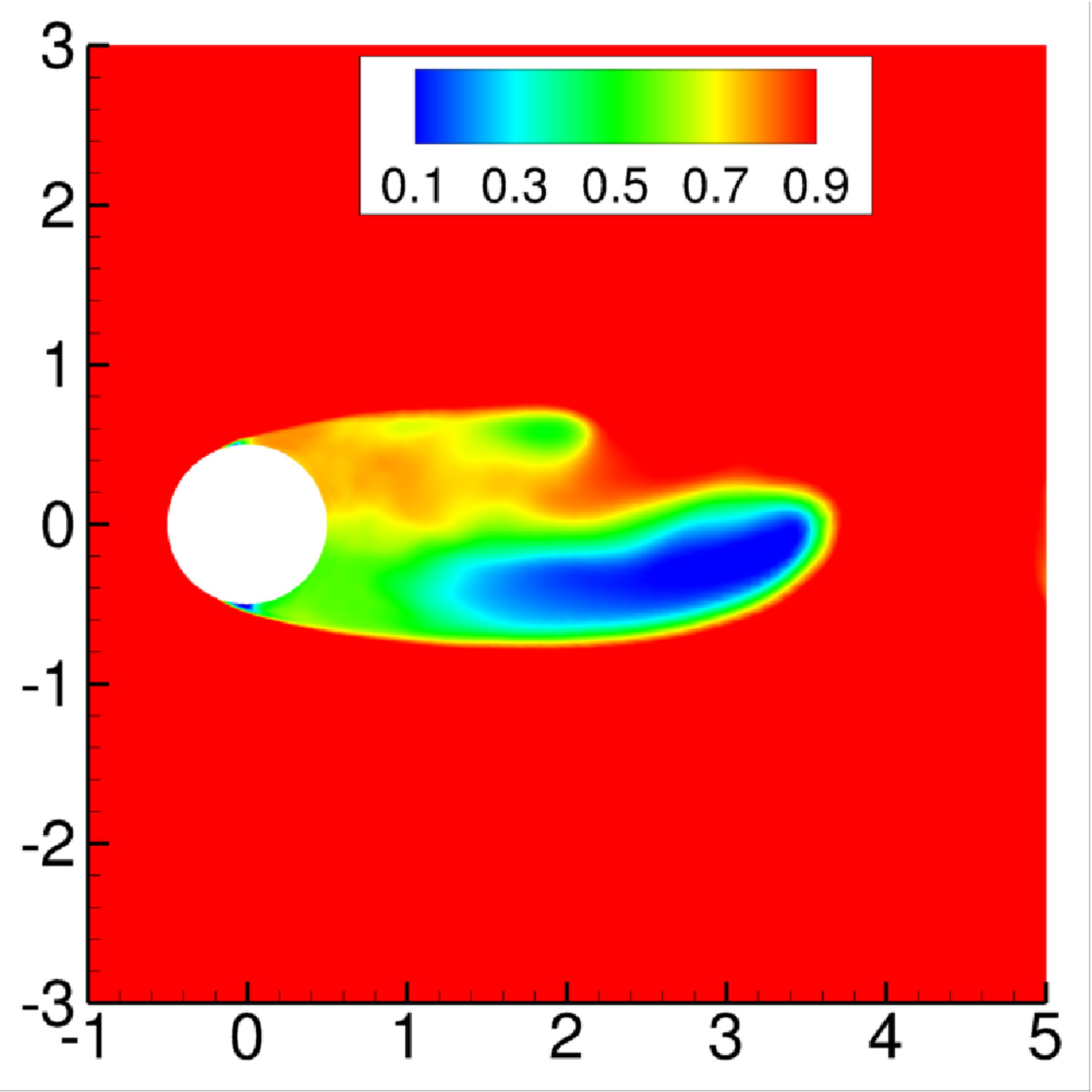}
\put(-77,-8){$x/D$}
\put(-165,70){$y/D$}
\put(-165,140){$(b)$}
\put(-70,100){{\textcolor{black}{\vector(0,-1){25}}}}
\put(-98,100){\fcolorbox{black}{white}{\parbox{20mm}{\footnotesize{$1^{st}$ front}}}}
\end{minipage}
\begin{minipage}{12pc}
\includegraphics[width=12pc, trim={0 0.1cm 0.1cm 0}, clip]{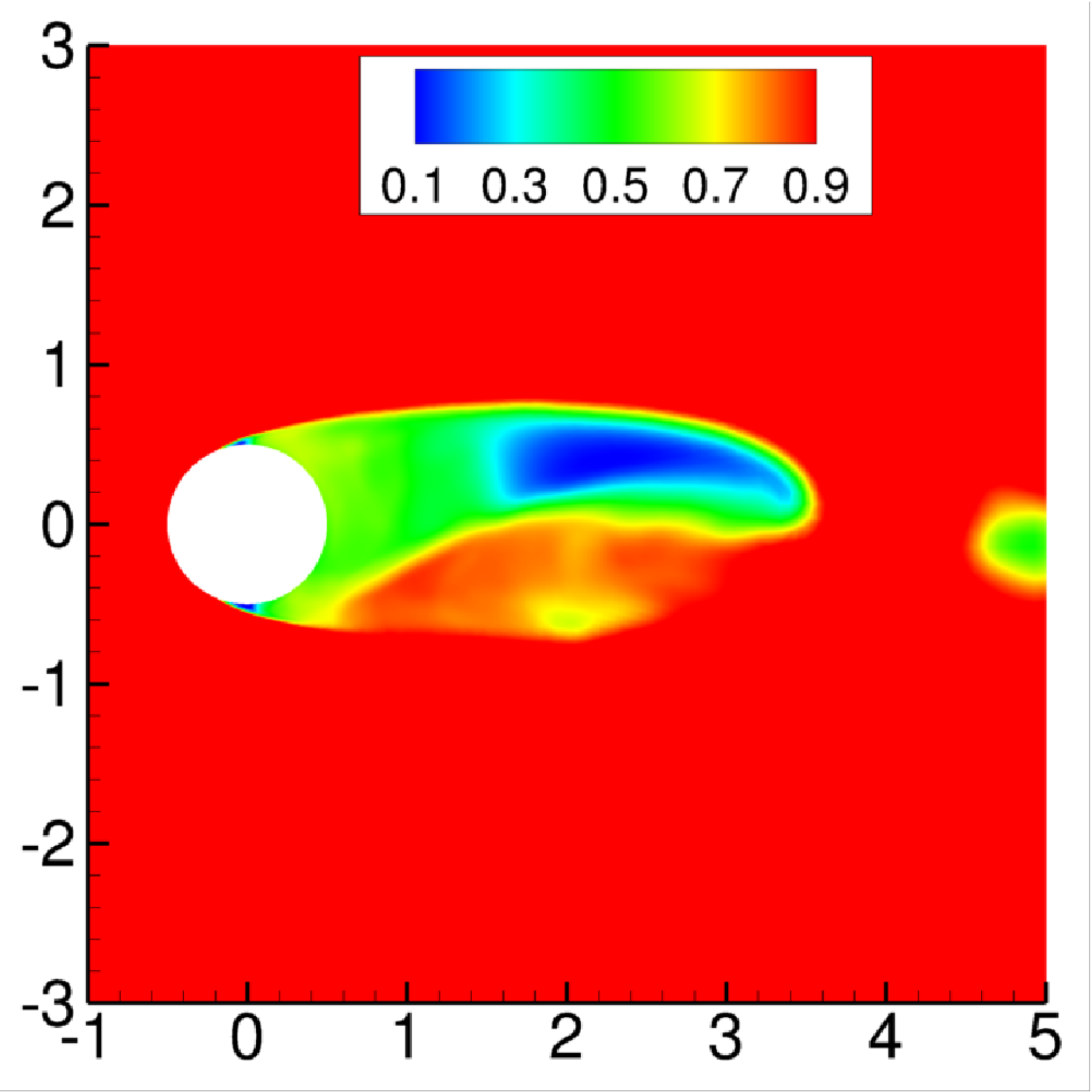}
\put(-77,-8){$x/D$}
\put(-165,70){$y/D$}
\put(-165,140){$(c)$}
\put(-78,57){{\textcolor{black}{\vector(-1,1){12}}}}
\put(-98,47){\fcolorbox{black}{white}{\parbox{20mm}{\footnotesize{$2^{nd}$ front}}}}
\put(-30,15){{\textcolor{black}{\textbf{$\overline{t}=1$}}}}
\end{minipage}\hspace{2pc}
\begin{minipage}{12pc}
\includegraphics[width=12pc, trim={0 0.1cm 0.1cm 0}, clip]{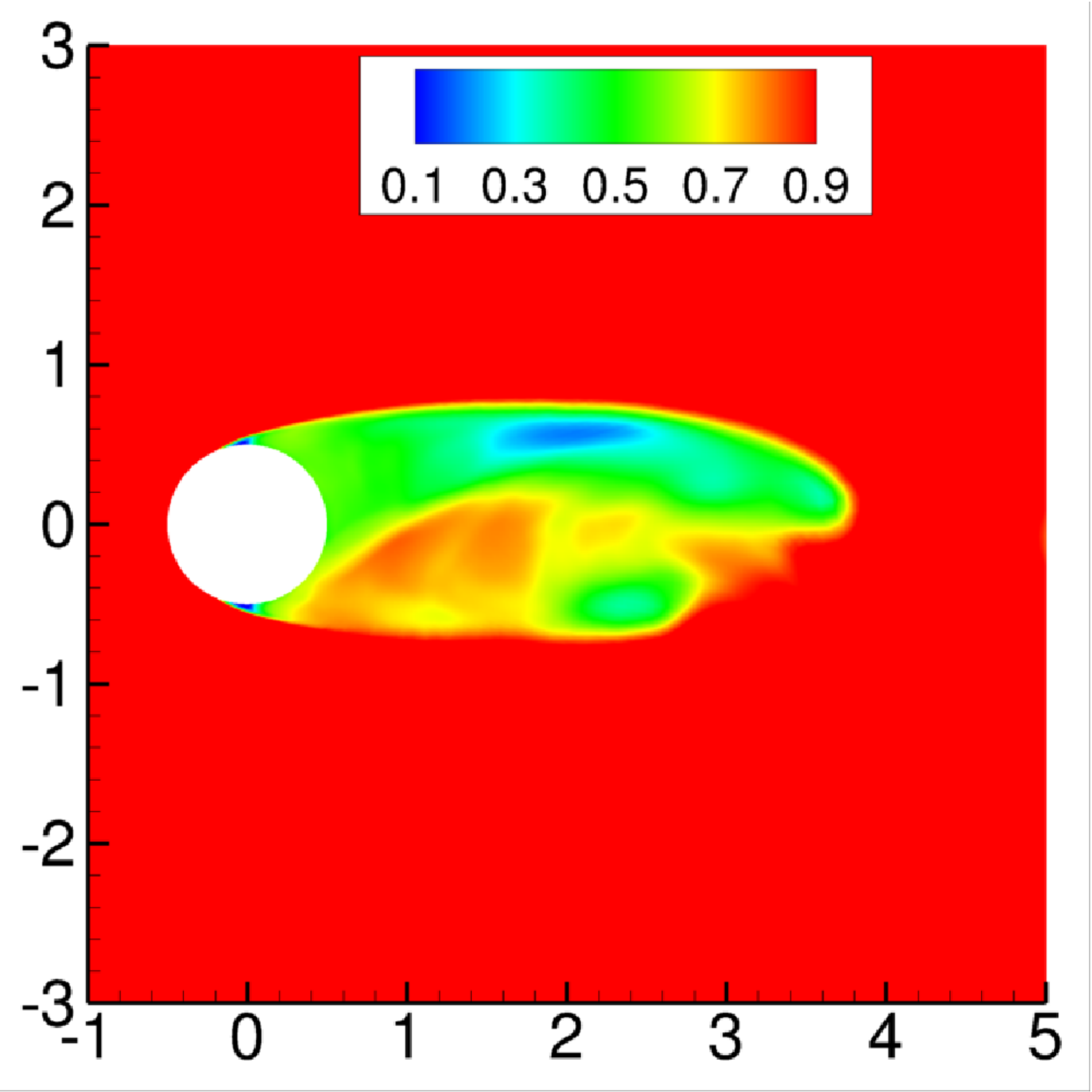}
\put(-77,-8){$x/D$}
\put(-165,70){$y/D$}
\put(-165,140){$(d)$}
\put(-78,57){{\textcolor{black}{\vector(-1,1){15}}}}
\put(-98,47){\fcolorbox{black}{white}{\parbox{20mm}{\footnotesize{$2^{nd}$ front}}}}
\put(-30,15){{\textcolor{black}{\textbf{$\overline{t}=4$}}}}
\end{minipage}
\caption{Spanwise average of density contours showing the propagation of a condensation front at $\sigma=0.7$ and $Re=3900$. Time increases from left to right.}
\label{shock-prop-highRe}
\end{figure}

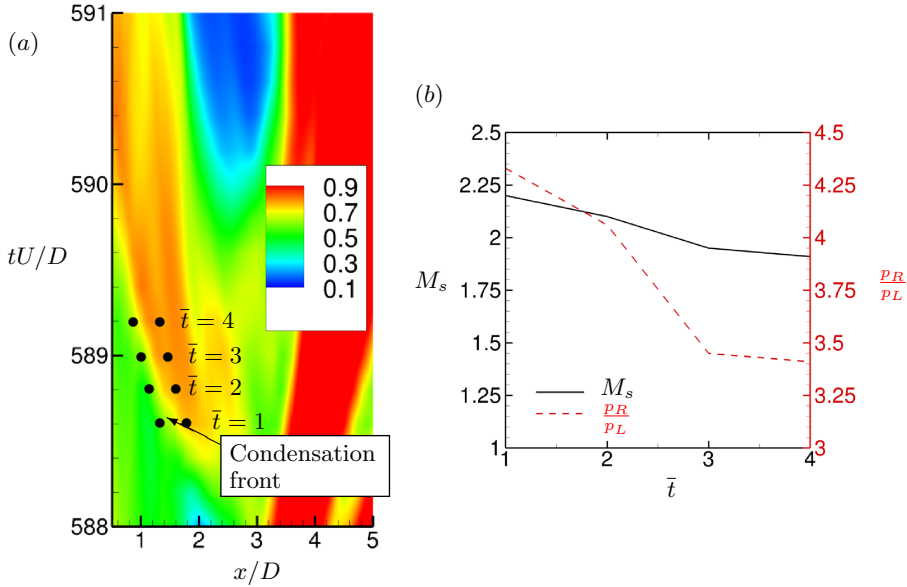
\begin{figure}
\centering
\begin{minipage}{10pc}
\includegraphics[width=10pc, trim={0 0.1cm 0.1cm 0}, clip]{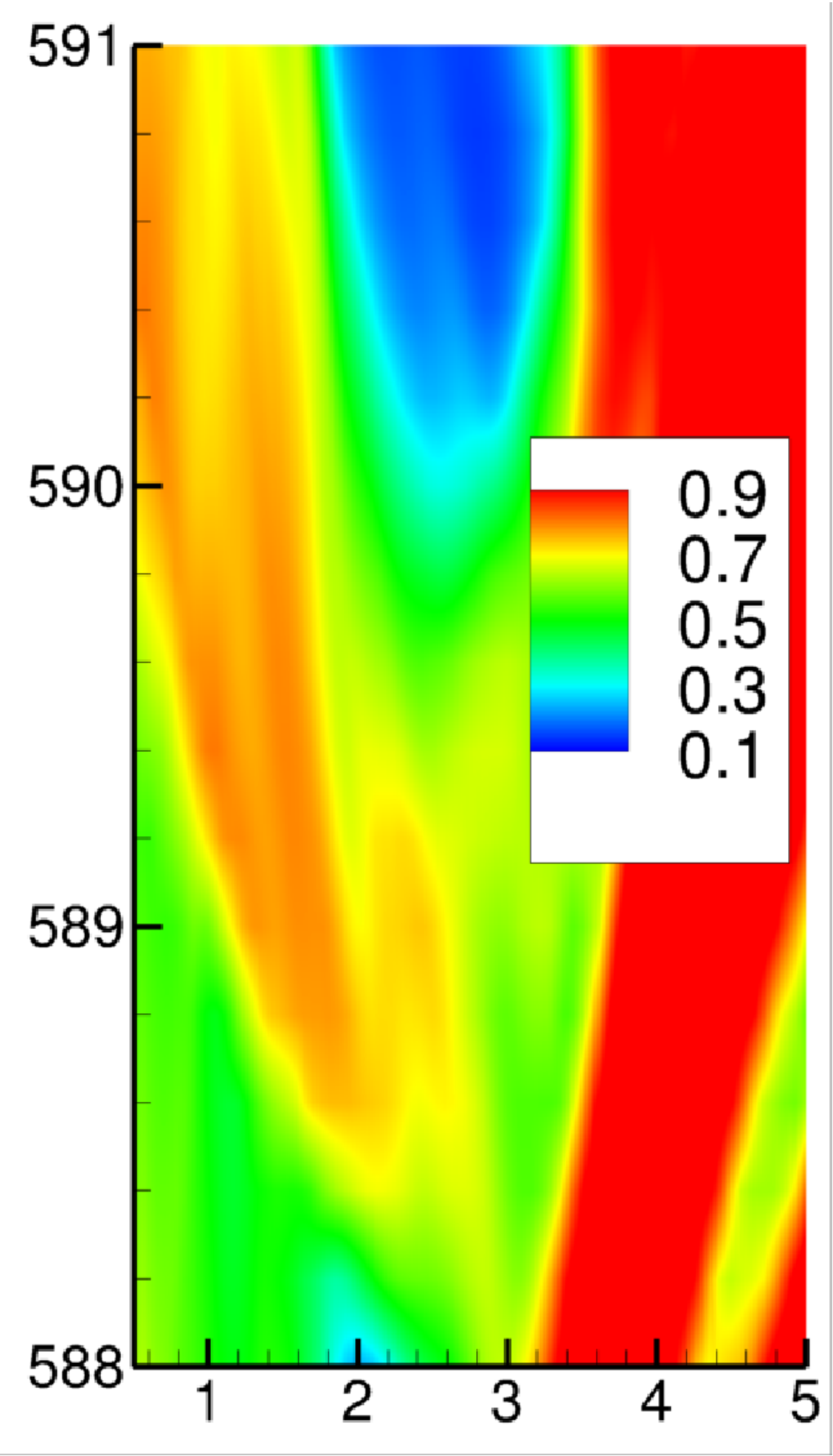}
\put(-55,-8){$x/D$}
\put(-140,110){$tU/D$}
\put(-140,190){$(a)$}
\put(-55,40){{\textcolor{black}{\vector(-2,1){25}}}}
\put(-60,32){\fcolorbox{black}{white}{\parbox{20mm}{\footnotesize{Condensation \\ front}}}}
\put(-85,48){$\bullet$}
\put(-75,48){$\bullet$}
\put(-63,48){{\textcolor{black}{\textbf{$\overline{t}=1$}}}}
\put(-89,61){$\bullet$}
\put(-79,61){$\bullet$}
\put(-72,61){{\textcolor{black}{\textbf{$\overline{t}=2$}}}}
\put(-92,73){$\bullet$}
\put(-82,73){$\bullet$}
\put(-72,73){{\textcolor{black}{\textbf{$\overline{t}=3$}}}}
\put(-95,86){$\bullet$}
\put(-85,86){$\bullet$}
\put(-75,86){{\textcolor{black}{\textbf{$\overline{t}=4$}}}}
\end{minipage}\hspace{2pc}
\begin{minipage}{13pc}
\includegraphics[width=13pc, trim={0 0.1cm 0.1cm 0}, clip]{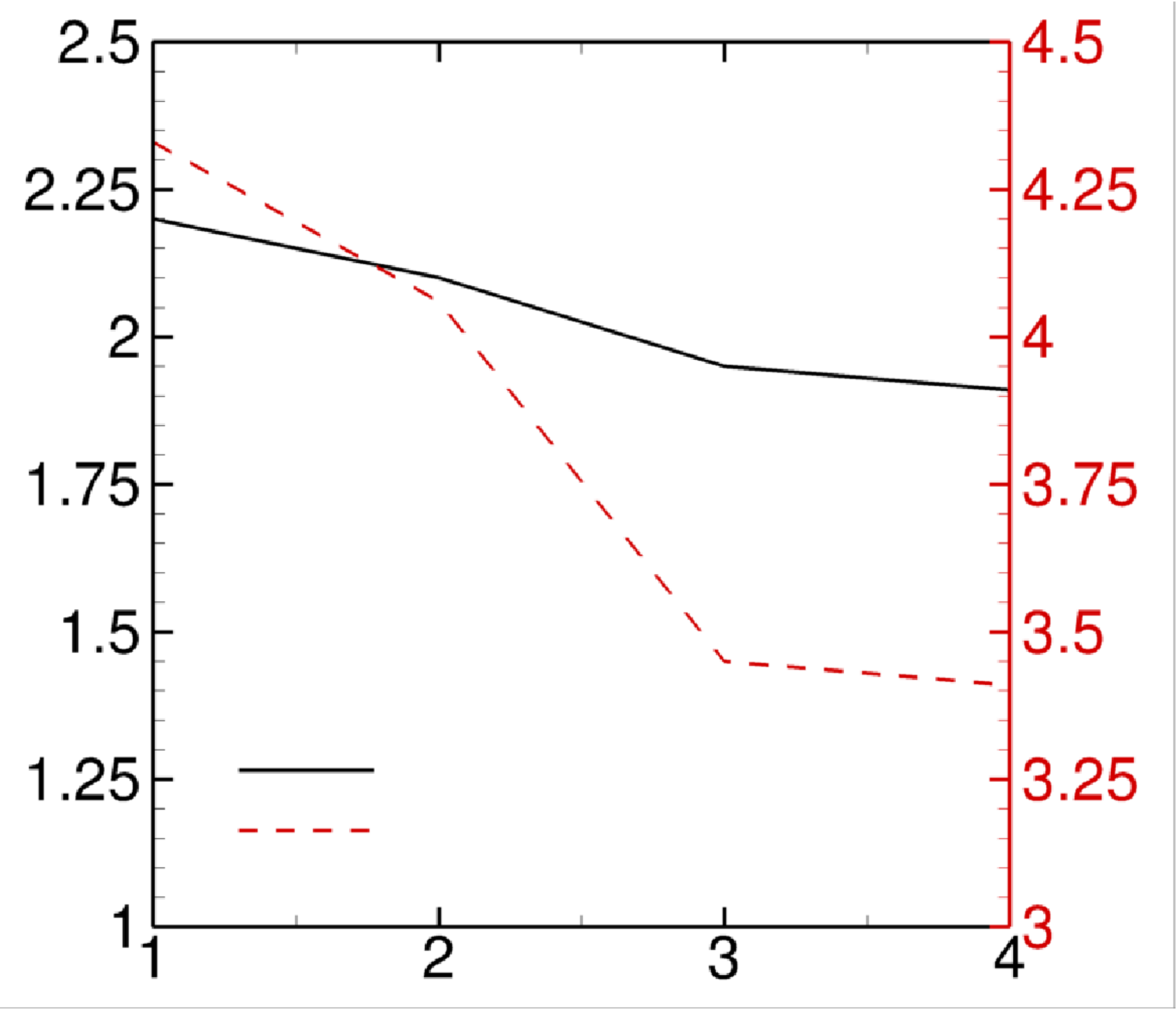}
\put(-75,-8){$\overline{t}$}
\put(-170,70){$M_s$}
\put(4,70){\textcolor{red}{$\frac{p_R}{p_L}$}}
\put(-170,140){$(b)$}
\put(-100,30){$M_s$}
\put(-100,20){\textcolor{red}{$\frac{p_R}{p_L}$}}
\end{minipage}
\caption{$x-t$ plot of spanwise averaged density $(a)$ and condensation front Mach number and pressure ratio $(b)$ at $\sigma=0.7$ and $Re=3900$.}
\label{Ms_pratio_highRe}
\end{figure}

In the transitional regime, at $Re=200$, we observed that in the presence of \textit{NCG} propagation of multiple condensation shocks leading to complete cavity detachment. The same is also observed at $Re=3900$ as illustrated in figure \ref{shock-prop-highRe} using spanwise averaged density contours at multiple time instances. 
Compared to $Re=200$, the condensation front at $Re=3900$ is oriented more vertically rather than towards the cylinder trailing edge. Note the first front moving in the direction of negative $y-axis$ and the second front moving in the direction of positive $y-axis$ and also oriented towards the cylinder surface. Time evolution of condensation front propagation is shown using $x-t$ diagram in figure \ref{Ms_pratio_highRe}$(a)$, only the section front moving towards the cylinder surface is considered. Initial and final time instances of the propagation are displayed in figure \ref{Ms_pratio_highRe}$(c)$ and \ref{Ms_pratio_highRe}$(d)$. Mach number and the pressure ratio across the condensation front are computed at the time instances indicated in figure \ref{Ms_pratio_highRe}$(a)$ and are displayed in figure \ref{Ms_pratio_highRe}$(b)$. Condensation front moves at supersonic speeds at all the instances considered, and consequently is a condensation shock wave. Also, note that the condensation shock is weakened as it propagates and approaches cylinder surface as indicated in figure \ref{Ms_pratio_highRe} by reduction in the pressure ratio.  

\section{Summary}\label{Summ}

The numerical method of \cite{AswinIJMF} to study cavitating flows based on homogeneous mixture of water--vapor is extended to include {\it NCG}. Cavitating flow over a circular cylinder is studied for a range of $\sigma$ showing both cyclic and transitional cavitation regimes at different Reynolds number and with different amounts of freestream vapor and gas volume fraction. {\it NCG} is introduced in the freestream as a free gas (prescribed in a similar way as vapor nuclei) and its effect on the flow field is discussed. In the cyclic regime, cavitation was observed in periodically shedding K\'arm\'an vortices; while in the transitional regime, the cavity shedding was observed due to propagating condensation shocks. 

$St$ based on cavity length when plotted with $\sigma$ shows a peak just before the onset of condensation shock wave induced cavity shedding. Cavity shedding frequency obtained from the drag history showed similar behavior in the presence of {\it NCG}. At $\sigma = 0.85$ (near transition), however, $FFT$ of pressure history did not show a secondary peak due to low frequency detachment in the presence of {\it NCG}, suggesting its influence on delaying the transition to low frequency shedding behavior. DMD used to investigate wake characteristics revealed that cavitation delays the first transition of the K\'arm\'an vortex street. Reduction in shedding frequency as $\sigma$ is lowered from non--cavitating to cavitating conditions is used to explain this behavior.

It was observed that vapor and gas uniformly introduced in the freestream, distribute themselves differently in the wake of cylinder depending upon local flow conditions, particularly at lower cavitation numbers as the pressure in the wake dropped below vapor pressure. This was explained using $\sigma_{local}$, to distinguish vapor production due to phase change as compared to expansion of vapor or gas. Vapor and {\it NCG} distribution in the boundary layer suggested that cavitation as a mass transfer process only occurs inside a fine layer in the near wall region, while the remaining of the boundary--layer only undergoes expansion of both vapor and gas. Freestream void fraction was shown to have a large impact on the mean gas volume fraction observed inside the cavity. However, mean vapor volume fraction seems relatively independent. The boundary--layer separation point for both laminar and turbulent flow was observed to move downstream as the freestream volume fraction is increased. 

Rankine--Hugoniot jump conditions were derived for the complete system. The shock speed obtained using the jump conditions is used to compute the Mach number of propagating condensation front to show that it is indeed supersonic for both $Re$ studied. It is however noted that the condensation shock propagates into a subsonic region of the cavity, based on local Mach numbers.
In the presence of {\it NCG}, it was shown that the strength of the condensation shock reduces as it approaches the cylinder surface. This results in multiple condensation shocks being necessary for detachment of cavity. Weakening of shock strength in the presence of gas was due to reduction in pressure ratio across the condensation shock as it approaches the cylinder surface. As the shock moves upstream towards the cylinder, it condenses the vapor along the way. However, the fact that it can not condense the {\it NCG} resulted in a lower pressure behind the shock, which was shown using the jump condition. The conditions necessary for the occurrence of supersonic condensation front is, then, assessed using its pressure ratio.

At $Re=3900$, it was observed that the location of maximum vapor production in the cyclic regime is shifted from the cylinder surface to the immediate wake, with no major changes to the levels of volume fraction inside the cavity. The mean {\it NCG} volume fraction, however, reduces by an order of magnitude inside the cavity when compared to the mean values at $Re=200$. The growth of a nearly two--dimensional cavity in the transitional regime, significantly reduces vortex stretching and baroclinic torque from the values observed in the cyclic regime. Finally, it is observed that at $Re=3900$ for the transitional regime, the cavity detaches after the passage of a more vertically oriented condensation shock, different from the horizontally oriented shock at $Re=200$.

\section*{Acknowledgment}\label{Ackn}

This work is supported by the United States Office of Naval Research under Grant ONR N00014--17-1--2676 with
Dr. Ki--Han Kim as the program manager. Computing resources were provided by the Minnesota Supercomputing
Institute (MSI) and the High Performance Computing Modernization Program (HPCMP).

\appendix 

\section{Analysis of condensation shock using Rankine--Hugoniot jump conditions}\label{RH}

Analysis of shock waves in bubbly flows of liquid--gas mixture was considered by \cite{Campbell}, who related shock propagation speed to the pressure in the high--pressure side of the shock, the density of the liquid, and relative proportions of gas and liquid. They also showed negligible temperature rise across steady condensation shock waves. Here, we consider the current system of homogeneous mixture with both vapor and gas mass transport closed with the mixture equation of state and consider a left--moving shock (moving upstream of the flow direction) in a frame of reference moving with the shock as described in figure \ref{shock-diag}. The velocities in the moving reference frame are
\begin{equation}
\label{shock}
\hat{u}_{L}=u_{L}-S \quad  \textrm{and} \quad \hat{u}_{R}=u_{R}-S.
\end{equation}
Here, S is the shock speed. ``\char`\^" is used to indicate quantities in moving reference frame. ``$L$" and ``$R$" subscripts are used respectively for the quantities at the left and at the right of the shock. Since we are considering a left--moving shock, left and right sides of the shock become ahead and behind the shock respectively.

\begin{figure}
\centering
\setlength{\unitlength}{1cm}
\thicklines
\begin{picture}(6,4)(0,0)

\put(1,0){\line(0,1){2.5}}
\put(0.9,2.8){S}
\put(1.5,2.6){\vector(-1,0){1}}
\put(0.5,2.2){$\rho_{L}$}
\put(0.5,1.7){$p_{L}$}
\put(0.5,1.2){$u_{L}$}
\put(0.4,0.7){$Y_{vL}$}
\put(0.4,0.2){$Y_{gL}$}

\put(1.2,2.2){$\rho_{R}$}
\put(1.2,1.7){$p_{R}$}
\put(1.2,1.2){$u_{R}$}
\put(1.2,0.7){$Y_{vR}$}
\put(1.2,0.2){$Y_{gR}$}
\put(-0.5,2.2){$(a)$}

\put(5,0){\line(0,1){2.5}}
\put(4.9,2.8){0}
\put(4.5,2.2){$\rho_{L}$}
\put(4.5,1.7){$p_{L}$}
\put(4.5,1.2){$\hat{u}_{L}$}
\put(4.4,0.7){$Y_{vL}$}
\put(4.4,0.2){$Y_{gL}$}

\put(5.2,2.2){$\rho_{R}$}
\put(5.2,1.7){$p_{R}$}
\put(5.2,1.2){$\hat{u}_{R}$}
\put(5.2,0.7){$Y_{vR}$}
\put(5.2,0.2){$Y_{gR}$}
\put(3.5,2.2){$(b)$}

\end{picture}
\caption{Left moving shock in a stationary frame of reference $(a)$ and in a frame of reference moving with the shock $(b)$.}
\label{shock-diag}
\end{figure}

The Rankine--Hugoniot jump conditions in a frame of reference moving with the shock are:
\begin{equation}
\label{cont_movingframe}
\rho_{R} \hat{u}_{R} = \rho_{L} \hat{u}_{L}, 
\end{equation}
\begin{equation}
\label{mome_movingframe}
\rho_{R} \hat{u}_{R}^2 + p_{R} = \rho_{L} \hat{u}_{L}^2 + p_{L}, 
\end{equation}
\begin{equation}
\label{ener_movingframe}
\rho_{R} \hat{u}_{R}(e_{R} + p_{R}/\rho_{R} + \hat{u}_{R}^2/2) = \rho_{L} \hat{u}_{L}(e_{L} + p_{L}/\rho_{L} + \hat{u}_{L}^2/2), 
\end{equation}
\begin{equation}
\label{vap_movingframe}
\rho_{R} Yv_{R} \hat{u}_{R} = \rho_{L} Yv_{L} \hat{u}_{L} \quad \textrm{and} 
\end{equation}
\begin{equation}
\label{ncg_movingframe}
\rho_{R} Yg_{R} \hat{u}_{R} = \rho_{L} Yg_{L} \hat{u}_{L}. 
\end{equation}
Here, subscripts \textit{v} and \textit{g} in equations (\ref{vap_movingframe}) and (\ref{ncg_movingframe}) denote vapor and \textit{NCG} respectively. Note that phase change between vapor and water is not explicitly modeled, however, its effects are implicitly calculated since vapor mass fraction will have different values across the shock.

Applying equation (\ref{cont_movingframe}) to equations (\ref{mome_movingframe}) and (\ref{ener_movingframe}) we have
\begin{equation}
\label{ur_square}
\hat{u}_{R}^2 = \frac{\rho_{L}}{\rho_{R}} \frac{(p_{R}-p_{L})}{(\rho_{R}-\rho_{L})}, 
\end{equation}
\begin{equation}
\label{ul_square}
\hat{u}_{L}^2 = \frac{\rho_{R}}{\rho_{L}} \frac{(p_{R}-p_{L})}{(\rho_{R}-\rho_{L})} \quad \textrm{and}
\end{equation}
\begin{equation}
\label{ener_movingframe_new}
e_{R} + p_{R}/\rho_{R} + \hat{u}_{R}^2/2 = e_{L} + p_{L}/\rho_{L} + \hat{u}_{L}^2/2. 
\end{equation}
Substituting equation (\ref{ur_square}) and (\ref{ul_square}) into equation (\ref{ener_movingframe_new}) we have an equation for the energy difference across the shock,
\begin{equation}
\label{ener_diff}
e_{R} - e_{L} = \frac{1}{2} \frac{(p_{R}+p_{L})(\rho_{R}-\rho_{L})}{\rho_{R}\rho_{L}}. 
\end{equation}
From equations (\ref{eos}) and (\ref{mixture internal energy}), the mixture internal energy can be written as
\begin{equation}
\label{ener_eos}
e = \frac{C_{vm} p^2 +[C_{vm}+(1-Y_v-Y_g)K_l] P_c p}{[\rho (Y_v R_v + Y_g R_g)(p + P_c) + \rho (1-Y_v-Y_g) K_l p]},
\end{equation}
and can be simplified as
\begin{equation}
\label{ener_simpl}
e = \frac{p}{\rho (\overline{\gamma}-1)},  \quad \textrm{where} 
\end{equation}
\begin{equation}
\label{gama_bar}
\frac{1}{\overline{\gamma}-1} = \frac{C_{vm} p +[C_{vm}+(1-Y_v-Y_g)K_l] P_c}{[(Y_v R_v + Y_g R_g)(p + P_c) + (1-Y_v-Y_g) K_l p]}. 
\end{equation}
Using equation (\ref{ener_simpl}) in equation (\ref{ener_diff}) followed by algebraic simplification we obtain an equation for the density ratio across the shock as 
\begin{equation}
\label{dens_ratio}
\frac{\rho_R}{\rho_L} = \frac{\frac{p_R}{p_L} \frac{\overline{\gamma}_R+1}{\overline{\gamma}_R-1} + 1}{\frac{p_R}{p_L} + \frac{\overline{\gamma}_L+1}{\overline{\gamma}_L-1}}. 
\end{equation}
With equations (\ref{shock}), (\ref{cont_movingframe}), (\ref{ul_square}) and (\ref{dens_ratio}) an equation for the shock speed can be derived:
\begin{equation}
\label{shock_speed}
S = u_L - \sqrt{\frac{(p_R - p_L)[\frac{p_R}{p_L} \frac{\overline{\gamma}_R+1}{\overline{\gamma}_R-1} + 1]}{(\rho_R - \rho_L)[\frac{p_R}{p_L} + \frac{\overline{\gamma}_L+1}{\overline{\gamma}_L-1}]}}. 
\end{equation}
Note that in the single--phase limit with $Y_g=1$, $Y_v=0$ and $1-Y_v-Y_g=0$, equation (\ref{shock_speed}) simplifies into the classical gasdynamics equation (e.g. \cite{Toro}). 

\section{Temperature jump relation across condensation shock}\label{app_tjump}
Is is known that temperature variations in hydrodynamic cavitation are mostly negligible due to the high specific heat capacity of liquid. However, temperature variations are observed to increase in developed cavitation involving mass transfer in large cavities \citep{holltemp}. Therefore, the propagation of a condensation shock might be expected to lead to larger temperature fluctuations. Here, we use results from the simulations at $Re=200$ and $\sigma=0.7$ along with an equation for the temperature ratio across the condensation shock, to show that temperature variation in this process is also negligible.

\begin{figure}
\centering
\includegraphics[width=12pc, trim={0 0.1cm 0.1cm 0}, clip]{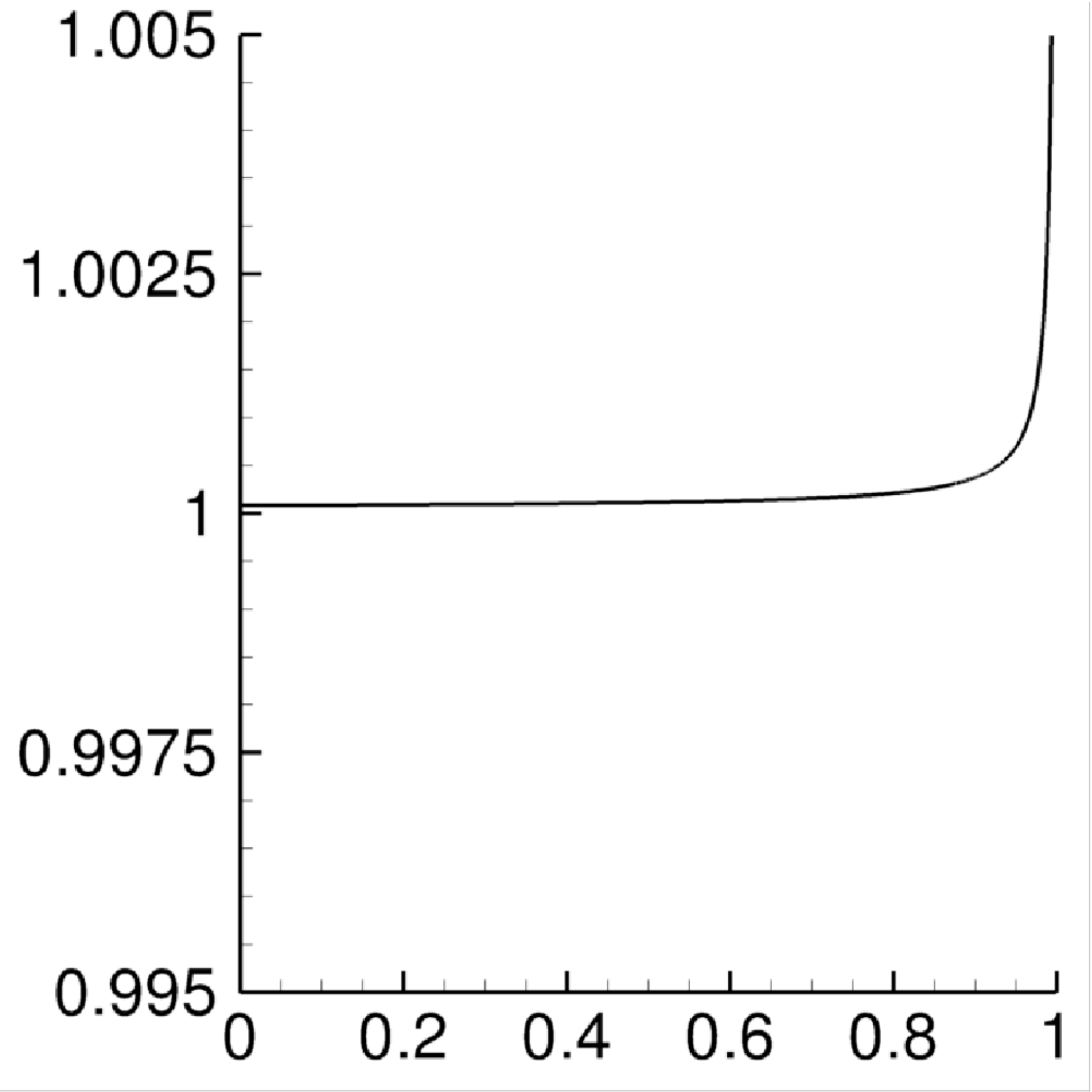}
\put(-65,-7){$\alpha_L$}
\put(-160,70){$\frac{T_R}{T_L}$}
\caption{Temperature ratio across a condensation shock for different amounts of gaseous phase ahead of shock.}
\label{tempratio_parametric}
\end{figure}

Temperature ratio across the condensation shock can be derived from the density relation given by equation (\ref{dens_ratio}) and the mixture equation of state (equation (\ref{eos})). It is given by

\begin{equation}
\label{temp_ratio}
\begin{aligned}
\frac{T_L}{T_R} = \frac{[\frac{2b_R}{a_R}+1+\frac{p_L}{p_r}](p_L+P_c)a_R}{[\frac{p_R}{p_L}+\frac{2b_L}{a_L}+1](p_R+P_c)a_L}, \quad \text{where} \\
a = (Y_v R_v + Y_g R_g)(p + P_c) + (1-Y_v-Y_g) K_l p \quad \textrm{and} \\
b = C_{vm} p +[C_{vm}+(1-Y_v-Y_g)K_l] P_c. \\ 
\end{aligned}
\end{equation}
Note from equation (\ref{temp_ratio}) that considering different values of vapor and gas mass fraction for left and right implies that, although the temperature jump is of comparable magnitude to pressure ratio when having completely gaseous phases on both sides, the ratio is nearly unity in the presence of liquid. We investigate this by considering a scenario where shock is moving from water to a mixture of water--vapor--gas with increasing amount of gaseous phase void fraction. Pressure in the liquid is considered 1 atm and pressure inside the mixture is chosen to be vapor pressure. This mimics the scenario of condensation shock propagating through the cavity. Temperature jump obtained from the equation (\ref{temp_ratio}) for increasing amount of gaseous phase void fraction is plotted in figure \ref{tempratio_parametric}. The figure shows that temperature ratio for any amount of gaseous void fraction ahead of the shock is negligible. The maximum ratio in the plot is ${T_R}/{T_L} =1.005$ for $\alpha_L = 0.994$ (which is much higher than the maximum void fractions reached in practical applications). The same conclusions would hold even in the presence of some amount of gaseous void fraction mixed with the liquid behind the shock (not shown here).  

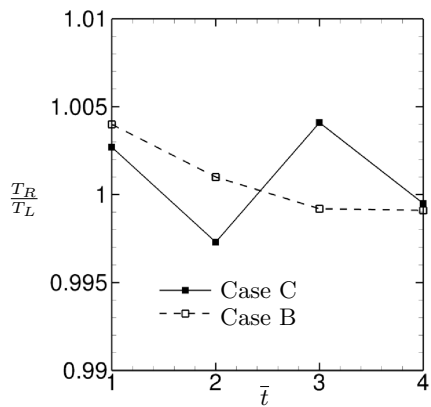
\begin{figure}
\centering
\includegraphics[width=12pc, trim={0 0.1cm 0.1cm 0}, clip]{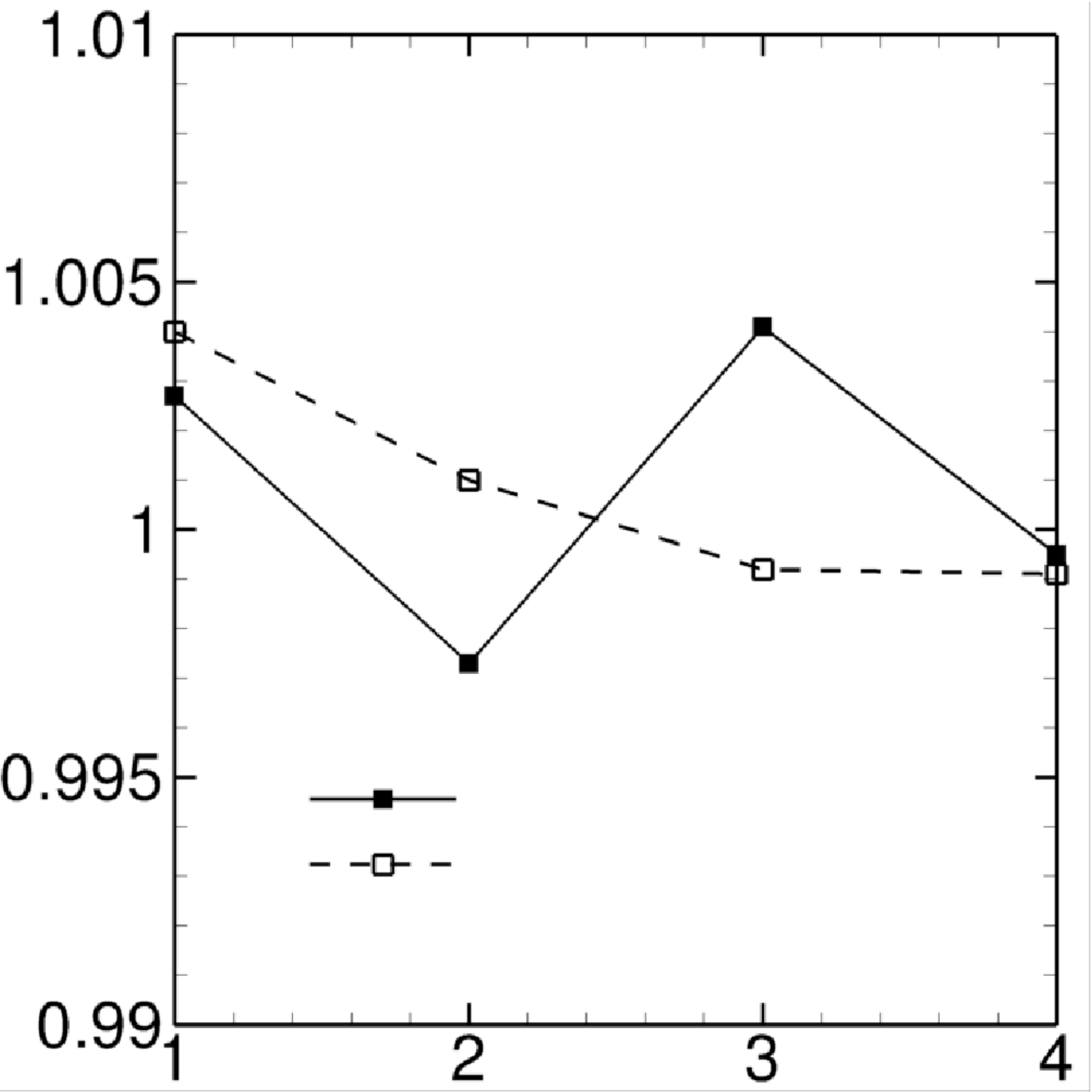}
\put(-160,70){$\frac{T_R}{T_L}$}
\put(-65,-5){$\overline{t}$}
\put(-80,35){Case C}
\put(-80,25){Case B}
\caption{Temperature ratio across a condensation shock for $\sigma=0.7$.}
\label{temp_ratio-sigma-0.7}
\end{figure}

The isothermal behavior of the condensation shock is confirmed in our simulations. Here, we obtain the temperature jump across the condensation shock observed for $\sigma=0.7$ at time instances mentioned in figure \ref{xt-rho-sigma-0.7}. Figure \ref{temp_ratio-sigma-0.7} shows the temperature ratio both in presence and absence of gas. It is evident that the condensation shock in the current calculation is nearly isothermal.  

\bibliography{jfm-rebuttal}
\bibliographystyle{jfm}

\end{document}